\documentclass[useAMS,usenatbib,usegraphicx]{mn2e}
\usepackage{amsmath}
\usepackage{pifont}
\usepackage{color}
\usepackage{multirow}

\newif\ifAMStwofonts
\AMStwofontstrue

\newcommand{\simlt}{\lower.5ex\hbox{$\; \buildrel < \over \sim \;$}}
\newcommand{\simgt}{\lower.5ex\hbox{$\; \buildrel > \over \sim \;$}}

\newcommand{\cmark}{\ding{51}}%
\newcommand{\xmark}{\ding{55}}%

\title[FIGS: Massive galaxies since Cosmic Noon]
{FIGS: Spectral fitting constraints on the star formation history of massive galaxies
since Cosmic Noon}
\author[I. Ferreras et al.]
{Ignacio Ferreras$^1$\thanks{E-mail: i.ferreras@ucl.ac.uk}, 
Anna Pasquali$^2$, Nor Pirzkal$^3$, John Pharo$^4$, \and
Sangeeta Malhotra$^{4,5}$, James Rhoads$^{4,5}$, Nimish Hathi$^3$, Rogier Windhorst$^4$,\and
Andrea Cimatti$^{6,7}$,  
Lise Christensen$^8$, 
Steven L. Finkelstein$^{9}$,
Norman Grogin$^3$,\and
Bhavin Joshi$^4$, 
Keunho Kim$^4$, 
Anton Koekemoer$^3$,
Robert O'Connell$^{10}$, \and
G\"oran \"Ostlin$^{11}$, 
Barry Rothberg$^{12,13}$, 
Russell Ryan$^3$
 \medskip\\
$^1$ Mullard Space Science Laboratory, University College London, 
Holmbury St Mary, Dorking, Surrey RH5 6NT, UK\\
$^2$ Astronomisches Rechen-Institut, Zentrum f\"ur Astronomie, Universit\"at Heidelberg, M\"onchhofstr. 12-14, D-69120 Heidelberg, Germany\\
$^3$ Space Telescope Science Institute, 3700 San Martin Drive, Baltimore, MD 21218, USA\\
$^4$ School of Earth \& Space Exploration, Arizona State University, Tempe, AZ 85287-1404, USA\\
$^5$ NASA/Goddard Space Flight Center, Astrophysics Science Division, Code 660, Greenbelt, MD 20771, USA\\
$^6$ Department of Physics and Astronomy (DIFA), Universit\`a di Bologna, via Gobetti 93/2,
I-40129 Bologna, Italy\\
$^7$ INAF, Osservatorio Astronomico di Arcetri, Largo E. Fermi 5, I-50125 Firenze, Italy\\
$^8$ Dark Cosmology Centre, Niels Bohr Institute, University of Copenhagen,
Juliane Maries Vej 30; DK-2100, Denmark\\
$^9$ Department of Astronomy, The University of Texas at Austin, Austin, TX 78712, USA\\
$^{10}$ The University of Virginia, Charlottesville, VA 22904-4325, USA\\
$^{11}$ Department of Astronomy, Stockholm University, AlbaNova University Centre, 106 91, Stockholm, Sweden\\
$^{12}$ Large Binocular Telescope Observatory, University of Arizona, AZ 85721, USA\\
$^{13}$ Department of Physics \& Astronomy, George Mason University,
MS 3F3, 4400 University Drive, Fairfax, VA 22030, USA
}

\begin{document}
\date{MNRAS, Accepted 2019 March 18. Received 2019 February 18; in original form 2018 May 2}
\pagerange{\pageref{firstpage}--\pageref{lastpage}} \pubyear{2019}
\maketitle
\label{firstpage}

%%\newif\ifAMStwofonts
%%\AMStwofontstrue

\begin{abstract}
We constrain the stellar population properties of a sample of 52
massive galaxies -- with stellar mass $\log($M$_s/$M$_\odot)\simgt
10.5$ -- over the redshift range $0.5<z<2$ by use of observer-frame
optical and near-infrared slitless spectra from {\sl Hubble Space
Telescope}'s ACS and WFC3 grisms. The deep exposures ($\sim$100\,ks)
allow us to target {\sl individual} spectra of massive galaxies to
F160W=22.5\,AB. Our spectral fitting approach uses a set of six base
models adapted to the redshift and spectral resolution of each
observation, and fits the weights of the base models, including
potential dust attenuation, via an MCMC method. Our sample comprises a
mixed distribution of quiescent (19) and star-forming galaxies
(33). We quantify the width of the age distribution ($\Delta t$) that is found to
dominate the variance of the retrieved parameters according to
Principal Component Analysis. The population parameters follow the
expected trend towards older ages with increasing mass, and
$\Delta t$ appears to weakly anti-correlate with stellar mass,
suggesting a more efficient star formation at the massive end.
As expected, the redshift dependence of the relative stellar age
(measured in units of the age of the Universe at the source) 
in the quiescent sample rejects the hypothesis of a single burst (aka
monolithic collapse). Radial colour gradients within each galaxy are also
explored, finding a wider scatter in the star-forming subsample, but
no conclusive trend with respect to the population
parameters.
\end{abstract} 

\begin{keywords}
galaxies: stellar content -- galaxies: high-redshift -- 
galaxies: evolution -- galaxies: formation
\end{keywords}

%%%%%%%%%%%%%%%%%%%%%%%%%%%%%%%%%%%%%%%%%%%%%%%%
\section{Introduction}
\label{Sec:Intro}

Massive galaxies represent one of the best probes to understand the
physical mechanisms governing galaxy formation and evolution, in
particular the interplay between structure growth -- mostly driven by
the dark matter density field -- and star formation -- regulated by
both gas infall/outflows and by feedback processes.
Observational constraints on the evolution of the massive
galaxy population over cosmic time \citep[see, e.g.,][and references therein]{Renzini:06}
reveals an early, intense and short-lived
star formation episode within relatively small volumes (``galaxy
cores'').  These compact massive cores are already found at z$\sim$1-3
\citep[e.g.][]{Daddi:05,Trujillo:06,AC:08,vdK:08}, featuring relatively
quiescent populations \citep{AC:04,I3}. The recent findings of a
non-standard initial mass function (IMF) in the central regions of
massive early-type galaxies \citep[e.g.][]{IMN:15,FLB:16} can be
related to a different mode of formation in the cores, following a
more efficient conversion of gas into stars that produces a highly
turbulent interstellar medium, leading to enhanced
fragmentation \citep[][]{Chab:14}. In contrast, the outer regions
(R$\simgt$R$_e$) feature a standard IMF \citep[see,
e.g.][]{FLB:17,vdK:17}. This core-envelope dichotomy has been
presented over the past few years as the two-stage paradigm of
formation \citep[e.g.][]{Oser:10}, whereby the stellar populations in
a galaxy are the product of both {\sl in-situ} formation, along with
an additional component of stars formed {\sl ex-situ}, 
incorporated into the galaxy via mergers. The study of massive
galaxies at the peak of galaxy formation activity, corresponding to
redshifts z$\sim$1--3~\citep{CSFH}, provides a unique
opportunity to probe this formation mechanism, by focusing on the {\sl
in-situ} component.

Over the past few years, a number of works have focused on the
analysis of high-redshift massive galaxies, including  deep
spectroscopy of 8 massive galaxies at z$\sim$1.5--2 with Keck/LRIS and
VLT/X-Shooter \citep{Bezanson:13,vdSande:13}. Strong Balmer absorption is found
in most of these galaxies, revealing a post-starbursting behaviour,
therefore representing systems recently quenched and on their way to
joining the red sequence \citep[see also][]{FW4871}. Their velocity dispersion and structural
properties, are indicative of an inside-out growth process, keeping a
relatively unevolved massive core
\citep[within the central $\sim$1\,kpc,][]{vdSande:13}.
Deep exposures with Subaru/MOIRCS targeted a sample of 24 massive galaxies
between z=1.25 and z=2.09, also finding the typical post-starburst
$\sim$1\,Gyr stellar ages when Balmer absorption is strongest, with
a tentative formation epoch around z$_{\rm FOR}\sim$2 \citep{Onodera:15}.
These authors also detected super-solar [Mg/Fe] abundance ratios, characteristic
of short-lived periods of star formation that prevent the later ($\simgt$1\,Gyr) contribution of
iron-rich type Ia supernova to the average stellar metallicity
\citep[see also][]{Kriek:16}.
\citet{Belli:15} explored a substantially larger sample of 62 massive
galaxies at z$\sim$1--1.6 with Keck/LRIS, finding a trend between age
and size, so that, at fixed mass, the younger galaxies were more
extended, analogously to the trends found at low
redshift \citep{Scott:17}.  A recent analysis of the underlying
stellar populations of massive galaxies at z=0.6--1.0 from the LEGA-C
survey \citep{Chauke:18} reinforce the idea of a strong age-mass
trend \citep[e.g.][]{Gallazzi:05,Gallazzi:14,LDG:18}; whereby the most
massive galaxies already undergo passive evolution by z$\sim$1, 
with sporadic rejuvenation events. During the refereeing process of this
paper, \citet{Estrada:18} presented their analysis of quiescent
massive galaxies with slitless grism spectroscopy from the CANDELS
Lyman-$\alpha$ Emission at Reionization survey, confirming the early
formation process expected of these galaxies, whereby most of them 
formed over 68\% of their stellar mass content
by redshift z$\simgt$2, with a prompt enrichment to solar abundances
by z$\sim$3.

At present, one of the best options to extract information from the stellar
populations of massive galaxies at these redshifts rely on slitless
grism spectra with high enough S/N in the continuum to produce
population constraints from spectral fitting.  This approach has been
exploited with the Advanced Camera for Surveys (ACS) and the Wide
Field Camera 3 (WFC3) on board the {\sl Hubble Space Telescope} ({\sl
HST}). Deep surveys, such as GRAPES \citep{AP:06} and
PEARS \citep{PEARS} allowed us to acquire a set of low-resolution
spectra of a number of early-type galaxies in the z$\simlt$1 redshift
window, consistently finding quiescent populations at the massive end,
a result that suggests an early and efficient phase of star formation
in these systems.  The Early Release Science data from the WFC3 NIR
grisms allowed us to study in detail a massive galaxy (FW4871, with
stellar mass $\simgt 10^{11}$M$_\odot$) at z=1.89 (\citealt{FW4871},
see also \citealt{vdKBram:10}), providing the best case to date of a
detailed spectrum of a massive and recently quenched post-starburst
galaxy. This paper builds upon our previous work by presenting a
combined analysis of the populations in massive galaxies via slitless
grism spectroscopy in the observer-frame optical (PEARS: ACS/G800L)
and NIR (FIGS: WFC3/G102 and G141) spectral windows.

In \S\ref{Sec:Data} we describe the data, giving details about both 
the slitless grism spectra as well as the surface brightness analysis.
\S\ref{Sec:Method} comments on the spectral fitting methodology
leading to the derivation of population parameters, that are presented
in \S\ref{Sec:PopTrends}, including 
a discussion about trends derived from Principal
Component Analysis (PCA).
Finally, \S\ref{Sec:Summ} summarizes our results. Throughout this
paper we quote magnitudes in the AB system \citep{ABmag}, and adopt a standard, flat
$\Lambda$CDM cosmology with $\Omega_m=0.3$ and
$H_0=70$\,km\,s$^{-1}$\,Mpc$^{-1}$.

%%%%%%%%%%%%%   SNR    %%%%%%%%%%%%%%%
%%%%%%%%%%%%%%%%%%%%%%%%%%%%%%%%%%%%%%%%%%%%%%%%
\begin{figure}
\begin{center}
\includegraphics[width=85mm]{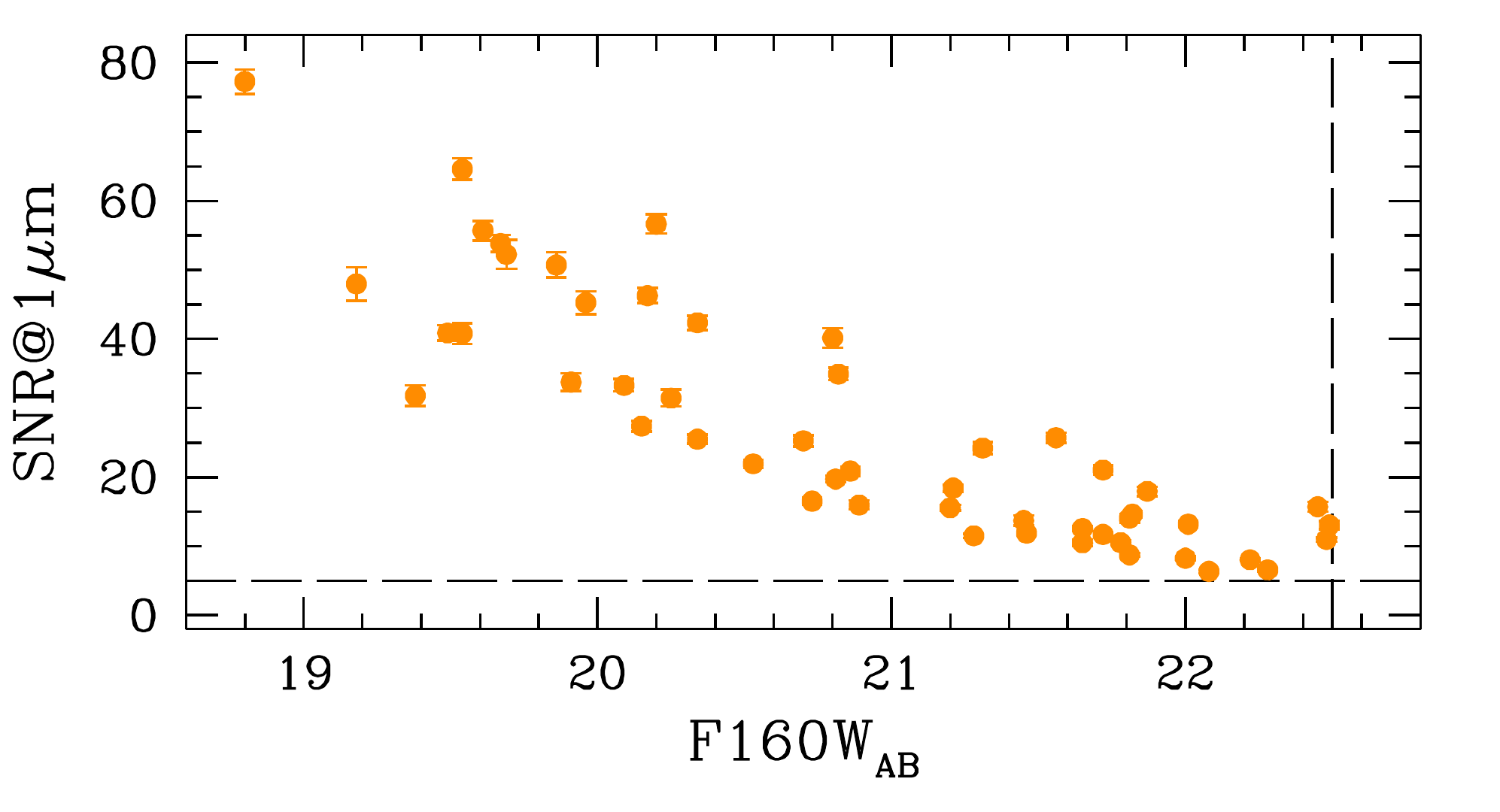}
\end{center}
\caption{Average signal to noise ratio per resolution element
of the G102 grism data corresponding to our massive
galaxy sample, evaluated within a 0.1$\mu$m window around
$\lambda=1\mu$m, plotted against F160W total magnitude. For reference,
the typical SNR expected from the WFC3 exposure time calculator for a
100\,ks exposure (i.e. the FIGS integration time per field) is 5 for an
unresolved source at F160W=22.5AB (dashed lines).}
\label{fig:SNR}
\end{figure}
%%%%%%%%%%%%%%%%%%%%%%%%%%%%%%%%%%%%%%%%%%%%%%%%

%%%%%%%%%%%%%%%%%%%%%%%%%%%%%%%%%%%%%%
%%%%%%%%%%   TABLE 1   %%%%%%%%%%%%%%%
%%%%%%%%%%%%%%%%%%%%%%%%%%%%%%%%%%%%%%
\begin{table*}
\caption{Details of the source selection: Col.~1 identifies the
FIGS field, with equatorial coordinates given in cols.~2 and 3.
Col.~4 gives the total number of FIGS grism sources in each field.
Col.~5 is the number of massive ($>10^{10.5}$M$_\odot$) galaxies,
with F160W$<$22.5AB and with redshift within the adopted $0.5<z<2.5$
window. Col.~6 is the number of massive galaxies used in this work
(i.e. with both ACS and WFC3 grism data available).
\label{tab:sample}}
\begin{center}
\begin{tabular}{cccrrr}
\hline
Field & RA & Dec & N$_{\rm TOT}$ & N$_{\rm massive}$ & N$_{\rm Sample}$\\
\hline
GN1 & 12$^h$36$^m$42.56$^s$ & $+$62$^{\rm o}$17$^\prime$16.89$^{\prime\prime}$ & 706 & 109 & 10\\
GN2 & 12$^h$37$^m$32.04$^s$ & $+$62$^{\rm o}$18$^\prime$26.06$^{\prime\prime}$ & 565 &  75 & 14\\
GS1 & 03$^h$32$^m$41.56$^s$ & $-$27$^{\rm o}$46$^\prime$38.80$^{\prime\prime}$ & 684 & 103 & 27\\
\end{tabular}
\end{center}
\end{table*}
%%%%%%%%%%%%%%%%%%%%%%%%%%%%%%%%%%%%%%

%%%%%%%%%%%%%%%%%%%%%%%%%%%%%%%%%%%%%%%%%%%%%%%%
\section{Data}
\label{Sec:Data}
Our sample selection starts with the catalogue of sources detected in
the Faint Infrared Grism Survey \citep[FIGS,][]{FIGS}. FIGS is a
160-orbit cycle 22 HST Treasury programme (Proposal ID: 13779, PI:
S. Malhotra), that observed four distinct fields at five position
angles, using the WFC3/G102 grism. We use v1.2 of the catalogue, where
the redshift information originates either from the available
spectroscopic measurements or from the photometric redshifts derived
by combining broadband photometry and grism data \citep{Pharo:18}.  We
note that these photometric redshifts achieve an accuracy of $\Delta
z/(1 + z) \sim$ 0.029 within redshifts $z$ = 0.3 and 3. We refer to
\citet{FIGS} for a detailed description of the FIGS data reduction and
spectral extraction methods.

In order to perform a homogeneous selection of the targets
based on stellar mass, we use the same photometric data in
all four FIGS pointings, available from the 3D-{\sl HST} survey
\citep{Skelton:14}. We select all targets within a redshift
range 0.5 $< z<$ 2.5 and derive stellar masses using the fluxes
in F606W, F775W, F850LP (from {\sl HST}/ACS), F125W,
F140W, F160W (from {\sl HST}/WFC3), as well as K$_s$
(from Subaru/MOIRCS in the North and VLT/ISAAC in the South),
and Spitzer/IRAC 3.6$\mu$m fluxes. We only select sources with F160W
$<$ 22.5 AB, as the grism data become very noisy at fainter
magnitudes. Note that in this paper we perform spectral fitting on
slitless grism data corresponding to {\sl individual} galaxies, rather
than relying on stacking large numbers of galaxies at low
SNR \citep[as in the 3D-{\sl HST} survey,][]{Fumagalli:16}.  As an
example, the WFC3 exposure time calculator predicts a S/N around 5 per
resolution element for an unresolved F160W=22.5\,AB source in the G102
grism, with the typical (100ks) exposures of the FIGS
fields. Fig.~\ref{fig:SNR} shows the observed SNR in the G102 grism
data (evaluated at $\lambda=1\mu$m) as a function of F160W total
magnitude.

%%%%%%%%%%%   Comparison lMs     %%%%%%%%%%%%%%%
%%%%%%%%%%%%%%%%%%%%%%%%%%%%%%%%%%%%%%%%%%%%%%%%
\begin{figure}
\begin{center}
\includegraphics[width=80mm]{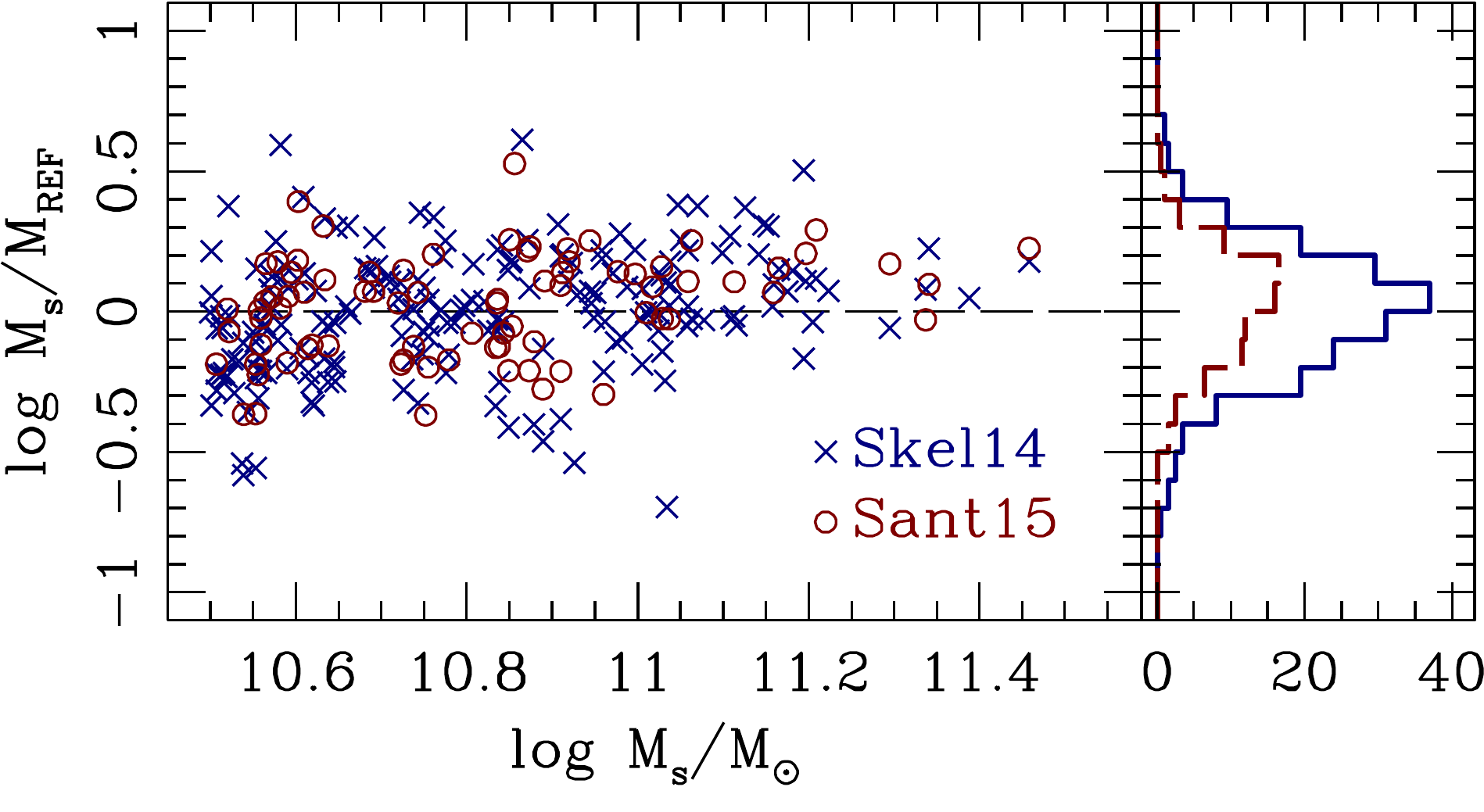}
\end{center}
\caption{Comparison of our starting set of stellar
masses with the 3D-{\sl HST} masses \citep[][labelled Skel14]{Skelton:14}, or the CANDELS
estimates \citep[][labelled Sant15]{Santini:15}, adopting the same (Chabrier) IMF.
The scatter, given as one half of the difference between the 75\% and the
25\% of the distribution is 0.15 in both cases. The panel on the right shows
the distribution of mass offsets.
}
\label{fig:lMs}
\end{figure}
%%%%%%%%%%%%%%%%%%%%%%%%%%%%%%%%%%%%%%%%%%%%%%%%

The stellar masses are derived from a comparison between the observed
photometric fluxes and a set of composite populations assembled from a base set of
3 $\times$ 4 simple stellar populations (SSPs) from the models
of \citet{BC03} for a
\citet{Chab:03} initial mass function (IMF). The set comprises
three metallicities ([Z/H]=\{-0.5,0.0,+0.3\})
and four ages (logarithmically spaced between 0.1\,Gyr and the
age of the Universe at the redshift of each galaxy).
An ensemble
modelling the underlying probability distribution function is created
with a Monte Carlo Markov Chain code based on the \citet{emcee} {\sc
emcee} sampler, where the free parameters are the weights of each of
the 12 SSPs, along with a reddening parameter, E(B$-$V), applied as 
a foreground screen, following the standard Milky Way extinction
law \citep{CCM:89}. We note that the uncertainties in the derivation
of stellar masses are significantly smaller than those related to the
other population parameters (such as age and metallicity)
at a fixed IMF \citep[see, e.g.][]{FSB:08}. Moreover, at this
stage we want to make sure we select all massive galaxies within
the adopted redshift range and flux limit.  Fig.~\ref{fig:lMs} shows a
comparison of our working stellar masses with the estimates of 3D-{\sl
HST} from \citet{Skelton:14}, and those from \citet{Santini:15} in the
CANDELS survey.  We restrict the comparison to our mass threshold,
$\log$M$_s$/M$_\odot\geq$10.5, although the agreement is equally good
down to stellar masses $\log$M$_s$/M$_\odot$=9.5. The SIQR
statistic\footnote{The SIQR (semi-interquartile range) is defined as
half the difference between the 75th and the 25th percentiles of the
distribution.} comparing our mass estimates with those from these
published studies is 0.15\,dex (3D-{\sl HST}) and 0.14 (CANDELS).
Although the derivation of stellar masses is not critical for our
purposes at this stage, Fig.~\ref{fig:lMs} suggests a potential
systematic trend (comparable to the observed scatter), which can
be attributed to the use of specific functional forms for the
star formation rates.

%%%%%%%%%%%%%   Sample distrib   %%%%%%%%%%%%%%%
%%%%%%%%%%%%%%%%%%%%%%%%%%%%%%%%%%%%%%%%%%%%%%%%
\begin{figure}
\begin{center}
\includegraphics[width=85mm]{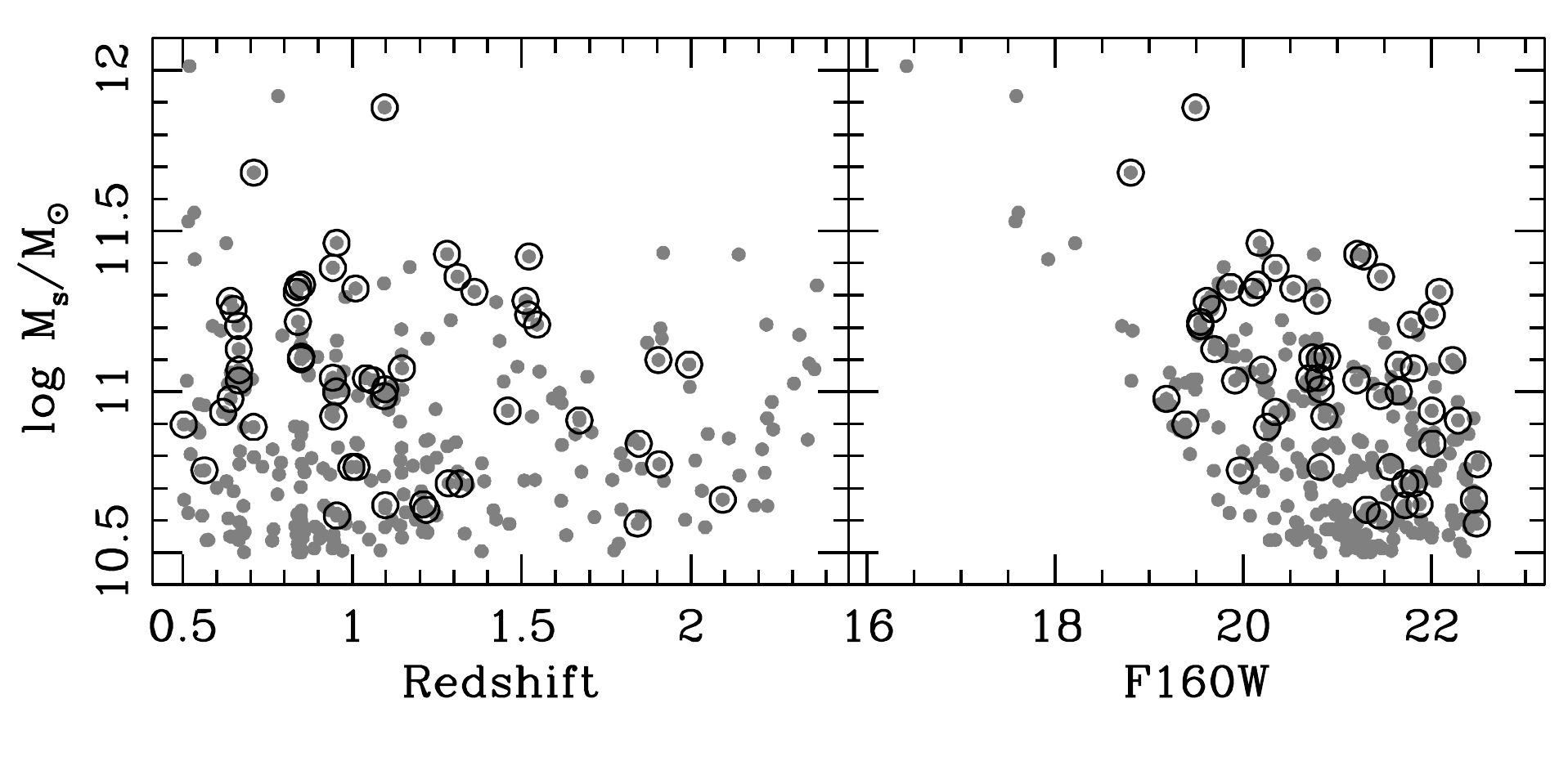}
\end{center}
\caption{Distribution of stellar mass with respect to redshift (left) and F160W magnitude
(right). The filled dots are all galaxies detected in FIGS, whereas
the open dots show our working sample of galaxies with good quality
FIGS + PEARS grism data for spectral fitting constraints.  }
\label{fig:zlMs}
\end{figure}
%%%%%%%%%%%%%%%%%%%%%%%%%%%%%%%%%%%%%%%%%%%%%%%%

%%%%%%%%%%%%%     Images      %%%%%%%%%%%%%%%%%%
%%%%%%%%%%%%%%%%%%%%%%%%%%%%%%%%%%%%%%%%%%%%%%%%
\begin{figure*}
\begin{center}
\includegraphics[width=170mm]{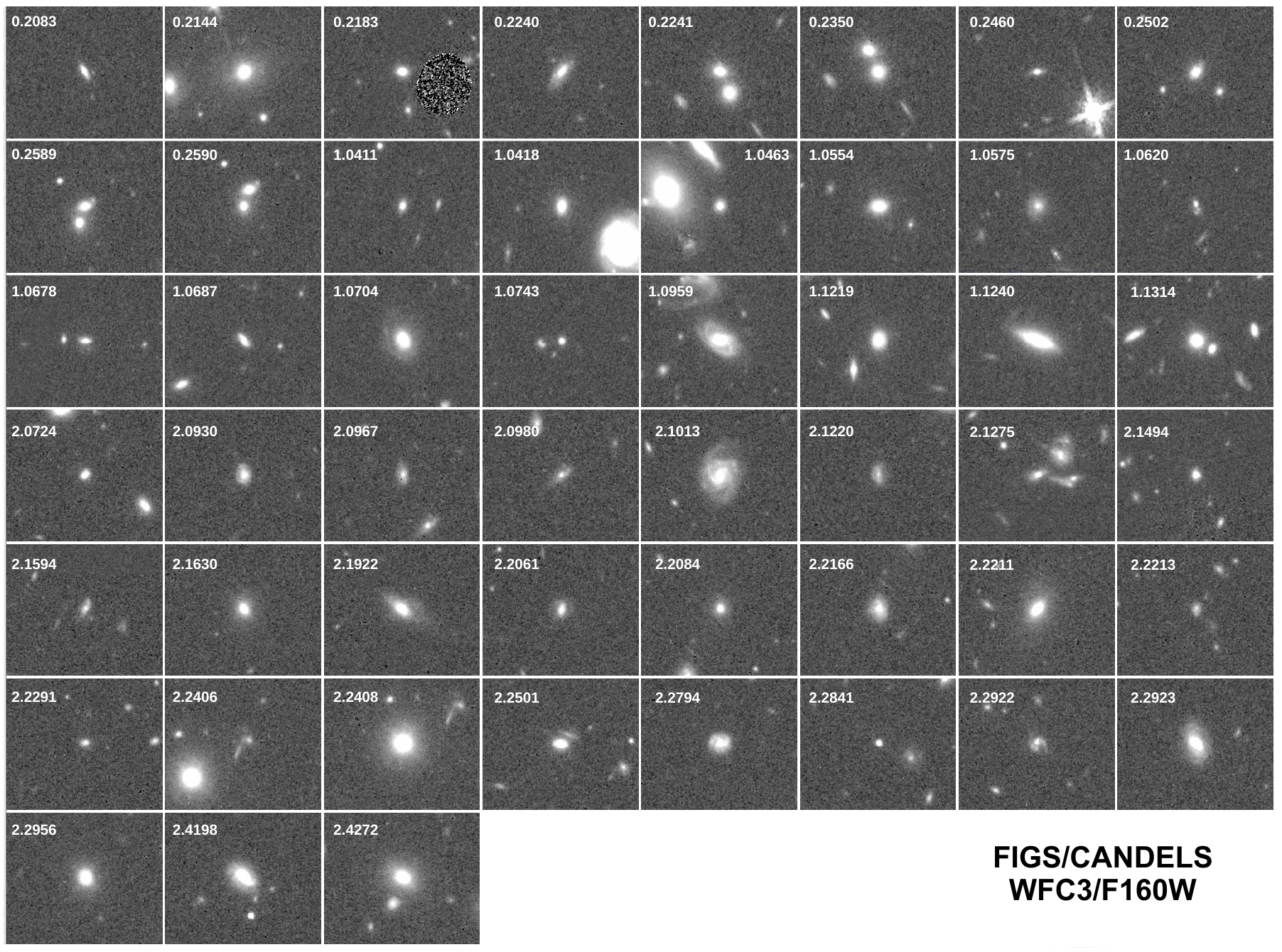}
\end{center}
\caption{Postage stamps of the 51 galaxies in our
FIGS sample. Each stamp is a WFC3/F160W image from CANDELS
\citep{CANDELS1,CANDELS2}, covering a
$16^{\prime\prime}\times 16^{\prime\prime}$ area.  The stamps are
labelled with the field: (0=GN1, 1=GN2, 2=GS1), and the corresponding ID.
A stamp of the additional galaxy in this sample, FW4871,
can be seen in Fig.~2 of \citet{FW4871}.}
\label{fig:f160w}
\end{figure*}
%%%%%%%%%%%%%%%%%%%%%%%%%%%%%%%%%%%%%%%%%%%%%%%%

The starting sample of massive galaxies is then matched to the
catalogue of FIGS WFC3/G102 spectra, as well as to the catalogue of
slitless spectra from the PEARS ACS/G800L survey (ID 10530, PI
Malhotra, e.g. \citealt{PEARS}).  The WFC3/G102 grism covers the
0.8-1.15\,$\mu$m spectral window at a resolution of R=210; and the ACS/G800L
(WFC) observations extend over the interval 0.55--1.05\,$\mu$m at R=100.  When
available, we add WFC3/G141 grism data analyzed as part of programme
AR\,13266 (PI Ryan). This grism provides a spectral coverage 1.075--1.7\,$\mu$m 
at resolution R=130.  From this starting sample we retain only those
galaxies for which {\sl both} PEARS (G800L) and FIGS (G102) spectra are available. We
note that only three of the four FIGS pointings (GN1, GN2 and GS1;
see Table~\ref{tab:sample})
overlap with PEARS data.  The FIGS GS2 pointing targets a parallel
CDFS field (HUDF-Par\,2), not included in the ACS grism programme.

For each target and grism dataset, we correct the individual spectra 
-- taken at a specific telescope roll angle -- for contamination from
nearby sources as computed in \citet{FIGS}. We combine 
the corrected spectra (excluding, in very few cases, those that
deviate more than 3\,$\sigma$ from the average), and compute the uncertainty associated
with the mean spectrum by propagating the errors of the individual
spectra. The average PEARS and FIGS spectra of each galaxy are
subsequently combined by scaling, in flux, the PEARS spectrum to the
FIGS one within their overlapping spectral range, making sure to avoid data 
at the edges, where the instrument sensitivity drops and the flux
calibration is not reliable. In the overlapping
spectral region, the FIGS/G102 mean spectrum is interpolated to the
dispersion of the scaled mean PEARS/G800L data. The two spectra are
averaged and their errors propagated.  The same procedure applies when
combining the mean FIGS/G102 and G141 spectra. In this case the mean,
G102 spectrum is interpolated to the dispersion of the
lower-resolution G141 spectrum. Moreover, the G141 data are scaled to
match the flux of G102.  We exclude from this processing all galaxies
whose spectra (either PEARS or FIGS) are truncated because the source
is located at the edge of the field of view.

Table~\ref{tab:sample} summarizes the source selection.
Fig.~\ref{fig:zlMs} shows the starting sample, as grey dots, and the
final sample of galaxies with good PEARS and FIGS data for the
spectral analysis presented below. Clearly, the combination of PEARS
and FIGS data provide a wide coverage for the spectral fitting
analysis, which in turn allows us to better constrain the star
formation histories of massive galaxies in the rest-frame optical
window.  We cross-correlated the sample with the X-ray 2\,Ms
catalogues in the CDFN \citep{XCDFN} and CDFS \citep{XCDFS}, and only
found three sources with a hard X ray detection (in the 2-8\,keV
band), namely galaxies 2144 and 2502 in GN1 and galaxy 980 in GS1,
with X-ray fluxes L$_X=\{0.01, 0.02, 0.35\}\times 10^{44}$erg\,s$^{-1}$,
respectively.  Given the low luminosity of these sources, we do not
expect the rest-frame optical fluxes to be contaminated by AGN
emission.

Fig.~\ref{fig:f160w} shows the WFC3/F160W images of the final set of
51 galaxies from CANDELS \citep{CANDELS1,CANDELS2}. We note that the
FIGS fields GN1 and GN2 are covered by the CANDELS GN05 GOODS-N
pointing, and that GS1 is fully covered by the CANDELS GSD01 GOODS-S
pointing. In the appendix, Table~\ref{tab:massive} shows the general
details of the sample, including visual morphology, stellar
coordinates, redshift and F160W magnitude, as well as stellar mass and
rest-frame U$-$V and V$-$J colours (derived from the spectral
analysis, see below). The morphology estimate is split into
early-types (E) and late-types (L), and follows 
\citet{HC:15}, who train a set of convolutional neural networks on 
the results from a visual classification in the $H$
band \citep{Kartaltepe:15} to produce a catalogue of ``visual-like''
classifications in the five CANDELS fields. We use their spheroid
fraction parameter to split our sample into early- ($f_{\rm Sph}\geq
0.5$) and late-types ($f_{\rm Sph}<0.5$). Next to the morphological
type (col.~3 of Table~\ref{tab:massive}) we include the quiescence (Q)
vs star-formation (S) flag based on the standard analysis on a
colour-colour diagram, as presented in \S\ref{Sec:PopTrends}. Note
that in addition to the 51 galaxies from the combined PEARS+FIGS
sample, we include the spectrum of massive galaxy FW4871
\citep[presented in][]{FW4871},
also produced from a combination of ACS and WFC3 grism data.

%%%%%%%%%%%%%%   SED example     %%%%%%%%%%%%%%%
%%%%%%%%%%%%%%%%%%%%%%%%%%%%%%%%%%%%%%%%%%%%%%%%
\begin{figure*}
\begin{center}
\includegraphics[width=75mm]{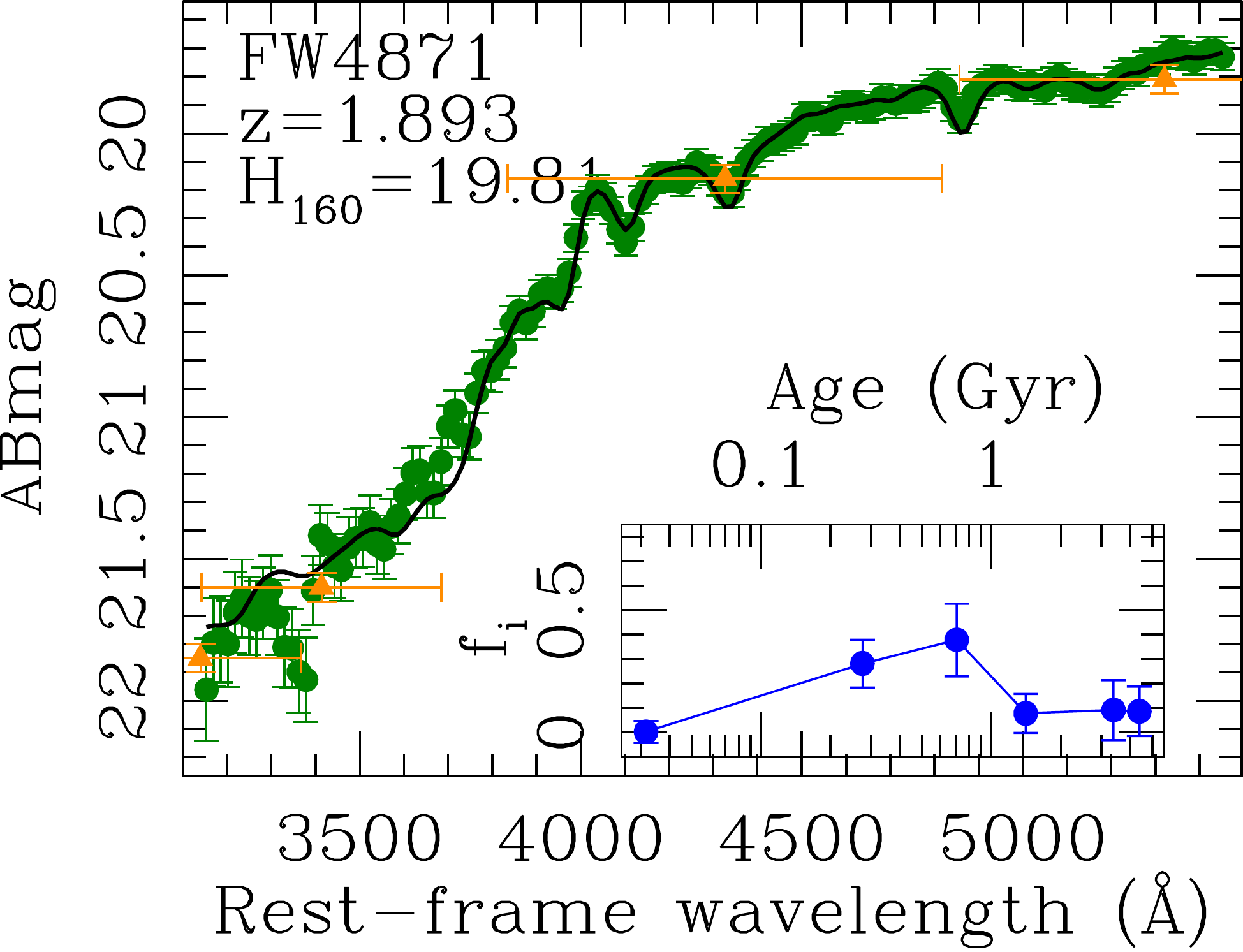}
\includegraphics[width=75mm]{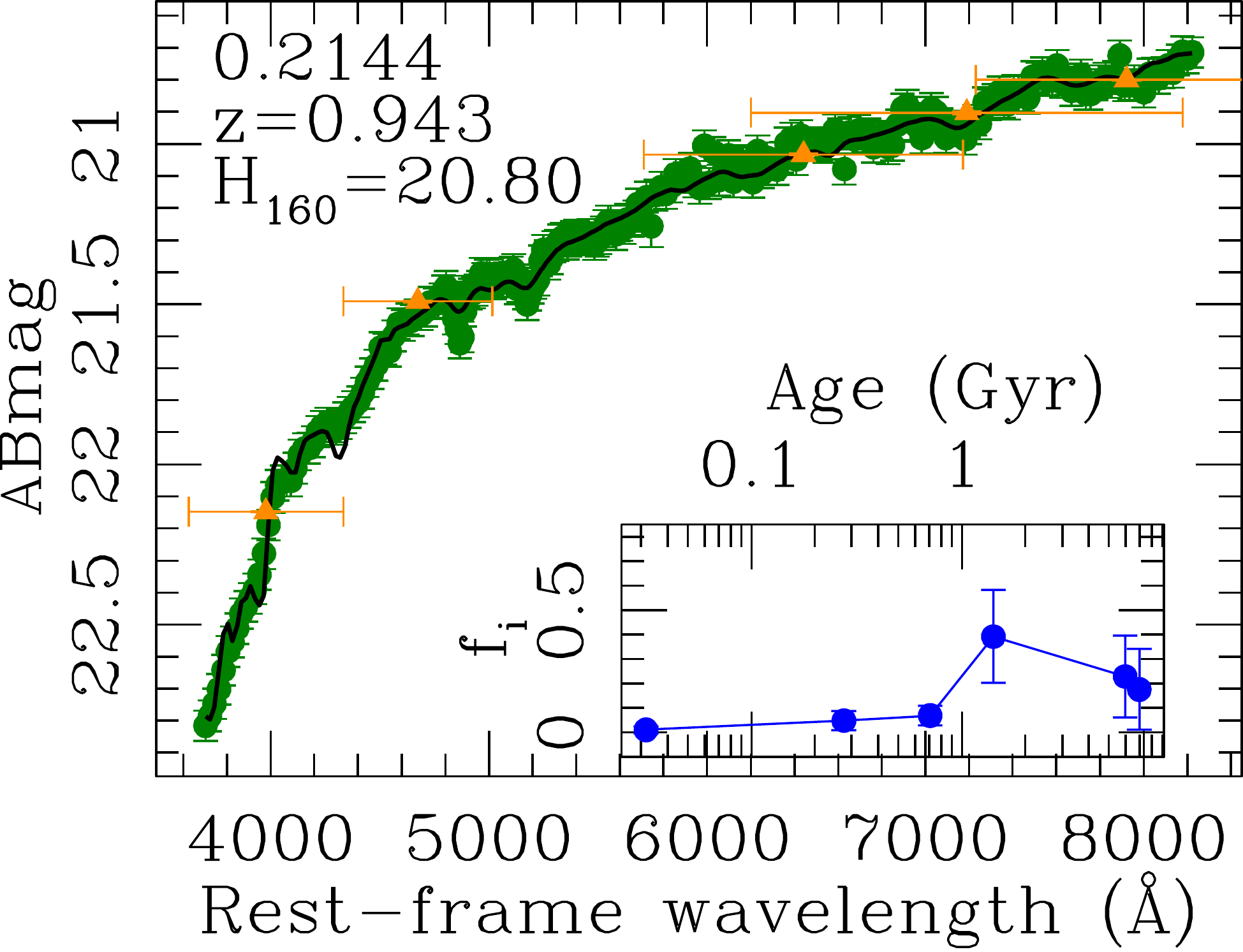}
\end{center}
\caption{Examples of the full spectral fitting results, combining the ACS and WFC3 grism data.
We show galaxies FW4871 (z=1.893, left) and GN1/2144 (z=0.943, right).
The slitless grism data are shown as green dots, and the best fit model is
given by the solid line. For reference, we include the broadband photometry
as orange triangles, with the horizontal error bars representing the FWHM
of the bandpasses. The insets show the fractional contribution {\sl by mass} of the six
base models, labelled with respect to their average stellar age.  The
extracted star formation history reveals a younger population in 4871
(average age of $\sim$1\,Gyr) with respect to 2144 ($\sim$2\,Gyr), as
reflected by the prominent Balmer absorption features
(see \S\ref{Sec:Method} for details). The rest of the spectral fits
are shown in the appendix.}
\label{fig:sed}
\end{figure*}
%%%%%%%%%%%%%%%%%%%%%%%%%%%%%%%%%%%%%%%%%%%%%%%%

%%%%%%%%%%%%%%%%%%%%%%%%%%%%%%%%%%%%%%%%%%%%%%%%
\subsection{Surface brightness fits and colour gradients}
\label{SSec:CGrads}

In addition to the low-resolution grism spectra, we perform a surface
brightness analysis, applying {\sc Galfit} \citep{galfit} to the
CANDELS WFC3 images in F125W and F160W \citep{CANDELS2}.  We consider
a single S\'ersic profile, and the sizes are quoted as the
circularized half-light radii, i.e. $R_e\equiv \sqrt{a_eb_e}$, where $a_e$ and $b_e$ are
the semi-major and semi-minor axes, respectively, engulfing half of the total light.
An image of the point spread function for
each pointing and field is created by median stacking a number of
stars in the CANDELS GN05 and GSD01 fields.  We inspected visually the
fits, making sure there were no significant residuals. Moreover, we
compared our results in the F160W band with the surface brightness
fits presented in \citet{vdWel12}, and find a difference in the
S\'ersic index of
$\Delta n_S\equiv n_{\rm S,FIGS} - n_{\rm S,vdWel}=-0.18\pm 1.04$ and
in the effective radius of
$\Delta R_e\equiv R_{\rm e,FIGS} - R_{\rm e,vdWel}=0.05\pm 0.18$\,arcsec.
We note that the CANDELS radii are quoted as the half-light semi-major
axis, so this comparison involves the raw {\sc Galfit} sizes
(also given as the semi-major axis).

The analytic surface brightness profiles, using the best fit
parameters from each band, are combined to create a
$C\equiv$F125W$-$F160W colour profile, from which we derive, via a
standard least squares linear fit, a slope of the (linear) radial
gradient: $\nabla_C\equiv\Delta C/\Delta\log R$.
Table~\ref{tab:massive2}, in the appendix, shows the results of the F160W surface
brightness fits and colour profiles.  As a test, we compared the
visual morphological classification (from col.~3 in
Table~\ref{tab:massive}) with the S\'ersic index (from col.~3 in
Table~\ref{tab:massive2}), finding an average value of
$n_S$=3.4$\pm$2.6 for the early-types, and $n_S$=1.3$\pm$0.7 for the
late-types. There does not seem to be a similar segregation in the
colour gradient with respect to visual morphology (average gradients
of $\nabla_C=-$0.07$\pm$0.16 and $-$0.04$\pm$0.27 for the early- and
late-types, respectively), but a potential variation if the sample is
segregated with respect to the S\'ersic index (average gradients of
$\nabla_C=+$0.03$\pm$0.05 and $-$0.11$\pm$0.24 for n$_S>$2.5 and
n$_S\leq$2.5, respectively).

%%%%%%%%%%%%%%%%%%%%%%%%%%%%%%%%%%%%%%%%%%%%%%%%
\section{Spectral fitting}
\label{Sec:Method}

The fitting procedure involves two stages. In the first stage we
perform an initial fit of the spectra using simple stellar populations
(SSPs) from the synthetic models of \citet{BC03} for a
\citet{Chab:03} initial mass function (IMF).
Although the use of SSPs is rather simplistic, the high level of
correlation of any spectra involving unresolved populations allows us
to assess whether a good fit is possible, and we also use this initial
fit to mask bad data, or strong emission lines, by applying a 4\,$\sigma$
clip. Moreover, we test the effective resolution of the spectra and
the spectral fitting range. We take into account the fiducial
resolution of the grisms -- namely R=100 @$\lambda=0.8\,\mu$m for ACS/WFC/G800L;
R=210 @$\lambda=1\,\mu$m for WFC3/G102, and
R=130 @$\lambda=1.4\,\mu$m for WFC3/G141, all valid for an unresolved source.
In slitless spectroscopy, the effective resolution depends on both
the actual resolution of the dispersion element, and the surface
brightness profile of the galaxy along the dispersion direction, since
the source acts as a slit. The spectral resolution of the grisms quoted above
correspond to an unresolved object, whereas an extended source will produce 
significantly lower values. We address this issue by taking into account
the S\'ersic surface brightness profiles presented in \S\S~\ref{SSec:CGrads}.
These profiles are convolved along the dispersion direction with the
synthetic data, in order to obtain spectra with the same effective
resolution as the observed data. This is done on a galaxy-by-galaxy basis. Once
the fit is satisfactory, the code creates a set of five ``base
models'' for each galaxy, with a constant star formation rate defined
within the age intervals as follows:

%%%%%%%%%%%%%%        CLs      %%%%%%%%%%%%%%%%%
%%%%%%%%%%%%%%%%%%%%%%%%%%%%%%%%%%%%%%%%%%%%%%%%
\begin{figure*}
\begin{center}
\includegraphics[width=160mm]{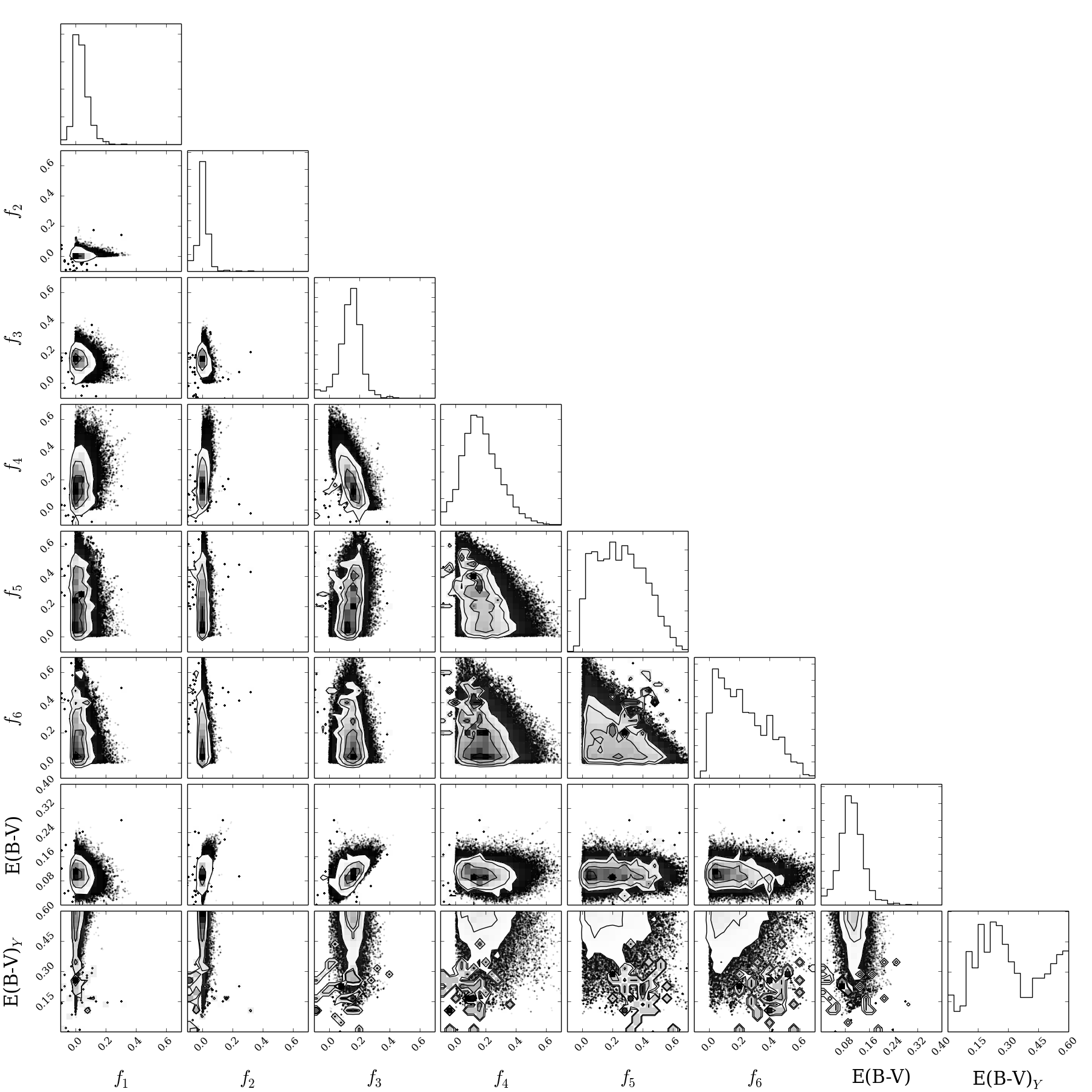}
\end{center}
\caption{Example of the parameter fits corresponding
to galaxy FW4871 (z=1.89, stellar mass $7.95\times 10^{11}$M$_\odot$).
A set of panels shows the 2D marginalised distribution of
the parameters (see text for details).
The spectral fit is shown in Fig.~\ref{fig:sed} (left panel).}
\label{fig:MCMC}
\end{figure*}
%%%%%%%%%%%%%%%%%%%%%%%%%%%%%%%%%%%%%%%%%%%%%%%%

\begin{itemize}
\item Base Model 1: $\log t/{\rm Gyr}\in [-2,-1]$
\item Base Model 2: $\log t/{\rm Gyr}\in [-1,-0.3]$
\item Base Model 3: $\log t/{\rm Gyr}\in [-0.3,0.0]$
\item Base Model 4: $\log t/{\rm Gyr}\in [0.0,0.3]$
\item Base Model 5: $\log t/{\rm Gyr}\in [0.3,lt_{\rm MAX}]$
\item Base Model 6: $\log t/{\rm Gyr}\in [0.3,lt_{\rm MAX}]$
\end{itemize}
where $lt_{\rm MAX}$ is the log$_{10}$ of the age of the Universe at
the redshift of the galaxy, i.e. corresponding to the oldest possible
age.  These base models have the same metallicity as the best-fit
value obtained during the first fitting stage. Note the 
sixth base model has the same age distribution as BM5,
but at a metallicity lower than the best fit value by
$-0.3$\,dex. Base Model 6 thus represents an old, metal-poor component
expected in formation histories with a low star formation
efficiency. Although this component should not dominate the budget in massive
galaxies \citep[e.g.][]{FS:00}, we include this potential contribution
as a free parameter. We note that the choice of six base models may
seem rather arbitrary. However, we point the reader to Sec.~\ref{SSec:PCA},
where PCA suggests most of the variance in the data can be encoded into
$\sim$4-5 components. Our use of five time components plus an additional
old and metal poor one is thus a good compromise to constrain the stellar
populations in these galaxies. In appendix~\ref{App:Tests}, we compare
our results with a new set of runs where seven base models are considered,
finding consistent constraints.

%%%%%%%%%%%%%   CCD VJ vs UV     %%%%%%%%%%%%%%%
%%%%%%%%%%%%%%%%%%%%%%%%%%%%%%%%%%%%%%%%%%%%%%%%
\begin{figure}
\begin{center}
\includegraphics[width=85mm]{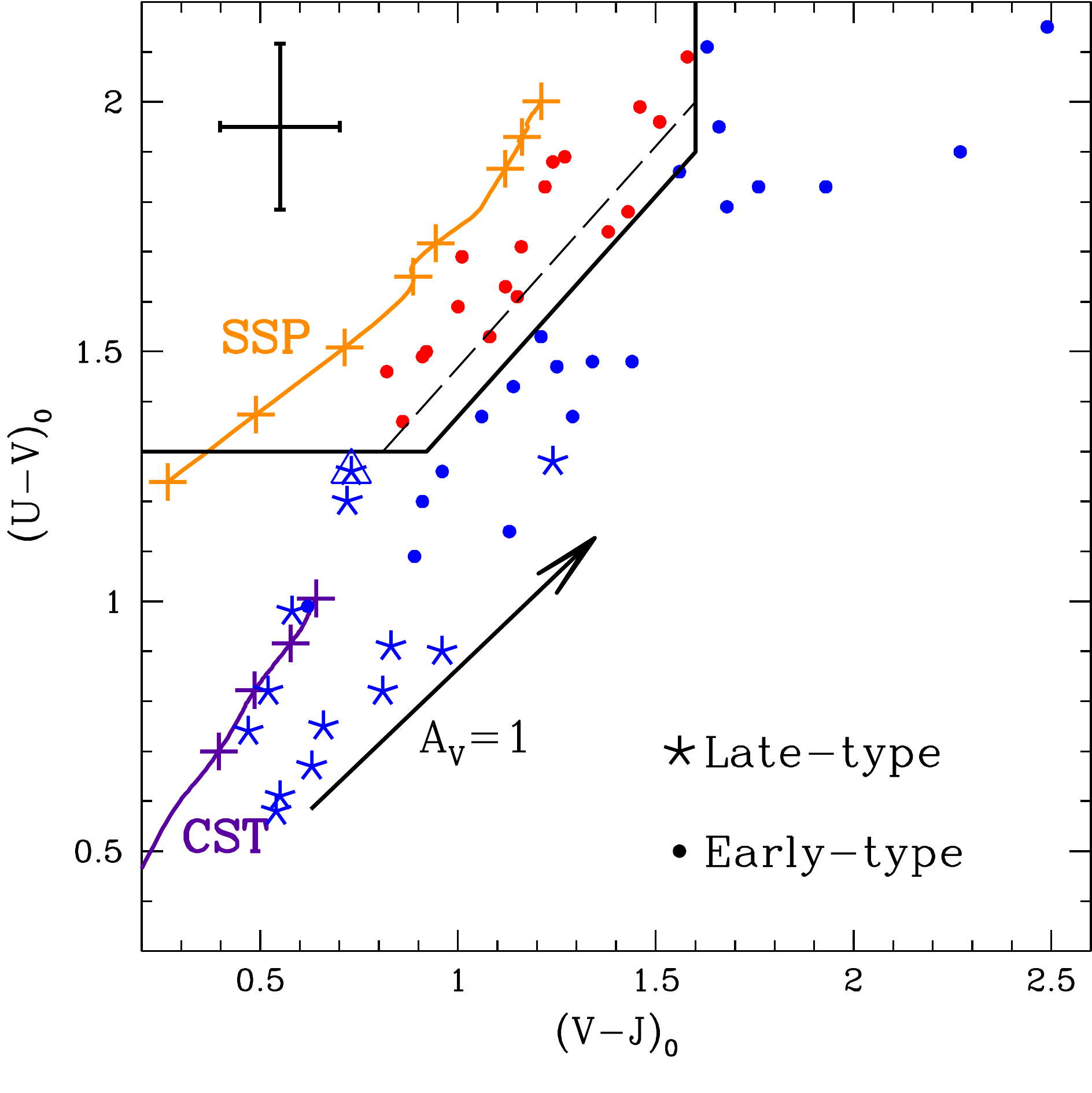}
\end{center}
\caption{Rest-frame colour-colour diagram, showing
the standard regions defining quiescent and star forming galaxies,
following \citet{Williams:09}.  The solid (dashed) line corresponds to
1$<$z$<$2 (0.5$<$z$<$1).  Our sample is split with respect to visual
morphology, with early-types shown as filled dots and late-types as
star symbols. The sample is colour coded, with red (blue) galaxies
representing quiescent (star-forming) galaxies, as shown in
Fig.~\ref{fig:age}. The open triangle represents galaxy FW4871.  A
characteristic error bar, at the 1\,$\sigma$ level, is shown in the
top-left corner. For reference, two tracks from the population
synthesis models of \citet{BC03} are shown, SSP for a quiescent
population, and CST for constant star formation (see text for
details). The arrow is the $A_V=1$ dust attenuation vector.  }
\label{fig:UVJ}
\end{figure}
%%%%%%%%%%%%%%%%%%%%%%%%%%%%%%%%%%%%%%%%%%%%%%%%

The second stage of the fitting procedure uses the six base models to
perform linear superpositions -- exploring a wide range of complex star formation
histories -- and including the presence of dust. We use the standard
extinction law of \citet{CCM:89} and consider two independent reddening
components -- each parametrised by a standard colour excess E(B$-$V).
One component is expected to originate from star-forming regions, and is
only applied to the two youngest components (BM 1 and 2). We note
that the typical timescales for the dispersion of dust clouds in star
forming regions is significantly shorter than the age of BM2 \citep[see, e.g.][]{CF:00}.
However, we are targeting the whole stellar distribution of these
galaxies as one composite population, and our simple phenomenological
model aims at assessing whether the populations from the younger stars 
in a potential post-starbursting system, are significantly more affected
by dust than the general stellar component.  A second dust parameter
traces the diffuse distribution and affects the whole spectrum. This
seven parameter model\footnote{Note each base model is weighted by
mass, but the normalization  -- $\sum_if_i=1$ -- removes one of these weights as a free
parameter.}  is fitted using an implementation of the Python MCMC
sampler {\sc emcee} \citep{emcee}. The models and data are normalized
in the observer frame $\lambda\sim 0.9-1.0\,\mu$m spectral window.

Fig.~\ref{fig:sed} illustrates two examples regarding the fitting results. The
panel on the left shows galaxy FW4871, whose WFC3 NIR
spectra in G102 and G141 were obtained during the WFC3 ERS programme
\citep{WFC3:ERS}, and
was combined with the optical ACS/G800L data \citep{FW4871}.  This
source is a z=1.89 galaxy, identified as a typical near-quiescent,
compact massive galaxy, potentially a progenitor of the cores found in
massive early-type galaxies at low redshift.  The panel on the right
shows GN1/2144, another massive galaxy, this time from the combined FIGS+PEARS
data. Each panel shows the observed fluxes as filled green circles with error
bars, along with the best fit model (solid line). Orange triangles give,
for reference, the fluxes in broadband filters covering the same
spectral window, from the available photometry in the \{F606W, F775W, F850LP,
F125W, F140W, F160W\} passbands.  The inset in each panel shows the
weight, along with error bars of each of the six base models with
respect to the age of each one, giving an estimate of the star
formation history. For ease of visualization, Base Model 6 (that has
the same age as BM5) is displaced by $+$1\,Gyr.
Similar plots for the whole sample are shown in
the appendix.

The confidence levels of the fitting parameters of
FW4871 are shown, for reference, in Fig.~\ref{fig:MCMC}, with contours
at the 1, 2, and 3\,$\sigma$ levels. For a comparison between this
free form, component-based fitting and a more standard approach with exponentially
decaying (or constant-plus-truncation) star formation histories, we
refer the interested reader to \citet{FW4871}, where a detailed
comparison is made. As a reference, we note that the average age
quoted here for FW4871 ($1.18\pm 0.19$\,Gyr) is compatible with those derived
from such generic functional forms: $0.72\pm 0.10$\,Gyr for an
exponentially decaying SFH, and $1.44\pm 0.20$\,Gyr for a constant
SFH, both derived from the spectrum extracted within the inner
$0^{\prime\prime}.64$ region of FW4871 (all quoted at the 1\,$\sigma$
level).  As discussed in detail in \citet{FW4871}, we emphasize that
the use of exponentially decaying functions can lead to significant
biases in the estimates of stellar age and formation timescale
\citep[see also][]{Simha:14}.

From the best fit models, we derive a number of properties, including
the best-fit metallicity, the average age, weighted according to the
mass fractions of each base model: ${f_1,\cdots f_6}$, the age of the
oldest 10\% stars ($t_{10}$), and a parameter that characterizes the
width of the age distribution ($\Delta t$), defined as the difference
between the average age and $t_{90}$, where $t_{90}$ is the age of
the youngest 10\% fraction (by mass) of the stellar component. We use
the subindex $90$ here as these stars represent a cumulative fraction 
 at the 90\% level (and to distinguish this parameter from
$t_{10}$, as defined above). The fitting parameters are listed
in Table~\ref{tab:massive} and \ref{tab:massive3}, including error
bars at the 1\,$\sigma$ level. The uncertainties are derived from the
MCMC sampling, taking the last 1000 points from the chains. The
parameter $t_{10}$ serves as a proxy of the formation time, with
higher values implying earlier formation. For instance, galaxy GN1/2083
has $t_{10}=2.9$\,Gyr, meaning that the oldest 10\% of its stellar
populations have ages older than 2.9\,Gyr. At the redshift of this
galaxy (z=0.953), this implies a formation redshift around $z_{\rm
FOR}\simgt 2.1$. We also define f$_Y\equiv$f$_1$+f$_2$ as the stellar
mass fraction in the youngest components (BM1 and BM2); and
f$_Z\equiv$f$_6$ as the mass fraction in low-metallicity stars
(i.e. BM6).

%%%%%%%%%%%%%   Mass vs size     %%%%%%%%%%%%%%%
%%%%%%%%%%%%%%%%%%%%%%%%%%%%%%%%%%%%%%%%%%%%%%%%
\begin{figure}
\begin{center}
\includegraphics[width=85mm]{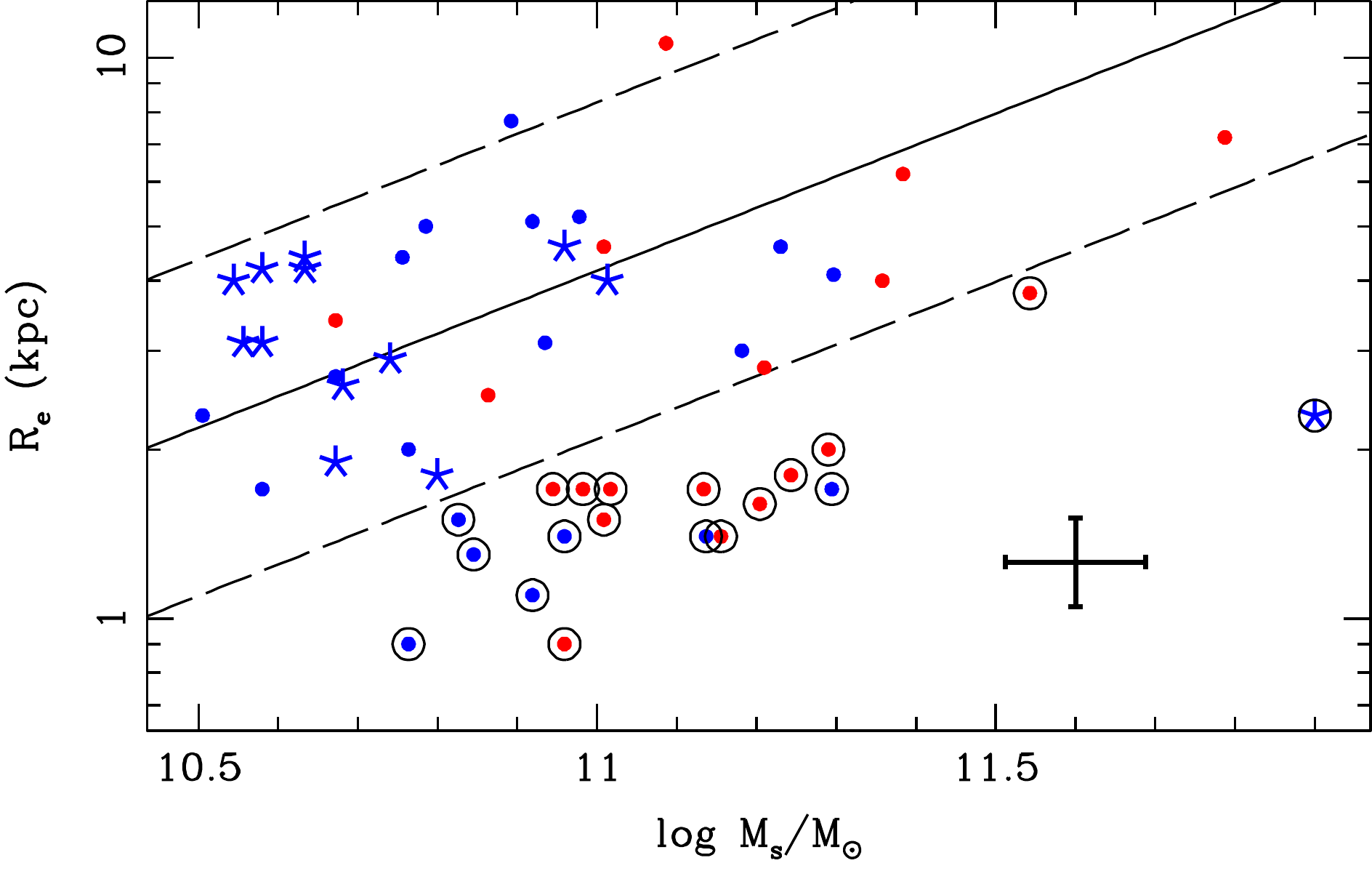}
\end{center}
\caption{Mass vs size relation of our sample of massive galaxies.
The symbol and colour coding is the same as in Fig.~\ref{fig:UVJ},
with red (blue) symbols representing quiescent (star-forming) galaxies
and filled dots (stars) coding the visual morphology as early-types
(late-types). The local relation from \citet{Shen:03} for early-type galaxies
is given by the
solid line, with the dashed lines marking a $\pm 0.3$\,dex region
about this fit. The compact galaxies are hereafter represented by
the larger open circles.}
\label{fig:MRe}
\end{figure}
%%%%%%%%%%%%%%%%%%%%%%%%%%%%%%%%%%%%%%%%%%%%%%%%

%%%%%%%%%%%%%%%%%%%%%%%%%%%%%%%%%%%%%%%%%%%%%%%%
\section{Population trends}
\label{Sec:PopTrends}

%%%%%%%%%%%%   Age indicators    %%%%%%%%%%%%%%%
%%%%%%%%%%%%%%%%%%%%%%%%%%%%%%%%%%%%%%%%%%%%%%%%
\begin{figure*}
\begin{center}
\includegraphics[width=88mm]{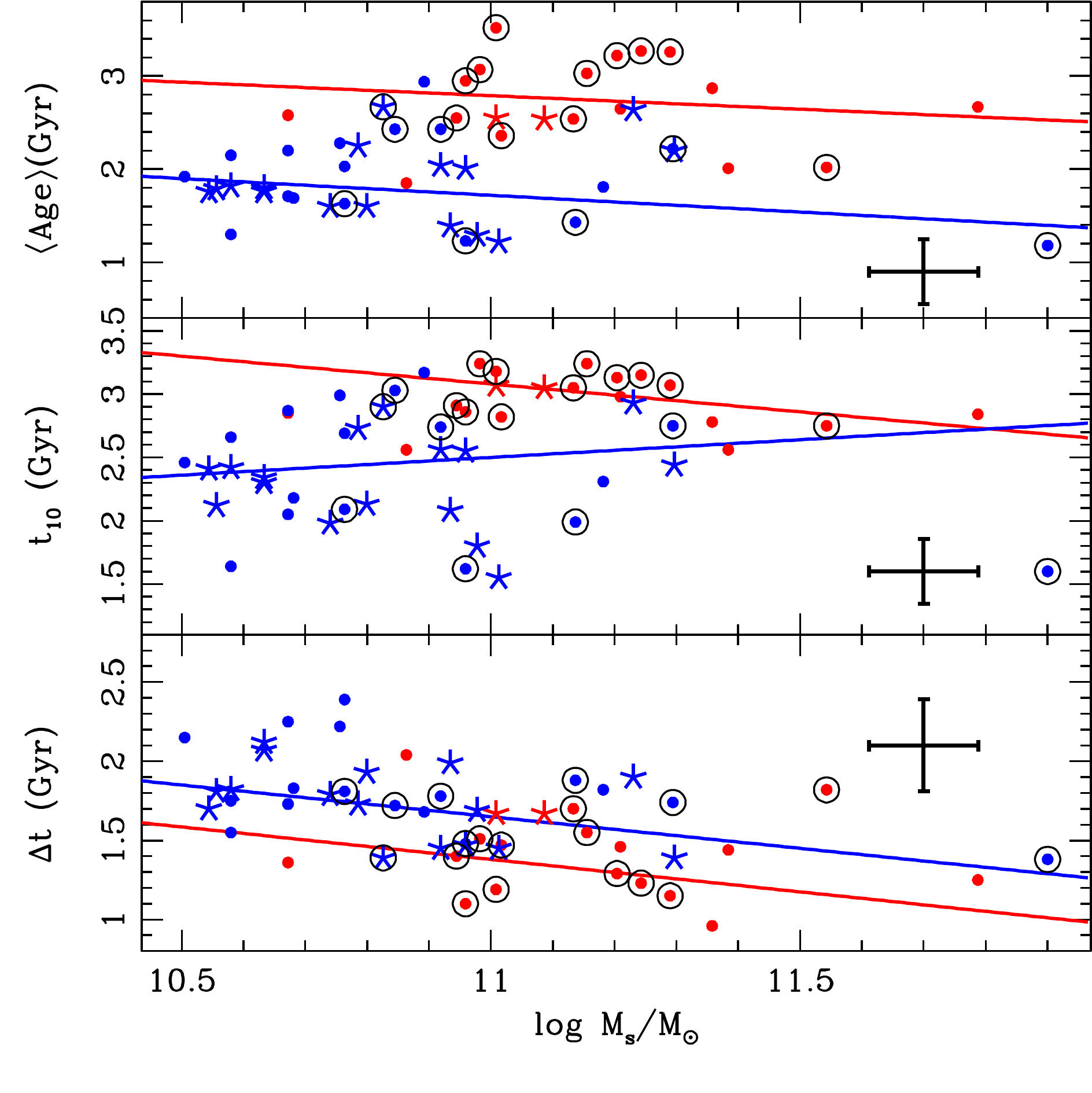}
\includegraphics[width=88mm]{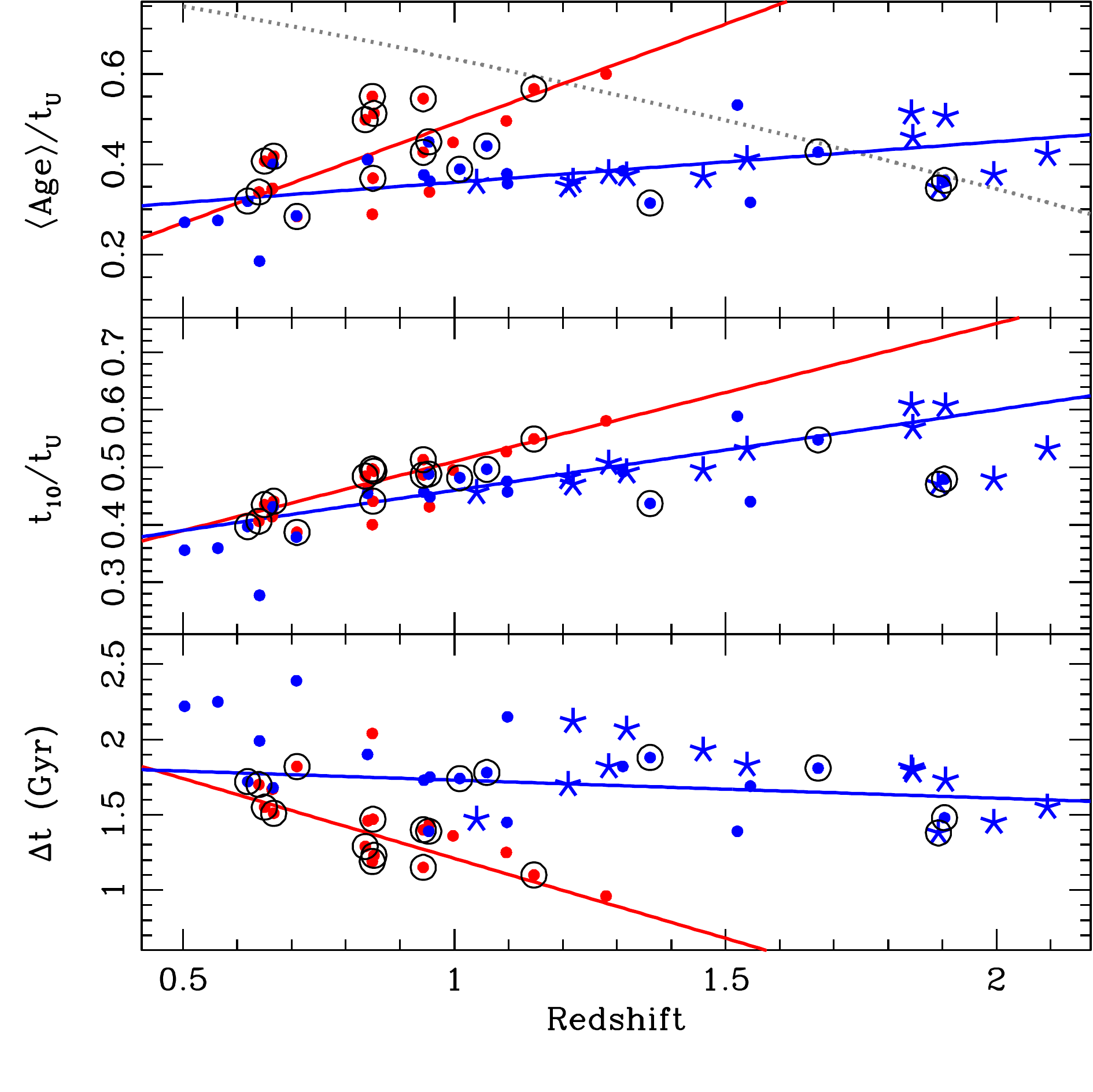}
\end{center}
\caption{The age-related parameters are plotted
with respect to stellar mass. The sample is
split between quiescent (red) and
star-forming galaxies (blue), and between
early-types (circles) and late-types (stars),
as defined in Fig.~\ref{fig:UVJ}. In each panel
we show a linear fit to each subsample. Typical
error bars are given at the 1\,$\sigma$ level.
Galaxies that appear more compact than
the local mass-size relation (see Fig.~\ref{fig:MRe})
include an open circle. The dotted grey line in the top-right
panel shows the expected trend in relative age (age/$t_U$) for a population
formed in a single burst at z$_{\rm FOR}$=3.}
\label{fig:age}
\end{figure*}
%%%%%%%%%%%%%%%%%%%%%%%%%%%%%%%%%%%%%%%%%%%%%%%%

Fig.~\ref{fig:UVJ} shows the rest-frame $(U-V)_0$ and $(V-J)_0$
colours, derived from the best-fit models. This colour-colour diagram
has become a standard tool when separating galaxy samples between quiescent and star-forming
systems \citep[e.g.][]{Williams:09}. The symbols split the sample
into late- and early-type galaxies, following our visual
classification, and we follow the colour criterion of \citet{Williams:09} to
define quiescent and star-forming galaxies. Hereafter, the
figures show these two subsamples in red and blue, respectively. Note
the strong correlation between the photometric selection and the
morphological one, where most star forming galaxies -- especially towards
the bottom-left part of the diagram -- display a late-type
morphology (star symbols) and all quiescent galaxies have an
early-type morphology (solid circles).  In addition, galaxy FW4871 is
shown as an open triangle. This galaxy is at the boundary between
star-forming and quiescent behaviour, as expected since its spectrum
shows strong Balmer absorption on a quiescent continuum (see
Fig.~\ref{fig:sed}, left panel), a typical feature of post-starburst
galaxies \citep{FW4871}. The fact that most of the early-type galaxies classified as 
star-forming appear in the transition region suggests a similar
type of post-starburst behaviour.
However, we should warn that the morphological classification may be
limited by the effect of dust.
For reference, two tracks from the population
synthesis models of \citet{BC03} are shown: the orange (labelled SSP)
corresponds to a quiescent population with ages marked by the crosses
-- from left to right: \{0.5, 0.75, 1, 1.5, 1.75, 2, 3, 4,
5\}\,Gyr. The purple line (labelled CST) is a constant star formation
history, with crosses marking the ages (also from left to right) \{2,
3, 4, 5\}\,Gyr. The dust vector for a \citet{CCM:89} attenuation law
with $A_V=1$ is shown as an arrow. The classification based on either
star formation activity or visual morphology is also presented, with the
same symbols and colour coding, on a mass vs size plane in Fig.~\ref{fig:MRe}.
For reference, the local relation observed in  early-type
galaxies \citep[from][]{Shen:03} is shown as a solid line, including a
$\pm 0.3$\,dex region accounting for the scatter, as dashed lines. A
significant fraction of our sample comprises compact systems, marked
with open circles. We will show in the figures below the same
identification to assess whether the compact galaxies in our sample present any
differences regarding their stellar populations. We emphasize that our
definition of the compactness criterion is rather simplistic, as we only use the
local relation of early-type galaxies and a 0.3\,dex offset.
This work does not aim at a detailed analysis of compact galaxies,
but is meant, instead, to roughly assess whether compact galaxies
display significant differences with respect to the general sample
of massive galaxies.

%%%%%%%%%%%%%%%%%%%%%%%%%%%%%%%%%%%%%%
%%%%%%%%%%   TABLE 2     %%%%%%%%%%%%%
%%%%%%%%%%%%%%%%%%%%%%%%%%%%%%%%%%%%%%
\begin{table}\phantom{...}
\caption{Linear regression to the results shown in Figs.~\ref{fig:age}
and \ref{fig:frac}, with respect to stellar mass. The model for
parameter $\pi$ is $\pi=a\log M_{11} + b$, where $M_{11}$ is the
stellar mass in units of $10^{11}$M$_\odot$ and $\pi$ corresponds to
the following: average age, $t_{10}$, $\Delta t$, $f_Y$, $f_Z$, or
metallicity.  Col.~1 identifies the parameter fit, col.~2
identifies the sample considered: Q for quiescent and SF for
star-forming.  Cols.~3 and 4 give the slope ($a$) and intercept
($b$) at $10^{11}$M$_\odot$, respectively. Col.~5 is the linear correlation
coefficient. The error bars, quoted at the 1\,$\sigma$ level, take
into account the individual uncertainties of the measurements.
\label{tab:fits1}}
\begin{center}
\begin{tabular}{c|cccc}
\hline
$\pi$ & Ty & $a$ & $b$ & $\rho_{xy}$\\
(1) & (2) & (3) & (4) & (5)\\
\hline
\multirow{2}{*}{$\langle{\rm Age}\rangle/t_U$} & Q & $-0.02\pm 0.10$  & $0.44\pm 0.03$  & $+0.08\pm  0.12$\\
%\cline{2-5}
 & SF & $-0.00\pm 0.04$  & $0.37\pm 0.01$  & $-0.03\pm 0.14$\\
\hline
\multirow{2}{*}{$t_{10}/t_U$} & Q & $+0.12\pm 0.05$  & $0.48\pm 0.02$  & $+0.11\pm  0.13$\\
 & SF & $+0.02\pm 0.04$  & $0.47\pm 0.01$  & $-0.05\pm 0.15$\\
\hline
\multirow{2}{*}{$\Delta t$} & Q & $-0.41\pm 0.26$ & $1.38\pm 0.07$ & $-0.17\pm 0.16$\\
%\cline{2-5}
 & SF & $-0.40\pm 0.16$ & $1.65\pm 0.05$ & $-0.24\pm 0.15$\\
\hline
\multirow{2}{*}{$f_Y$} & Q & $-0.01\pm 0.04$  & $0.06\pm 0.01$  & $-0.09\pm  0.16$\\
 & SF & $+0.01\pm 0.09$  & $0.22\pm 0.03$  & $+0.01\pm 0.14$\\
\hline
\multirow{2}{*}{$f_Z$} & Q & $-0.06\pm 0.07$  & $0.23\pm 0.02$  & $-0.04\pm  0.19$\\
 & SF & $-0.04\pm 0.03$  & $0.18\pm 0.01$  & $+0.01\pm 0.13$\\
\hline
\multirow{2}{*}{[Z/H]}  & Q & $+0.11\pm 0.11$  & $-0.10\pm 0.03$  & $+0.19\pm  0.10$\\
 & SF & $+0.17\pm 0.07$  & $-0.08\pm 0.02$  & $+0.23\pm 0.09$\\
\hline
\end{tabular}
\end{center}
\end{table}
%%%%%%%%%%%%%%%%%%%%%%%%%%%%%%%%%%%%%%

%%%%%%%%%%   Other indicators    %%%%%%%%%%%%%%%
%%%%%%%%%%%%%%%%%%%%%%%%%%%%%%%%%%%%%%%%%%%%%%%%
\begin{figure*}
\begin{center}
\includegraphics[width=88mm]{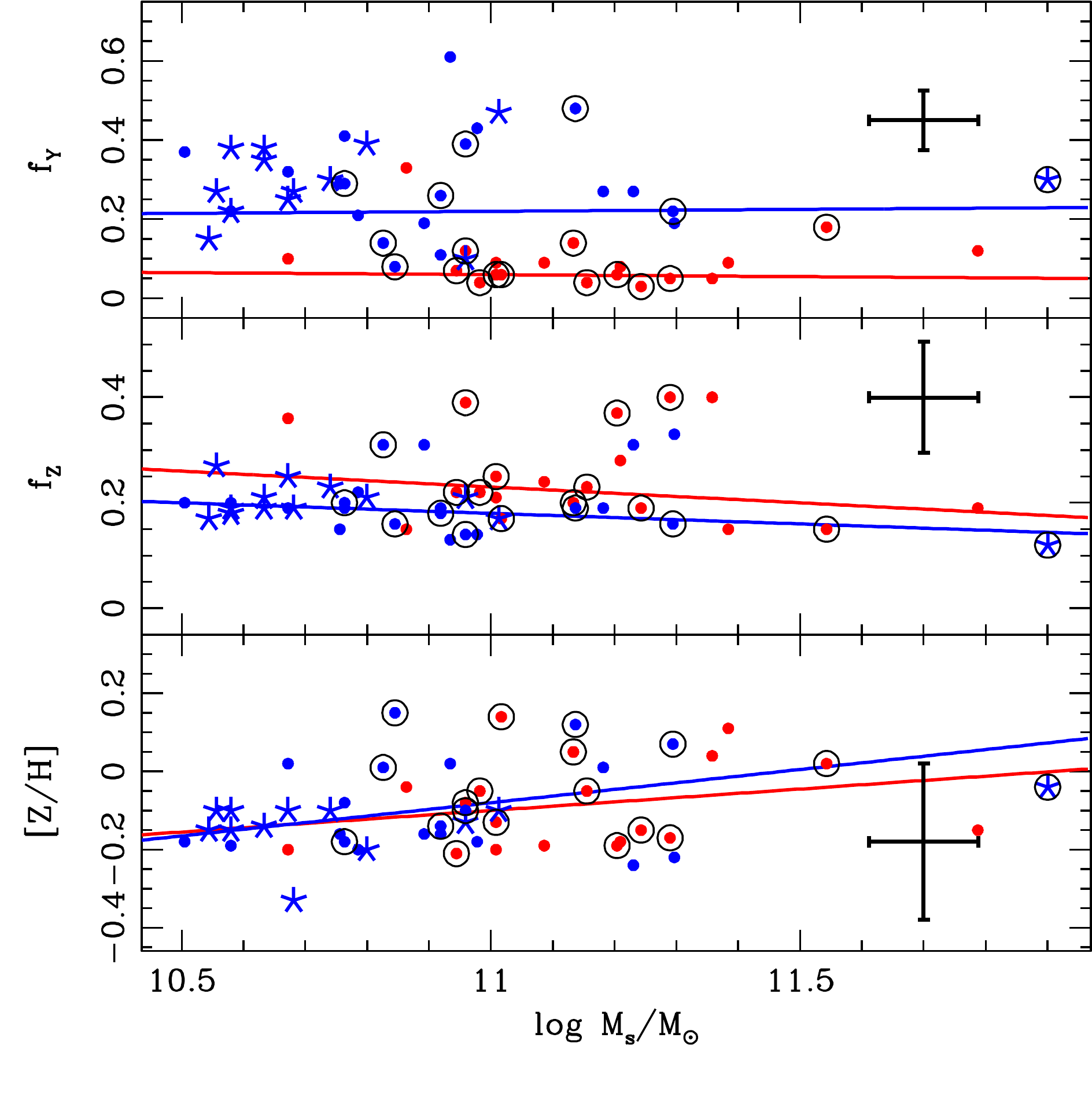}
\includegraphics[width=88mm]{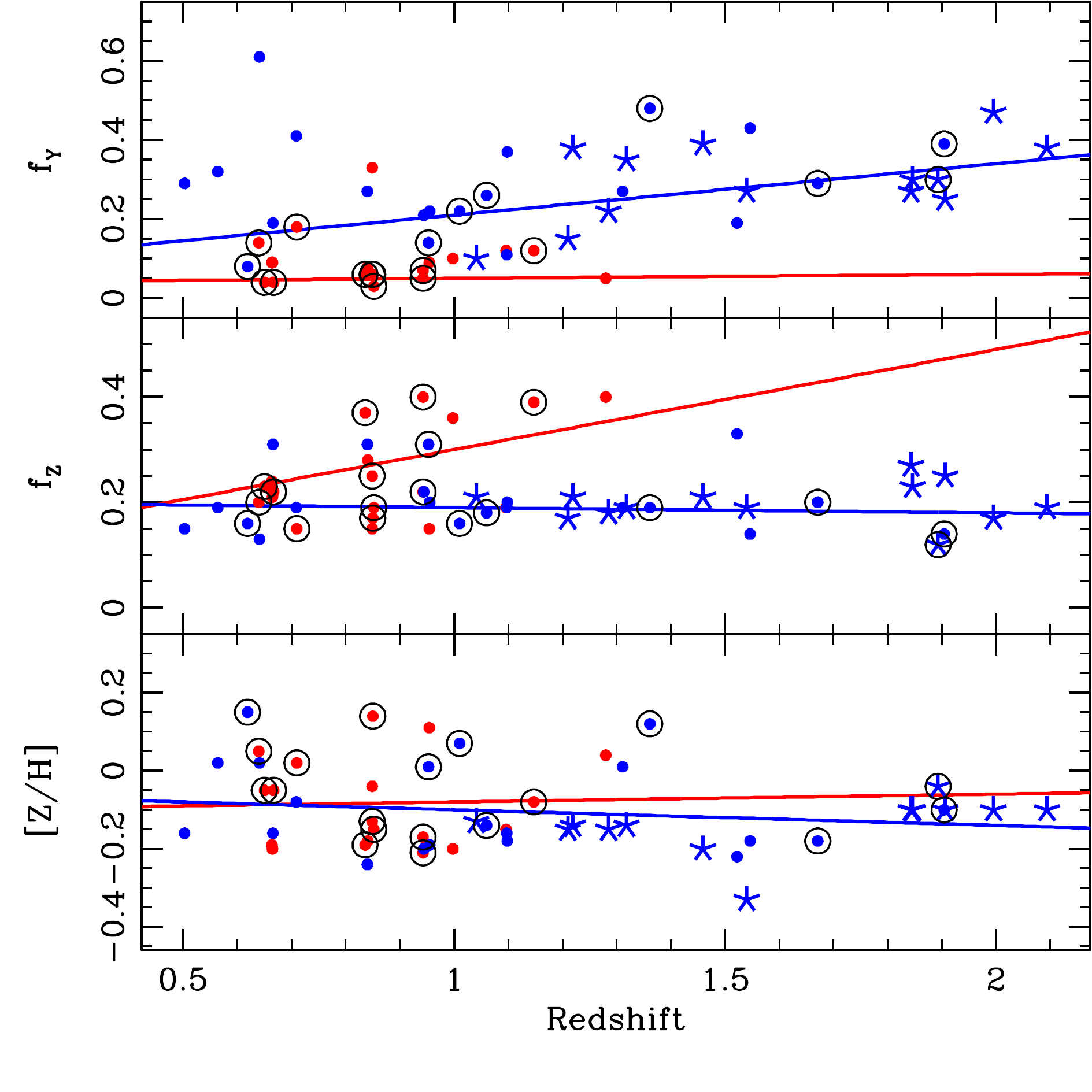}
\end{center}
\caption{This is the equivalent of Fig.~\ref{fig:age}
for the other spectral fitting parameters, from top to bottom:
fraction of mass in young stars (f$_Y$), fraction of mass in
low-metallicity stars (f$_Z$) and average metallicity ([Z/H]), The
sample is split between quiescent (red solid) and star-forming
galaxies (blue open), as defined in Fig.~\ref{fig:UVJ}. In each panel
we show a linear fit to each subsample. Typical
error bars are given at the 1\,$\sigma$ level.
Galaxies that appear more compact than
the local mass-size relation (see Fig.~\ref{fig:MRe})
include an open circle.}
\label{fig:frac}
\end{figure*}
%%%%%%%%%%%%%%%%%%%%%%%%%%%%%%%%%%%%%%%%%%%%%%%%

%%%%%%%%%%%%%%%%%%%%%%%%%%%%%%%%%%%%%%
%%%%%%%%%%   TABLE 3     %%%%%%%%%%%%%
%%%%%%%%%%%%%%%%%%%%%%%%%%%%%%%%%%%%%%
\begin{table}\phantom{...}
\caption{Equivalent of Table~\ref{tab:fits1} using redshift to perform the correlation analysis. 
Linear regression to the results shown in Figs.~\ref{fig:age}
and \ref{fig:frac}. The model for parameter $\pi$ is $\pi=a(z-1) + b$, where $\pi$ corresponds
to the following: average age, $t_{10}$, $\Delta t$, $f_Y$, $f_Z$, or metallicity.
Col.~1 identifies the parameter fit, col.~2 identifies the
sample considered: Q for quiescent and SF for star-forming.
Cols.~3 and 4 give the slope ($a$) and intercept ($b$)
at z=0, respectively. Col.~5 is the linear correlation coefficient. The
error bars, quoted at the 1\,$\sigma$ level, take into account the individual
uncertainties of the measurements.
\label{tab:fits2}}
\begin{center}
\begin{tabular}{c|cccc}
\hline
$\pi$ & Ty & $a$ & $b$ & $\rho_{xy}$\\
(1) & (2) & (3) & (4) & (5)\\
\hline
\multirow{2}{*}{$\langle{\rm Age}\rangle/t_U$} & Q & $+0.44\pm 0.11$  & $0.49\pm 0.10$  & $+0.57\pm 0.11$\\
%\cline{2-5}
 & SF & $+0.09\pm 0.02$  & $0.36\pm 0.03$  & $+0.41\pm 0.11$\\
\hline
\multirow{2}{*}{$t_{10}/t_U$} & Q & $+0.24\pm 0.03$  & $0.51\pm 0.02$  & $+0.72\pm 0.10$\\
 & SF & $+0.14\pm 0.02$  & $0.46\pm 0.02$  & $+0.58\pm 0.14$\\
\hline
\multirow{2}{*}{$\Delta t$} & Q & $-1.06\pm 0.20$ & $1.21\pm 0.18$ & $-0.58\pm 0.11$\\
%\cline{2-5}
 & SF & $-0.12\pm 0.10$ & $1.73\pm 0.12$ & $-0.28\pm 0.14$\\
\hline
\multirow{2}{*}{$f_Y$} & Q & $+0.01\pm 0.04$  & $0.05\pm 0.03$  & $-0.03\pm 0.16$\\
 & SF & $+0.13\pm 0.05$  & $0.21\pm 0.06$  & $+0.19\pm 0.11$\\
\hline
\multirow{2}{*}{$f_Z$} & Q & $+0.19\pm 0.11$  & $0.25\pm 0.10$  & $+0.30\pm 0.20$\\
 & SF & $-0.01\pm 0.02$  & $0.19\pm 0.03$  & $+0.00\pm 0.14$\\
\hline
\multirow{2}{*}{[Z/H]}  & Q & $+0.02\pm 0.15$  & $-0.08\pm 0.13$  & $-0.01\pm  0.19$\\
 & SF & $-0.04\pm 0.04$  & $-0.10\pm 0.05$  & $-0.10\pm 0.15$\\
\hline
\end{tabular}
\end{center}
\end{table}
%%%%%%%%%%%%%%%%%%%%%%%%%%%%%%%%%%%%%%

The trends of the population parameters with stellar mass
are presented in Figs.~\ref{fig:age} and \ref{fig:frac},
following the same notation regarding symbol shape and colour as in
Fig.~\ref{fig:UVJ}.
%%%%%%%%%%
We note that the redshift range covered by our sample maps into a
large interval of cosmic time, between 3.2 and 8.4\,Gyr (quoted as the
age of the Universe at z=2 and z=0.5, respectively). We emphasize that
this paper is not meant to look for one-to-one evolutionary paths of massive galaxies.
At the redshifts covered, these galaxies could have a wide and disjoint
range of potential progenitors \citep[e.g.][]{Choi:14}. We want to study instead, the general properties
of massive galaxies over a period that encompasses the peak of galaxy formation.
These properties reflect the complex mixture of evolutionary trends.
To mitigate the large redshift range covered, the age-related
population parameters that are expected to vary with lookback time are
factored by the age of the Universe at the redshift of the galaxy
($t_U$). Therefore, the average stellar age is replaced by the {\sl
relative} age, defined as the fraction age/$t_U$. For instance, in a
monolithic formation scenario, the old quiescent populations will vary
with redshift similarly to the age of the Universe: a galaxy formed
instantaneously at z$_{\rm FOR}$=3 will have an age/$t_U$ parameter
varying from 0.35 at z=2 to 0.75 at z=0.5 (see dotted line on the
top-right panel of Fig.~\ref{fig:age}).  More recent (earlier)
formation redshifts will result in a wider (narrower) range of
relative ages.  Variations of this parameter will therefore suggest
differences in the stellar age distribution.  Figs.~\ref{fig:age}
and \ref{fig:frac} show the parameters extracted from our methodology
as a function of stellar mass ({\sl left}) and redshift ({\sl right}).

%%%%%%%%%%   Colour gradient     %%%%%%%%%%%%%%%
%%%%%%%%%%%%%%%%%%%%%%%%%%%%%%%%%%%%%%%%%%%%%%%%
\begin{figure*}
\begin{center}
\includegraphics[width=160mm]{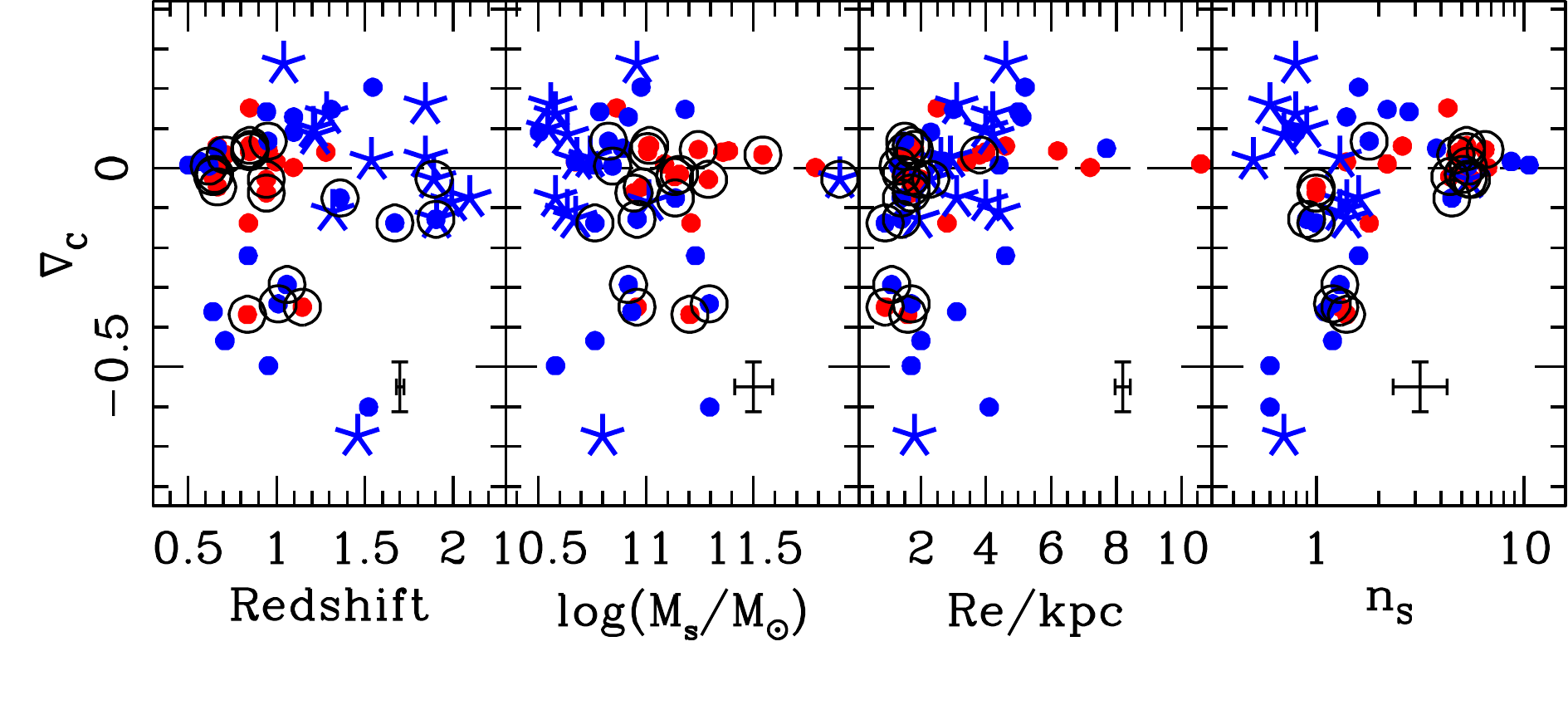}
\includegraphics[width=160mm]{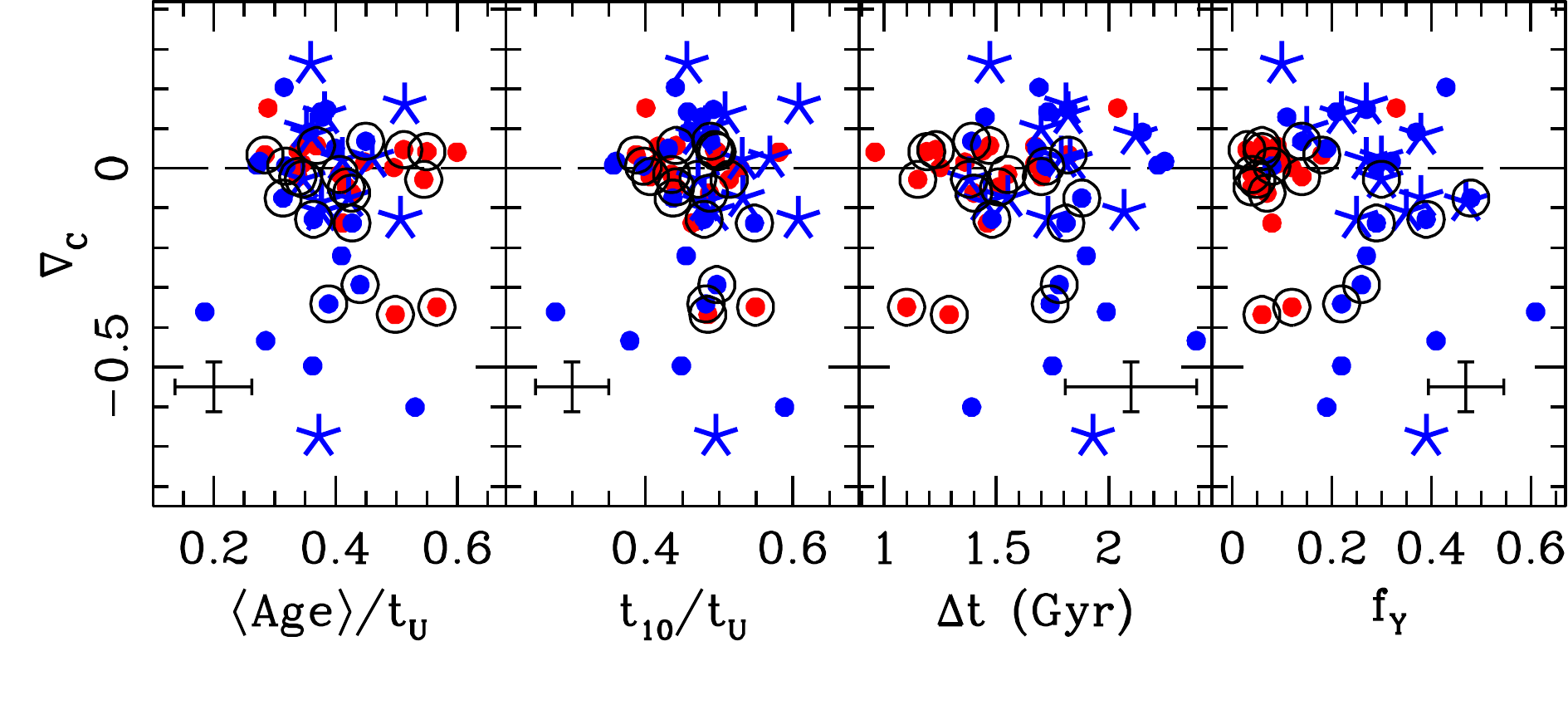}
\end{center}
\caption{Radial colour gradients (defined as $\nabla_C\equiv\Delta C/\Delta\log R$,
where $C\equiv$F125W--F160W), plotted as a function of (top from left to right):
redshift, stellar mass, effective radius in physical units, and
S\'ersic index; (bottom, from left to right): average stellar age,
and $t_{10}$, both relative to the age of the Universe, $\Delta t$,
and mass fraction in young stars The symbols follow the
previous figures, with red/blue representing quiescent/star forming
populations and dot/star plotting early-/late-types,
respectively. Typical error bars are shown at the 1\,$\sigma$
level. Galaxies that appear more compact than the local mass-size
relation (see Fig.~\ref{fig:MRe}) include an open circle.}
\label{fig:clrgrad}
\end{figure*}
%%%%%%%%%%%%%%%%%%%%%%%%%%%%%%%%%%%%%%%%%%%%%%%%

The results from a linear regression analysis of the data presented in these two figures are 
shown in Table~\ref{tab:fits1} (with respect to the logarithm of
stellar mass) and Table~\ref{tab:fits2} (with respect to redshift),
giving the slope, the best fit value at a fiducial point
($10^{11}$M$_\odot$ in mass and z=1 in redshift), as well as the
correlation coefficient $\rho_{xy}$. The errors are quoted at the
1\,$\sigma$ level, as derived by the {\sc SciPy} Orthogonal Distance
Regression package \citep[ODR,][]{ODR}. The errors in the correlation
coefficient -- derived via the {\sc SciPy} stats.linregress package --
are estimated from a Monte Carlo sampling of 100 realizations
produced by adding Gaussian noise as expected from the parameter uncertainties.
The analysis takes into account the uncertainties
in the individual data points, as quoted in the pertinent tables. The
solid lines in Figs.~\ref{fig:age} and \ref{fig:frac} represent the
best fits.  Note the typical mass-related trend such that quiescent
galaxies are more massive than star-forming systems. Moreover, at
fixed stellar mass, the age/$t_U$ ratio is younger in the latter subset,
supporting the use of the UVJ colour-colour diagram to classify
quiescent and star-forming galaxies \citep{Labbe:05}. The lack of quiescent
galaxies at z$\simgt$1.5 cannot be explained by the flux limit of our sample:
an SSP with solar metallicity, formed at redshift z$_{\rm FOR}$=5
\citep[using the models of][]{BC03} has F160W=21.5 at z=1.5 or 22.1 at z=2.0, within the
range of our observations (see Fig.~\ref{fig:SNR}).

%%%%%%%%%%   MIUSCAT comparison  %%%%%%%%%%%%%%%
%%%%%%%%%%%%%%%%%%%%%%%%%%%%%%%%%%%%%%%%%%%%%%%%
\begin{figure}
\begin{center}
\includegraphics[width=85mm]{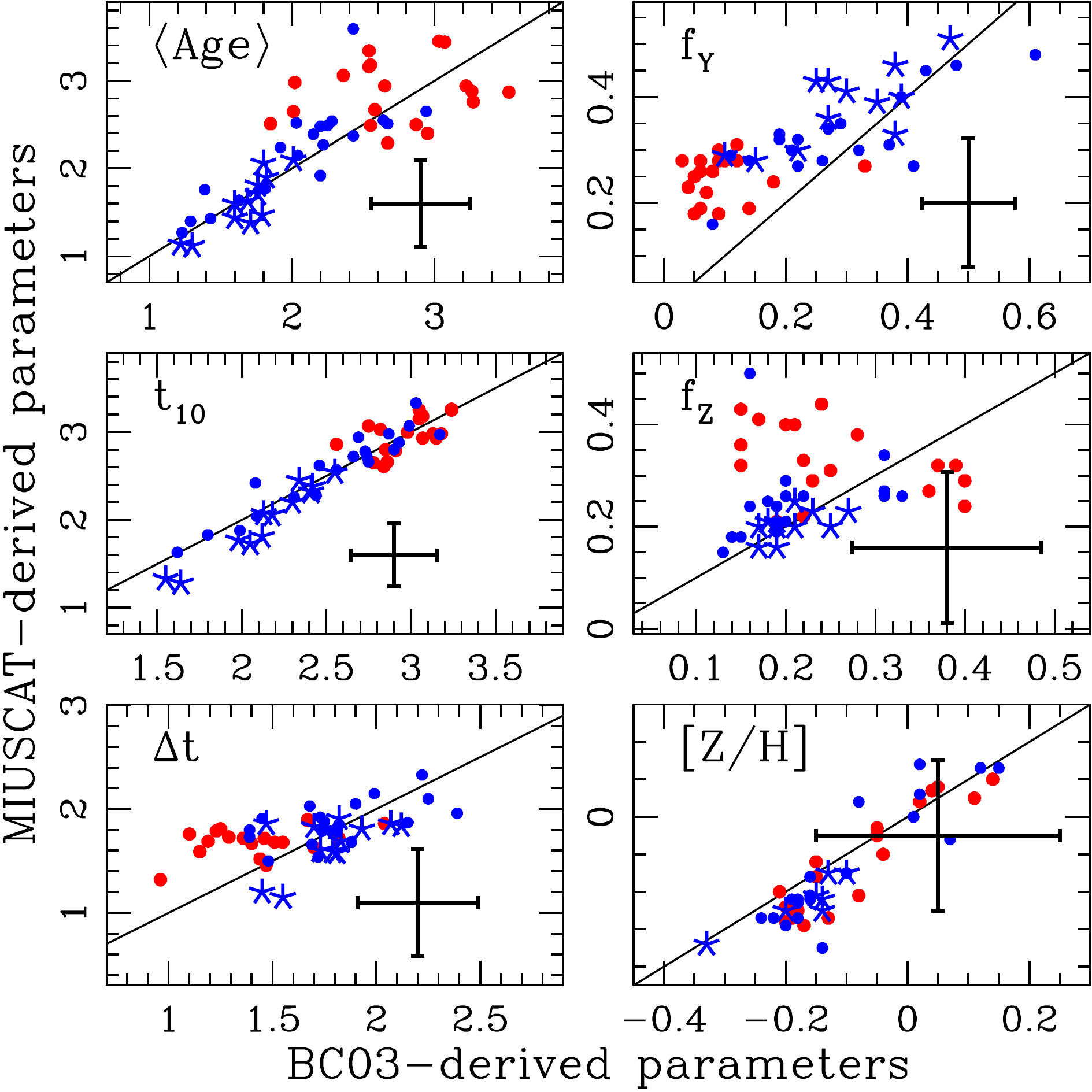}
\end{center}
\caption{Comparison between the model parameters derived from the
fiducial set of BC03 models \citep[][horizontal axes]{BC03} and those extracted
from the MIUSCAT models \citep[][vertical axes]{MIUSCAT}. For reference, a 1:1
straight line is included in all panels, as well as a typical error bar. The notation
of the symbols follow the previous figures. Timescales are shown in Gyr.}
\label{fig:MIUS}
\end{figure}
%%%%%%%%%%%%%%%%%%%%%%%%%%%%%%%%%%%%%%%%%%%%%%%%

Neither average age/$t_U$ nor $t_{10}$/$t_U$ give robust correlations
with stellar mass. The width of the age distribution ($\Delta t$,
bottom panel of Fig.~\ref{fig:age}, left) shows a weak level of correlation in the star-forming subsample,
with a decreasing trend with stellar mass. The quiescent sample appear
to follow a similar decreasing trend, but the $\rho_{xy}$ coefficient
(col.~5 of Table~\ref{tab:fits1}) is, however, compatible with no
correlation, mostly due to the limited mass range of the quiescent
subsample. At fixed mass, the star-forming galaxies have slightly
longer values of $\Delta t$ than the quiescent galaxies.
We emphasize that this sample
is restricted to the massive end, where age-mass trends tend to level
out (see \citealt{Gallazzi:05} for the mass-age trend at low redshift,
or \citealt{Gallazzi:14} at z$\simlt$0.7). With respect to redshift,
significantly increasing trends are found in age/$t_U$ and $t_{10}$/$t_U$.
The top-right panel of Fig.~\ref{fig:age} shows, as a dotted grey line, the
expected evolution of the relative age for a single burst population formed at
z$_{\rm FOR}$=3 (earlier redshifts will shift the curve to higher values).
Such trend -- posited by the traditional monolithic collapse scenario -- is
at odds with the observations, that suggest the opposite behaviour, namely
as time evolves, younger populations are being incorporated into massive
galaxies, decreasing their relative age. The same trend is obtained in the
$t_{10}$/$t_U$ parameter, reinforcing the idea of a continuous
contribution of additional populations. In such a scenario, the
parameter $\Delta t$  should also be expected to produce larger values
(more extended age distributions) at lower redshift, as shown in the
bottom-right panel of Fig.~\ref{fig:age}.
A similar trend is found in a large sample ($\sim$8,500) of 
quiescent galaxies from the ALHAMBRA medium band
survey \citep{LDG:18}, and our results are consistent with the
spectral fitting of stacked SDSS data presented by \citet{Choi:14}.

A significant difference is unsurprisingly found in the distribution
of the fraction in young stars (f$_Y$, top panel of
Fig.~\ref{fig:frac}, left) between the quiescent and the star forming
sample. No trend is noted with stellar mass in either subsample, but a
large scatter is found in star-forming galaxies. The mass fraction in
low-metallicity stars (f$_Z$, middle panel) does not show a
significant difference between these two sets, giving an average
around 20\% of the total stellar mass content in stars with lower (by
a factor of 2) metallicity with respect to the fiducial metallicity of
the best-fit value. However, we note the error bars are larger for
this parameter, and may be more affected by systematic effects
(see \S\S\ref{SSec:Sys}). Finally, the average metallicity (bottom
panel) shows the usual positive correlation with mass, with a large
scatter, although we note the model comparisons produce rather large
uncertainties on the metallicity. As regards to the compact galaxy subsample 
(encircled galaxies in Figs.~\ref{fig:age} and \ref{fig:frac}), no apparent
difference is found, supporting the idea that the compactness criterion,
at fixed mass and redshift, does not segregate the populations with
respect to age \citep{I3}. We emphasize this trend is not at odds with the
age variations found on the mass-size plane at low redshift \citep{Scott:17}, as
the analysis of these galaxies will be affected by the ``second'', {\sl ex-situ} step of
growth within the 2-stage formation paradigm. Namely, the additional material
incorporated via mergers, will potentially introduce a systematic trend making
extended galaxies, at fixed mass, younger.
Regarding the redshift evolution of these
parameters, the star forming subsample features an increasing
trend of $f_Y$ with redshift, as expected from the higher star formation
activity towards cosmic noon. The quiescent sample also features an intriguing 
increasing trend of $f_Z$ with redshift, but the scatter and the low number of
galaxies makes this correlation rather weak. No redshift trend is found with
respect to metallicity.

Fig.~\ref{fig:clrgrad} plots the overall properties of the sample with
respect to the F125W--F160W colour gradient, showing that most of the gradients
are very small, especially in the quiescent, early-type dominated sample.
Some of the star-forming galaxies with late-type morphology have slightly
positive colour gradients (i.e. blue cores), an aspect that may reflect
a central episode of star formation \citep[see, e.g.][]{eGOODS}. However,
the fraction in young stars ($f_Y$) appears not to correlate with colour
gradients, so that our sample includes systems with star formation
taking place either inside or outside of the core. It is also worth pointing out that
the compact subsample (encircled symbols) have either flat or negative colour
gradients (i.e. red cores), suggesting that the bulk of the stellar populations
is located centrally, from a characteristic early, {\sl in-situ} process.

%%%%%%%%%%%%%%%%%%%%%%%%%%%%%%%%%%%%%%%%%%%%%%%%
\subsection{Systematic effects related to population synthesis models}
\label{SSec:Sys}

We explore the potential systematic effects on the derivation of
population parameters by running the same method described
above with base models created from 
the stellar population synthesis models
MIUSCAT \citep{MIUSCAT}, instead of \citet{BC03}. These models use different sets of
prescriptions, interpolation schemes and stellar libraries, so a comparison
allows us to assess the robustness in our extracted parameters and the
resulting error bars. We note that the only two differences in the methodology
are: 1) the youngest base model (BM1), which originally comprises a constant star formation history
between 10 and 100\,Myr in our fiducial models is now restricted to the range
60--100\,Myr as younger ages are not available in MIUSCAT; and 2) the stellar
IMF used is Kroupa Universal \citep{Kr:01}, instead of \citet{Chab:03} for the BC03 models.

Fig.~\ref{fig:MIUS} shows a comparison of the parameters extracted from  these
two different population synthesis models, showing an overall concordance,
especially within error bars. We note that f$_Z$ gives the most discrepant
results, although the expected uncertainties are also rather large.
The comparison also shows a higher mismatch at low f$_Y$ and short $\Delta t$,
but always compatible with the error bars. Therefore, we conclude that,
as a ``lowest-order'' effect, the results found are resilient to variations
among stellar population models.

%%%%%%%%%%%%%%%%%%%%%%%%%%%%%%%%%%%%%%
%%%%%%%%%%   TABLE 4   %%%%%%%%%%%%%%%
%%%%%%%%%%%%%%%%%%%%%%%%%%%%%%%%%%%%%%
\begin{table*}
\caption{Principal Component Analysis: The table shows the first four principal components
along with their projections along the stellar population parameters of this analysis.
Col.~1 is the principal component rank, col.~2 gives the eigenvalue (as a percentage of
total variance), and cols.~3-9 are the PC coefficients $\{c_i\}_{i=1}^7$, corresponding
to the variables listed underneath.
\label{tab:pca}}
\begin{center}
\begin{tabular}{ccccccccc}
\hline
Component & $\lambda$ & $c_1$ & $c_2$ & $c_3$ & $c_4$ & $c_5$ & $c_6$ & $c_7$\\
          & \%        & $\langle{\rm Age}\rangle/t_U$ & $t_{10}/t_U$ & $\Delta t$ & $f_Y$ & $f_Z$ &  E(B--V) & E$_Y$(B--V)\\
\hline
PC1 & 64.8 & $-$0.18610 & $-$0.09052 & $+$0.91275 & $+$0.29944 & $-$0.11724 & $+$0.10723 & $-$0.09569\\
PC2 & 21.6 & $+$0.08181 & $+$0.16539 & $-$0.03391 & $+$0.04958 & $+$0.00385 & $-$0.37349 & $-$0.90710\\
PC3 &  6.1 & $-$0.07380 & $+$0.03626 & $-$0.35153 & $+$0.89635 & $-$0.14246 & $+$0.21271 & $-$0.02610\\
PC4 &  4.2 & $-$0.57993 & $-$0.45080 & $-$0.19841 & $-$0.25330 & $-$0.46815 & $+$0.27040 & $-$0.25425\\
\hline
\end{tabular}
\end{center}
\end{table*}
%%%%%%%%%%%%%%%%%%%%%%%%%%%%%%%%%%%%%%

\subsection{Looking for the driver of population variations with PCA}
\label{SSec:PCA}
We can assess the distribution of the variance in the results with
respect to the various population parameters by applying Principal
Component Analysis (PCA) to the results. PCA consists of creating
linear combinations of the model parameters for the sample so that
these combinations (the principal components, PCs) are
decorrelated. Moreover, these components are commonly sorted in
decreasing order of variance, allowing us to determine which
parameters are most responsible for the variance found in the
sample. Table~\ref{tab:pca} shows the results for the first four
principal components. Note that since the uncertainty in metallicity
is rather large, we opted not to include this parameter in the analysis.
Col.~2 gives the fractional contribution to
the total variance, showing that these four components amount to
over 96\% of the total. The rest of the columns define the PCs
as the coefficients corresponding to each of the model parameters.
The first component (PC1, 64.8\% of variance) is mostly dependent on 
$\Delta t$, and the second one (PC2, 21.6\% of variance) mostly
depends on the dust attenuation (both the diffuse component and the
one only affecting young stars). The third component (PC3, 6.1\%) is
mainly dependent on the mass fraction in young stars, and PC4 is
just shown to illustrate that at lower levels of variance, all model
parameters contribute in a similar way (achieving some sort of noise
level). Therefore, we can say that our analysis mostly discriminates
with respect to the width of the age distribution, $\Delta t$,
the dust attenuation and to a
lesser degree, the fraction in recently formed stars.

%%%%%%%%%%%%%%%%%%%%%%%%%%%%%%%%%%%%%%%%%%%%%%%%
\section{Summary}
\label{Sec:Summ}

By use of the WFC3/NIR slitless grism spectra from the FIGS
survey \citep{FIGS}, we compile a sample of 51 + FW4871 = 52 massive
galaxies (with stellar mass $\log($M$_s/$M$_\odot)\simgt 10.5$) over a
redshift interval corresponding to the peak of galaxy formation
activity ($0.5<z<2$). The NIR spectra are combined with the
observer-frame optical spectra from the PEARS
campaign \citep[e.g.,][]{PEARS}, using the ACS/G800L grism, and
studied by comparing with population synthesis models, adapted to the
resolution of each source, effectively given by a combination of the
resolving power of the grism and the surface brightness profile of the
galaxy.  Our sample comprises a mixture of 19 quiescent and 33
star-forming galaxies (Fig.~\ref{fig:UVJ}).  We find the expected
segregation with respect to stellar age between these two groups, but
no variation with respect to stellar mass -- noting that we are
dealing with massive galaxies, where age and metallicity trends
``level out''. In contrast, we find a significant trend of $\Delta t$
-- a parameter that describes the width of the stellar age distribution -- 
with respect to mass (Fig.~\ref{fig:age}).  Regarding redshift trends,
we find -- consistently with previous work in the literature
\citep[see, e.g.,][]{Stanford:04,Kaviraj:05,Conselice:08,vdK:08,Guo:11} --
that quiescent galaxies do not form following a simple
monolithic collapse, but more stellar populations are added with time
since formation as in the case of star-forming galaxies. The latter
are characterized by a fraction of mass in young stars that keeps
increasing towards the epoch of cosmic noon.  Tables~\ref{tab:fits1}
and \ref{tab:fits2} quantify these relations with respect to stellar
mass and redshift, respectively, including the correlation
coefficient, showing that the trend between mass and $\Delta t$ is the
most conspicuous one.  With respect to redshift, we find that
quiescent galaxies do not form following a simple monolithic collapse,
but more stellar populations are added with time since formation, as
in the case of star-forming galaxies. The latter are characterized by
a fraction of mass in young stars that keeps increasing with z towards
the epoch of cosmic noon.  In order to relate the population
properties with the presence of internal gradients, we explore
potential trends with radial colour gradients, finding no significant
correlation, except for a marked difference between quiescent galaxies
-- with very small colour gradients -- and star-forming galaxies --
that show a much wider range of gradients, both positive (blue cores)
and negative (blue outer envelopes).  The compact massive subsample
has either flat or negative colour gradients, i.e. displaying red
cores (Fig.~\ref{fig:clrgrad}), a result of its {\sl in-situ}, early
formation.

\section*{Acknowledgements}
AC acknowledges grants ASI n.I/023/12/0 and MIUR PRIN 2015 ``Cosmology
and Fundamental Physics: Illuminating the Dark Universe with Euclid''.
The anonymous referee is gratefully acknowledged for useful and
constructive criticism.  Based on observations made with the NASA/ESA
{\sl Hubble Space Telescope}, obtained from the Data Archive at the
Space Telescope Science Institute, which is operated by the
Association of Universities for Research in Astronomy, Inc., under
NASA contract NAS 5-26555. These observations are associated with
program \#13779.  Support for program
\#13779 was provided by NASA through a grant from the Space Telescope
Science Institute, which is operated by the Association of
Universities for Research in Astronomy, Inc., under NASA contract NAS
5-26555.

%%%%%%%%%%%%%%%%%%%%%%%%%%%%%%%%%%%%%%%%%%%%%%%%
%%%%%%%%%%%%%%%   REFERENCES   %%%%%%%%%%%%%%%%%
%%%%%%%%%%%%%%%%%%%%%%%%%%%%%%%%%%%%%%%%%%%%%%%%

%%%%%%%%%%%%%%%%%%%%%%

%%%%%%%%%%%%%%%%%%%%%%%%%%%%%%%%%%%%%%%%%%%%%%%%
\appendix
%%%%%%%%%%%%%%%%%%%%%%%%%%%%%%%%%%%%%%%%%%%%%%%%
\section{Tables}
\label{App:Tables}

This appendix shows three tables with the properties of the
full set of 52 massive galaxies studied in this paper. Table~\ref{tab:massive}
shows the general properties; Table~\ref{tab:massive2} gives the results from the
surface brightness fits and the colour gradients; and Table~\ref{tab:massive3}
presents the results from the stellar population analysis described in \S\ref{Sec:Method}.
All the measurements that require a fit are given as probability-weighted
averages, including, in brackets, the uncertainty at the 1\,$\sigma$ level.

%%%%%%%%%%%%%%%%%%%%%%%%%%%%%%%%%%%%%%
%%%%%%%%%%   TABLE A1   %%%%%%%%%%%%%%%
%%%%%%%%%%%%%%%%%%%%%%%%%%%%%%%%%%%%%%
\begin{table*}
\vskip-0.3truecm
\caption{Properties of the FIGS massive galaxy sample. Col.~1 shows the ID
of the galaxy. Col.~2 flags the presence of G141 grism data, and
col.~3 gives the  morphological type (E=early-type;
L=late-type), and the star formation classification (Q=quiescent; SF=star forming). 
Cols.~4 to 6 give the RA, Dec and redshift of the
galaxy. Col.~7 is the age of the Universe at the redshift of the source.
Col.~8 is the total apparent magnitude in the WFC3/F160W band. Col.~9 is the
best fit stellar mass in units of $10^{11}$M$_\odot$, and cols.~10-11 give the rest-frame U--V and V--J
colours. Cols.~9-11 include the 1\,$\sigma$ uncertainties in
brackets.
\label{tab:massive}}
\begin{center}
\begin{tabular}{ccccccccccc}
\hline
ID & G141? & Ty & RA & Dec & z & $t_U$ & F160W & M/M$_\odot$ & $(U-V)_0$ & $(V-J)_0$\\
   &       & & deg  & deg &  & Gyr & AB &  $\times 10^{11}M_\odot$ & AB & AB\\
(1) & (2) & (3) & (4) & (5) & (6) & (7) & (8) & (9) & (10) & (11)\\
\hline
\multicolumn{11}{|c|}{GN1}\\
\hline
 0.2083 & \cmark & E/SF & 189.191681 & $+$62.283558 & 0.953 & 5.938 & 21.65 & 0.67 (0.13) & 1.95 (0.20) & 1.66 (0.18)\\
 0.2144 & \cmark & E/Q & 189.167313 & $+$62.282146 & 0.943 & 5.980 & 20.80 & 0.88 (0.10) & 1.46 (0.14) & 0.82 (0.07)\\
 0.2183 & \cmark & E/SF & 189.180557 & $+$62.281265 & 0.944 & 5.976 & 20.86 & 0.61 (0.13) & 1.26 (0.17) & 0.96 (0.14)\\
 0.2240 & \cmark & E/Q & 189.155624 & $+$62.279949 & 0.943 & 5.980 & 20.34 & 1.95 (0.29) & 1.63 (0.14) & 1.12 (0.09)\\
 0.2241 & \cmark & E/Q & 189.155060 & $+$62.279434 & 0.852 & 6.384 & 20.15 & 1.75 (0.21) & 1.69 (0.06) & 1.01 (0.08)\\
 0.2350 & \cmark & E/SF & 189.198288 & $+$62.277798 & 1.361 & 4.554 & 22.08 & 1.37 (0.46) & 2.15 (0.24) & 2.49 (0.28)\\
 0.2460 & \xmark & E/SF & 189.193085 & $+$62.274826 & 0.503 & 8.400 & 19.38 & 0.57 (0.14) & 1.53 (0.21) & 1.21 (0.22)\\
 0.2502 & \cmark & E/Q & 189.145264 & $+$62.274536 & 0.849 & 6.398 & 20.81 & 0.73 (0.10) & 1.74 (0.14) & 1.38 (0.13)\\
 0.2589 & \cmark & E/Q & 189.163361 & $+$62.273373 & 0.849 & 6.398 & 20.73 & 1.02 (0.15) & 1.61 (0.17) & 1.15 (0.12)\\
 0.2590 & \cmark & E/Q & 189.163696 & $+$62.272980 & 0.850 & 6.394 & 20.89 & 1.04 (0.12) & 2.09 (0.09) & 1.58 (0.08)\\
\hline
\multicolumn{11}{|c|}{GN2}\\
\hline
 1.0411 & \cmark & E/Q & 189.397415 & $+$62.329533 & 1.147 & 5.207 & 21.81 & 0.91 (0.19) & 1.78 (0.20) & 1.43 (0.19)\\
 1.0418 & \cmark & E/SF & 189.360031 & $+$62.329098 & 1.010 & 5.708 & 20.53 & 1.97 (0.38) & 1.86 (0.19) & 1.56 (0.17)\\
 1.0463 & \cmark & E/SF & 189.388565 & $+$62.326694 & 1.060 & 5.517 & 21.20 & 0.83 (0.32) & 1.48 (0.35) & 1.44 (0.36)\\
 1.0554 & \cmark & E/Q & 189.397171 & $+$62.321640 & 0.836 & 6.460 & 20.09 & 1.60 (0.25) & 1.53 (0.16) & 1.08 (0.10)\\
 1.0575 & \cmark & E/SF & 189.357513 & $+$62.320595 & 1.522 & 4.144 & 21.28 & 1.98 (0.60) & 1.43 (0.24) & 1.14 (0.20)\\
 1.0620 & \cmark & L/SF & 189.350510 & $+$62.318043 & 2.094 & 3.081 & 22.45 & 0.38 (0.15) & 0.61 (0.13) & 0.55 (0.11)\\
 1.0678 & \cmark & L/SF & 189.367935 & $+$62.315254 & 1.459 & 4.297 & 22.00 & 0.63 (0.24) & 1.28 (0.22) & 1.24 (0.21)\\
 1.0687 & \xmark & E/SF & 189.418457 & $+$62.314888 & 0.955 & 5.930 & 21.45 & 0.38 (0.09) & 1.37 (0.18) & 1.06 (0.18)\\
 1.0704 & \cmark & E/Q & 189.356262 & $+$62.313892 & 0.841 & 6.436 & 19.86 & 1.62 (0.20) & 1.49 (0.12) & 0.91 (0.09)\\
 1.0743 & \cmark & E/SF & 189.381577 & $+$62.311573 & 1.671 & 3.815 & 22.28 & 0.58 (0.21) & 1.09 (0.25) & 0.89 (0.23)\\
 1.0959 & \xmark & E/SF & 189.398026 & $+$62.301456 & 0.840 & 6.441 & 19.54 & 1.70 (0.46) & 1.48 (0.22) & 1.34 (0.24)\\
 1.1219 & \xmark & E/SF & 189.400070 & $+$62.290546 & 0.709 & 7.110 & 20.25 & 0.58 (0.18) & 1.83 (0.25) & 1.93 (0.28)\\
 1.1240 & \xmark & E/SF & 189.393906 & $+$62.289795 & 0.641 & 7.501 & 19.18 & 0.86 (0.21) & 1.90 (0.27) & 2.27 (0.32)\\
 1.1314 & \xmark & E/SF & 189.360809 & $+$62.287090 & 0.564 & 7.984 & 19.96 & 0.47 (0.13) & 1.83 (0.26) & 1.76 (0.29)\\
\hline
\multicolumn{11}{|c|}{GS1}\\
\hline
 2.0724 & \cmark & L/SF & 53.172264 & $-$27.760622 & 1.540 & 4.102 & 21.82 & 0.48 (0.16) & 0.74 (0.16) & 0.47 (0.11)\\
 2.0930 & \cmark & L/SF & 53.181194 & $-$27.765678 & 1.219 & 4.972 & 21.31 & 0.43 (0.14) & 0.82 (0.13) & 0.81 (0.13)\\
 2.0967 & \cmark & L/SF & 53.166328 & $-$27.768587 & 1.210 & 5.000 & 21.87 & 0.35 (0.06) & 0.98 (0.13) & 0.58 (0.08)\\
 2.0980 & \cmark & E/SF & 53.165573 & $-$27.769794 & 1.546 & 4.088 & 21.78 & 0.95 (0.33) & 1.37 (0.23) & 1.29 (0.23)\\
 2.1013 & \cmark & E/Q & 53.169926 & $-$27.771027 & 0.664 & 7.365 & 19.54 & 1.22 (0.13) & 1.59 (0.11) & 1.00 (0.09)\\
 2.1220 & \cmark & L/SF & 53.176052 & $-$27.773706 & 1.285 & 4.770 & 21.81 & 0.38 (0.09) & 0.82 (0.17) & 0.52 (0.10)\\
 2.1275 & \cmark & E/Q & 53.152771 & $-$27.775288 & 0.998 & 5.755 & 21.56 & 0.47 (0.07) & 1.36 (0.14) & 0.86 (0.09)\\
 2.1494 & \cmark & E/SF & 53.145237 & $-$27.777905 & 1.098 & 5.378 & 21.72 & 0.32 (0.10) & 1.14 (0.20) & 1.13 (0.20)\\
 2.1594 & \cmark & L/SF & 53.155647 & $-$27.779299 & 1.846 & 3.480 & 22.01 & 0.55 (0.20) & 0.67 (0.13) & 0.63 (0.11)\\
 2.1630 & \cmark & E/SF & 53.161633 & $-$27.780252 & 0.619 & 7.634 & 20.34 & 0.70 (0.08) & 2.11 (0.10) & 1.63 (0.10)\\
 2.1922 & \cmark & E/Q & 53.160347 & $-$27.784008 & 0.954 & 5.934 & 20.17 & 2.42 (0.20) & 1.99 (0.09) & 1.46 (0.09)\\
 2.2061 & \cmark & E/SF & 53.176579 & $-$27.785448 & 1.311 & 4.694 & 21.46 & 1.52 (0.45) & 1.79 (0.26) & 1.68 (0.25)\\
 2.2084 & \cmark & E/Q & 53.165165 & $-$27.785872 & 1.280 & 4.785 & 21.21 & 2.28 (0.27) & 1.83 (0.07) & 1.22 (0.07)\\
 2.2166 & \cmark & E/SF & 53.166176 & $-$27.787518 & 1.097 & 5.382 & 20.82 & 0.83 (0.11) & 0.99 (0.14) & 0.62 (0.08)\\
 2.2211 & \cmark & E/Q & 53.172523 & $-$27.788107 & 0.640 & 7.507 & 19.61 & 1.36 (0.15) & 1.96 (0.12) & 1.51 (0.13)\\
 2.2213 & \cmark & L/SF & 53.161667 & $-$27.787436 & 1.843 & 3.485 & 22.48 & 0.36 (0.19) & 0.58 (0.18) & 0.54 (0.11)\\
 2.2291 & \cmark & L/SF & 53.149296 & $-$27.788527 & 1.906 & 3.376 & 22.49 & 0.47 (0.21) & 0.75 (0.19) & 0.66 (0.12)\\
 2.2406 & \cmark & L/SF & 53.153847 & $-$27.790684 & 1.318 & 4.674 & 21.72 & 0.43 (0.14) & 0.91 (0.16) & 0.83 (0.14)\\
 2.2408 & \cmark & E/Q & 53.155449 & $-$27.791491 & 0.710 & 7.105 & 18.80 & 3.49 (0.43) & 1.96 (0.14) & 1.51 (0.16)\\
 2.2501 & \cmark & E/Q & 53.169449 & $-$27.791927 & 0.667 & 7.348 & 20.20 & 0.96 (0.12) & 1.89 (0.07) & 1.27 (0.07)\\
 2.2794 & \cmark & L/SF & 53.176189 & $-$27.796133 & 1.041 & 5.588 & 20.70 & 0.91 (0.12) & 1.20 (0.13) & 0.72 (0.07)\\
 2.2841 & \cmark & E/SF & 53.158806 & $-$27.797157 & 1.904 & 3.379 & 22.22 & 0.91 (0.27) & 1.20 (0.20) & 0.91 (0.17)\\
 2.2922 & \cmark & L/SF & 53.166897 & $-$27.798733 & 1.995 & 3.231 & 21.65 & 1.03 (0.36) & 0.90 (0.14) & 0.96 (0.15)\\
 2.2923 & \cmark & E/SF & 53.180233 & $-$27.798927 & 0.666 & 7.354 & 19.91 & 0.78 (0.16) & 1.47 (0.18) & 1.25 (0.17)\\
 2.2956 & \cmark & E/Q & 53.163414 & $-$27.799547 & 0.650 & 7.447 & 19.67 & 1.43 (0.19) & 1.88 (0.07) & 1.24 (0.08)\\
 2.4198 & \cmark & E/Q & 53.178375 & $-$27.768240 & 0.665 & 7.359 & 19.69 & 1.02 (0.12) & 1.50 (0.14) & 0.92 (0.10)\\
 2.4272 & \cmark & E/Q & 53.154968 & $-$27.768909 & 1.096 & 5.385 & 19.49 & 6.13 (0.95) & 1.71 (0.11) & 1.16 (0.15)\\
\hline
 FW4871 & \cmark & L/SF & 53.062442 & $-$27.706903 & 1.893 & 3.398 & 19.81 & 7.95 (1.57) & 1.26 (0.15) & 0.73 (0.14)\\
\hline
\end{tabular}
\end{center}
\end{table*}
%%%%%%%%%%%%%%%%%%%%%%%%%%%%%%%%%%%%%%%%%%%%%%%%

%%%%%%%%%%%%%%%%%%%%%%%%%%%%%%%%%%%%%%
%%%%%%%%%%   TABLE A2   %%%%%%%%%%%%%%
%%%%%%%%%%%%%%%%%%%%%%%%%%%%%%%%%%%%%%
\begin{table}
\vskip-0.3truecm
\caption{Continuation of Table~\ref{tab:massive}, listing the 
properties related to the surface brightness (F160W) and colour distribution.
Col.~1 is the galaxy ID. Col.~2 is the circularized effective radius, in physical
units. Col.~3 is the S\'ersic index, and col.~4 is
the colour gradient ($\nabla_C\equiv\Delta C/\Delta\log R$,
where $C\equiv$F125W--F160W). Values
in brackets denote the 1\,$\sigma$ uncertainties.
\label{tab:massive2}}
\begin{center}
\begin{tabular}{cccc}
\hline
ID & R$_e$ & n$_S$ & $\nabla_C$\\
   & kpc   &  &  AB\\
(1) & (2) & (3) & (4)\\
\hline
\multicolumn{4}{|c|}{GN1}\\
\hline
 0.2083 &  1.5 (0.2) &  1.8 (0.1) & $+$0.068 (0.040) \\
 0.2144 &  1.7 (0.2) &  1.0 (0.1) & $-$0.062 (0.073) \\
 0.2183 &  5.0 (0.3) &  2.8 (0.1) & $+$0.142 (0.028) \\
 0.2240 &  2.0 (0.2) &  5.6 (0.1) & $-$0.028 (0.017) \\
 0.2241 &  1.8 (0.2) &  6.5 (0.1) & $+$0.047 (0.040) \\
 0.2350 &  1.4 (0.2) &  4.5 (0.2) & $-$0.075 (0.023) \\
 0.2460 &  4.4 (0.2) & 10.6 (0.2) & $+$0.008 (0.018) \\
 0.2502 &  2.5 (0.2) &  4.3 (0.1) & $+$0.152 (0.016) \\
 0.2589 &  1.5 (0.2) &  5.3 (0.1) & $+$0.042 (0.033) \\
 0.2590 &  1.7 (0.2) &  5.3 (0.1) & $+$0.057 (0.047) \\
\hline
\multicolumn{4}{|c|}{GN2}\\
\hline
 1.0411 &  0.9 (0.3) &  1.3 (0.4) & $-$0.349 (0.248) \\
 1.0418 &  1.7 (0.2) &  1.2 (0.1) & $-$0.341 (0.067) \\
 1.0463 &  1.1 (0.2) &  1.3 (0.2) & $-$0.293 (0.081) \\
 1.0554 &  1.6 (0.2) &  1.4 (0.1) & $-$0.368 (0.049) \\
 1.0575 &  4.1 (0.3) &  0.6 (0.1) & $-$0.601 (0.112) \\
 1.0620 &  3.1 (0.2) &  1.6 (0.1) & $-$0.074 (0.047) \\
 1.0678 &  1.8 (0.3) &  0.7 (0.2) & $-$0.674 (0.259) \\
 1.0687 &  1.7 (0.2) &  0.6 (0.1) & $-$0.497 (0.152) \\
 1.0704 &  2.8 (0.2) &  1.8 (0.1) & $-$0.138 (0.035) \\
 1.0743 &  0.9 (0.3) &  1.0 (0.8) & $-$0.138 (0.035) \\
 1.0959 &  4.6 (0.3) &  1.6 (0.1) & $-$0.220 (0.043) \\
 1.1219 &  2.0 (0.2) &  1.2 (0.1) & $-$0.434 (0.063) \\
 1.1240 &  3.1 (0.2) &  1.1 (0.1) & $-$0.361 (0.070) \\
 1.1314 &  2.7 (0.2) &  8.7 (0.2) & $+$0.017 (0.022) \\
\hline
\multicolumn{4}{|c|}{GS1}\\
\hline
 2.0724 &  2.6 (0.2) &  0.5 (0.1) & $+$0.021 (0.151) \\
 2.0930 &  4.2 (0.3) &  0.7 (0.1) & $+$0.083 (0.103) \\
 2.0967 &  4.0 (0.2) &  0.9 (0.1) & $+$0.100 (0.089) \\
 2.0980 &  5.2 (0.3) &  1.6 (0.1) & $+$0.204 (0.052) \\
 2.1013 & 10.6 (0.3) &  2.2 (0.1) & $+$0.011 (0.034) \\
 2.1220 &  4.2 (0.3) &  0.8 (0.1) & $+$0.136 (0.100) \\
 2.1275 &  3.4 (0.2) &  1.4 (0.1) & $+$0.016 (0.055) \\
 2.1494 &  2.3 (0.2) &  0.8 (0.1) & $+$0.091 (0.094) \\
 2.1594 &  2.9 (0.2) &  1.3 (0.1) & $+$0.026 (0.065) \\
 2.1630 &  1.3 (0.1) &  5.4 (0.1) & $+$0.006 (0.012) \\
 2.1922 &  6.2 (0.3) &  5.1 (0.1) & $+$0.044 (0.017) \\
 2.2061 &  3.0 (0.2) &  2.2 (0.1) & $+$0.148 (0.033) \\
 2.2084 &  4.0 (0.3) &  4.9 (0.1) & $+$0.041 (0.028) \\
 2.2166 &  5.1 (0.3) &  1.4 (0.1) & $+$0.129 (0.047) \\
 2.2211 &  1.7 (0.2) &  4.4 (0.1) & $-$0.021 (0.012) \\
 2.2213 &  3.1 (0.2) &  0.6 (0.1) & $+$0.160 (0.136) \\
 2.2291 &  1.9 (0.2) &  1.4 (0.1) & $-$0.127 (0.049) \\
 2.2406 &  4.4 (0.3) &  1.3 (0.1) & $-$0.109 (0.064) \\
 2.2408 &  3.8 (0.2) &  4.7 (0.1) & $+$0.034 (0.010) \\
 2.2501 &  1.7 (0.2) &  1.0 (0.1) & $-$0.047 (0.078) \\
 2.2794 &  4.6 (0.3) &  0.8 (0.1) & $+$0.263 (0.091) \\
 2.2841 &  1.4 (0.2) &  0.9 (0.1) & $-$0.128 (0.111) \\
 2.2922 &  4.0 (0.2) &  1.4 (0.1) & $-$0.085 (0.062) \\
 2.2923 &  7.7 (0.4) &  3.8 (0.1) & $+$0.050 (0.022) \\
 2.2956 &  1.4 (0.2) &  5.1 (0.1) & $-$0.015 (0.012) \\
 2.4198 &  4.6 (0.3) &  2.6 (0.1) & $+$0.056 (0.062) \\
 2.4272 &  7.2 (0.3) &  6.7 (0.1) & $+$0.002 (0.010) \\
\hline
 FW4871 &  2.3 (0.2) &  5.4 (0.2) & $-$0.027 (0.036) \\
\hline
\end{tabular}
\end{center}
\end{table}
%%%%%%%%%%%%%%%%%%%%%%%%%%%%%%%%%%%%%%%%%%%%%%%%

%%%%%%%%%%%%%%%%%%%%%%%%%%%%%%%%%%%%%%
%%%%%%%%%%   TABLE A3  %%%%%%%%%%%%%%%
%%%%%%%%%%%%%%%%%%%%%%%%%%%%%%%%%%%%%%
\begin{table*}
\vskip-0.3truecm
\caption{Continuation of Table~\ref{tab:massive2}, with an additional set of
population properties. Col.~1 is the ID of the galaxy. Col.~2 is the
best-fit metallicity (used in all base model, except BM6 (which has a
metallicity reduced by -0.3\,dex). Col.~3 is the average stellar age.
Col.~4 is the stellar mass fraction
in young stars (BM1 and BM2). Col.~5 is the mass fraction in
low-metallicity stars (BM6). Col.~6 is the time when the initial 10\% of the total
stellar mass was formed, and col.~7 is the width of the age distribution (defined
as $\langle{\rm Age}\rangle-t_{90}$). Col.~8 and 9 are the colour excess of the global population,
and the young components, respectively.
\label{tab:massive3}}
\begin{center}
\begin{tabular}{ccccccccc}
\hline
ID & [Z/H] & $\langle$Age$\rangle$ & f$_Y$ & f$_Z$ & t$_{10}$ & $\Delta$t & E(B--V) & E(B--V)$_Y$\\
   &       & Gyr &  &  & Gyr & Gyr & AB & AB\\
(1) & (2) & (3) & (4) & (5) & (6) & (7) & (8) & (9)\\
\hline
\multicolumn{9}{|c|}{GN1}\\
\hline
 0.2083 & $+$0.01 (0.18) & 2.67 (0.37) & 0.14 (0.06) & 0.31 (0.13) & 2.90 (0.20) & 1.39 (0.18) & 0.27 (0.05) & 0.54 (0.15)\\
 0.2144 & $-$0.21 (0.21) & 2.55 (0.31) & 0.07 (0.03) & 0.22 (0.11) & 2.91 (0.13) & 1.40 (0.12) & 0.02 (0.02) & 0.35 (0.13)\\
 0.2183 & $-$0.20 (0.21) & 2.25 (0.37) & 0.21 (0.08) & 0.22 (0.11) & 2.73 (0.25) & 1.73 (0.33) & 0.06 (0.04) & 0.36 (0.09)\\
 0.2240 & $-$0.17 (0.21) & 3.26 (0.30) & 0.05 (0.02) & 0.40 (0.15) & 3.07 (0.06) & 1.15 (0.10) & 0.12 (0.03) & 0.36 (0.10)\\
 0.2241 & $-$0.15 (0.25) & 3.27 (0.35) & 0.03 (0.02) & 0.19 (0.10) & 3.15 (0.08) & 1.23 (0.11) & 0.05 (0.03) & 0.64 (0.18)\\
 0.2350 & $+$0.12 (0.11) & 1.43 (0.35) & 0.48 (0.12) & 0.19 (0.10) & 1.99 (0.43) & 1.88 (0.44) & 0.39 (0.08) & 0.69 (0.13)\\
 0.2460 & $-$0.16 (0.25) & 2.28 (0.54) & 0.29 (0.11) & 0.15 (0.09) & 2.99 (0.39) & 2.22 (0.37) & 0.13 (0.07) & 0.43 (0.15)\\
 0.2502 & $-$0.04 (0.22) & 1.85 (0.37) & 0.33 (0.11) & 0.15 (0.08) & 2.56 (0.36) & 2.04 (0.36) & 0.24 (0.04) & 0.35 (0.07)\\
 0.2589 & $-$0.13 (0.22) & 3.52 (0.25) & 0.06 (0.03) & 0.25 (0.13) & 3.18 (0.04) & 1.19 (0.08) & 0.08 (0.04) & 0.43 (0.14)\\
 0.2590 & $+$0.14 (0.09) & 2.36 (0.39) & 0.06 (0.04) & 0.17 (0.09) & 2.82 (0.34) & 1.47 (0.22) & 0.27 (0.03) & 0.48 (0.14)\\
\hline
\multicolumn{9}{|c|}{GN2}\\
\hline
 1.0411 & $-$0.08 (0.22) & 2.95 (0.24) & 0.12 (0.06) & 0.39 (0.14) & 2.86 (0.06) & 1.10 (0.19) & 0.12 (0.06) & 0.64 (0.18)\\
 1.0418 & $+$0.07 (0.14) & 2.22 (0.36) & 0.22 (0.09) & 0.16 (0.10) & 2.75 (0.16) & 1.74 (0.39) & 0.26 (0.06) & 0.45 (0.14)\\
 1.0463 & $-$0.14 (0.22) & 2.43 (0.40) & 0.26 (0.10) & 0.18 (0.10) & 2.74 (0.18) & 1.78 (0.56) & 0.22 (0.09) & 0.38 (0.16)\\
 1.0554 & $-$0.19 (0.22) & 3.22 (0.40) & 0.06 (0.03) & 0.37 (0.16) & 3.13 (0.11) & 1.29 (0.13) & 0.12 (0.03) & 0.30 (0.09)\\
 1.0575 & $-$0.22 (0.19) & 2.20 (0.30) & 0.19 (0.09) & 0.33 (0.12) & 2.44 (0.17) & 1.39 (0.46) & 0.11 (0.06) & 0.43 (0.13)\\
 1.0620 & $-$0.10 (0.26) & 1.30 (0.25) & 0.38 (0.11) & 0.19 (0.09) & 1.64 (0.33) & 1.55 (0.33) & 0.03 (0.03) & 0.10 (0.04)\\
 1.0678 & $-$0.20 (0.20) & 1.60 (0.35) & 0.39 (0.12) & 0.21 (0.09) & 2.13 (0.35) & 1.93 (0.39) & 0.18 (0.06) & 0.34 (0.09)\\
 1.0687 & $-$0.19 (0.23) & 2.15 (0.36) & 0.22 (0.08) & 0.20 (0.11) & 2.66 (0.25) & 1.75 (0.36) & 0.05 (0.05) & 0.45 (0.1)\\
 1.0704 & $-$0.18 (0.24) & 2.65 (0.40) & 0.08 (0.04) & 0.28 (0.13) & 2.98 (0.23) & 1.46 (0.18) & 0.05 (0.03) & 0.33 (0.11)\\
 1.0743 & $-$0.18 (0.20) & 1.63 (0.28) & 0.29 (0.10) & 0.20 (0.10) & 2.09 (0.23) & 1.81 (0.29) & 0.07 (0.06) & 0.28 (0.10)\\
 1.0959 & $-$0.24 (0.18) & 2.64 (0.41) & 0.27 (0.09) & 0.31 (0.12) & 2.93 (0.19) & 1.90 (0.53) & 0.10 (0.07) & 0.54 (0.11)\\
 1.1219 & $-$0.08 (0.18) & 2.03 (0.47) & 0.41 (0.11) & 0.19 (0.09) & 2.69 (0.36) & 2.39 (0.41) & 0.31 (0.09) & 0.54 (0.12)\\
 1.1240 & $+$0.02 (0.11) & 1.39 (0.41) & 0.61 (0.10) & 0.13 (0.07) & 2.08 (0.61) & 1.99 (0.62) & 0.37 (0.10) & 0.56 (0.13)\\
 1.1314 & $+$0.02 (0.17) & 2.20 (0.49) & 0.32 (0.11) & 0.19 (0.10) & 2.87 (0.40) & 2.25 (0.40) & 0.20 (0.09) & 0.68 (0.17)\\
\hline
\multicolumn{9}{|c|}{GS1}\\
\hline
 2.0724 & $-$0.33 (0.10) & 1.69 (0.31) & 0.27 (0.10) & 0.19 (0.10) & 2.18 (0.29) & 1.83 (0.35) & 0.03 (0.03) & 0.12 (0.04)\\
 2.0930 & $-$0.14 (0.26) & 1.80 (0.37) & 0.38 (0.11) & 0.21 (0.10) & 2.34 (0.35) & 2.12 (0.40) & 0.09 (0.04) & 0.18 (0.05)\\
 2.0967 & $-$0.15 (0.25) & 1.76 (0.28) & 0.15 (0.06) & 0.17 (0.09) & 2.41 (0.23) & 1.70 (0.25) & 0.01 (0.02) & 0.15 (0.05)\\
 2.0980 & $-$0.18 (0.22) & 1.29 (0.30) & 0.43 (0.12) & 0.14 (0.07) & 1.80 (0.41) & 1.69 (0.38) & 0.21 (0.07) & 0.35 (0.10)\\
 2.1013 & $-$0.19 (0.22) & 2.54 (0.39) & 0.09 (0.05) & 0.24 (0.12) & 3.05 (0.25) & 1.67 (0.15) & 0.10 (0.03) & 0.33 (0.09)\\
 2.1220 & $-$0.15 (0.26) & 1.82 (0.28) & 0.22 (0.09) & 0.18 (0.09) & 2.42 (0.19) & 1.82 (0.30) & 0.02 (0.02) & 0.09 (0.04)\\
 2.1275 & $-$0.20 (0.22) & 2.58 (0.36) & 0.10 (0.05) & 0.36 (0.14) & 2.85 (0.16) & 1.36 (0.18) & 0.03 (0.02) & 0.32 (0.08)\\
 2.1494 & $-$0.18 (0.23) & 1.92 (0.39) & 0.37 (0.11) & 0.20 (0.10) & 2.46 (0.34) & 2.15 (0.45) & 0.16 (0.06) & 0.29 (0.08)\\
 2.1594 & $-$0.10 (0.26) & 1.60 (0.28) & 0.30 (0.11) & 0.23 (0.11) & 1.98 (0.29) & 1.79 (0.33) & 0.04 (0.03) & 0.11 (0.03)\\
 2.1630 & $+$0.15 (0.08) & 2.43 (0.43) & 0.08 (0.04) & 0.16 (0.10) & 3.03 (0.32) & 1.72 (0.18) & 0.30 (0.03) & 0.50 (0.15)\\
 2.1922 & $+$0.11 (0.11) & 2.01 (0.25) & 0.09 (0.05) & 0.15 (0.08) & 2.56 (0.31) & 1.44 (0.21) & 0.23 (0.03) & 0.45 (0.11)\\
 2.2061 & $+$0.01 (0.18) & 1.81 (0.30) & 0.27 (0.09) & 0.19 (0.09) & 2.31 (0.31) & 1.82 (0.35) & 0.23 (0.07) & 0.57 (0.16)\\
 2.2084 & $+$0.04 (0.15) & 2.87 (0.23) & 0.05 (0.02) & 0.40 (0.16) & 2.78 (0.05) & 0.96 (0.10) & 0.08 (0.03) & 0.67 (0.18)\\
 2.2166 & $-$0.16 (0.24) & 2.04 (0.29) & 0.11 (0.04) & 0.19 (0.10) & 2.56 (0.26) & 1.45 (0.21) & 0.01 (0.02) & 0.12 (0.05)\\
 2.2211 & $+$0.05 (0.17) & 2.54 (0.46) & 0.14 (0.07) & 0.20 (0.11) & 3.05 (0.36) & 1.70 (0.25) & 0.25 (0.04) & 0.46 (0.12)\\
 2.2213 & $-$0.10 (0.26) & 1.79 (0.28) & 0.27 (0.11) & 0.27 (0.12) & 2.12 (0.17) & 1.81 (0.42) & 0.02 (0.02) & 0.07 (0.03)\\
 2.2291 & $-$0.10 (0.26) & 1.71 (0.27) & 0.25 (0.10) & 0.25 (0.11) & 2.05 (0.22) & 1.73 (0.36) & 0.04 (0.03) & 0.12 (0.05)\\
 2.2406 & $-$0.14 (0.26) & 1.76 (0.31) & 0.35 (0.10) & 0.19 (0.10) & 2.30 (0.28) & 2.07 (0.33) & 0.08 (0.04) & 0.19 (0.05)\\
 2.2408 & $+$0.02 (0.20) & 2.02 (0.34) & 0.18 (0.08) & 0.15 (0.09) & 2.75 (0.35) & 1.82 (0.26) & 0.28 (0.05) & 0.45 (0.10)\\
 2.2501 & $-$0.05 (0.21) & 3.07 (0.49) & 0.04 (0.02) & 0.22 (0.12) & 3.24 (0.17) & 1.51 (0.13) & 0.16 (0.02) & 0.59 (0.18)\\
 2.2794 & $-$0.13 (0.25) & 2.01 (0.29) & 0.10 (0.05) & 0.21 (0.11) & 2.55 (0.32) & 1.47 (0.23) & 0.01 (0.01) & 0.21 (0.06)\\
 2.2841 & $-$0.10 (0.26) & 1.23 (0.23) & 0.39 (0.11) & 0.14 (0.07) & 1.62 (0.36) & 1.48 (0.32) & 0.08 (0.05) & 0.27 (0.09)\\
 2.2922 & $-$0.10 (0.26) & 1.22 (0.26) & 0.47 (0.11) & 0.17 (0.08) & 1.55 (0.39) & 1.45 (0.40) & 0.11 (0.04) & 0.21 (0.05)\\
 2.2923 & $-$0.16 (0.21) & 2.94 (0.46) & 0.19 (0.08) & 0.31 (0.11) & 3.17 (0.20) & 1.68 (0.24) & 0.17 (0.05) & 0.38 (0.09)\\
 2.2956 & $-$0.05 (0.21) & 3.03 (0.49) & 0.04 (0.02) & 0.23 (0.12) & 3.24 (0.18) & 1.55 (0.14) & 0.15 (0.03) & 0.60 (0.17)\\
 2.4198 & $-$0.20 (0.22) & 2.55 (0.42) & 0.09 (0.05) & 0.21 (0.11) & 3.07 (0.24) & 1.67 (0.16) & 0.09 (0.03) & 0.28 (0.10)\\
 2.4272 & $-$0.15 (0.25) & 2.67 (0.31) & 0.12 (0.06) & 0.19 (0.11) & 2.84 (0.09) & 1.25 (0.21) & 0.07 (0.04) & 0.82 (0.25)\\
\hline
 FW4871 & $-$0.04 (0.21) & 1.18 (0.19) & 0.30 (0.10) & 0.12 (0.07) & 1.60 (0.35) & 1.38 (0.32) & 0.03 (0.05) & 0.24 (0.09)\\
\hline
\end{tabular}
\end{center}
\end{table*}
%%%%%%%%%%%%%%%%%%%%%%%%%%%%%%%%%%%%%%%%%%%%%%%%

%%%%%%%%%%%%%%%%%%%%%%%%%%%%%%%%%%%%%%%%%%%%%%%%
\section{Spectral fits}
\label{App:Fits}
For reference, we show in Figs.~\ref{fig:appFits} and \ref{fig:appFits2} 
the spectral fits and resulting star formation histories of the complete sample,
following the same format as in Fig.~\ref{fig:sed}, with points in red
being masked out of the fitting procedure (\S\ref{Sec:Method}). The red points may represent
either line emission from the objects or a potential residual contamination from
neighbouring sources.

% uncomment below if you want to avoid appendix figures to be compiled
%\iffalse
%%%%%%%%%%   APPENDIX FIGURES    %%%%%%%%%%%%%%%
%%%%%%%%%%%%%%%%%%%%%%%%%%%%%%%%%%%%%%%%%%%%%%%%
%%%%%%%%%%%%%%%%%%%%%%%%%%%%%%%%%%%%%%%%%%%%%%%%
\begin{figure*}
\begin{center}
\includegraphics[width=42mm]{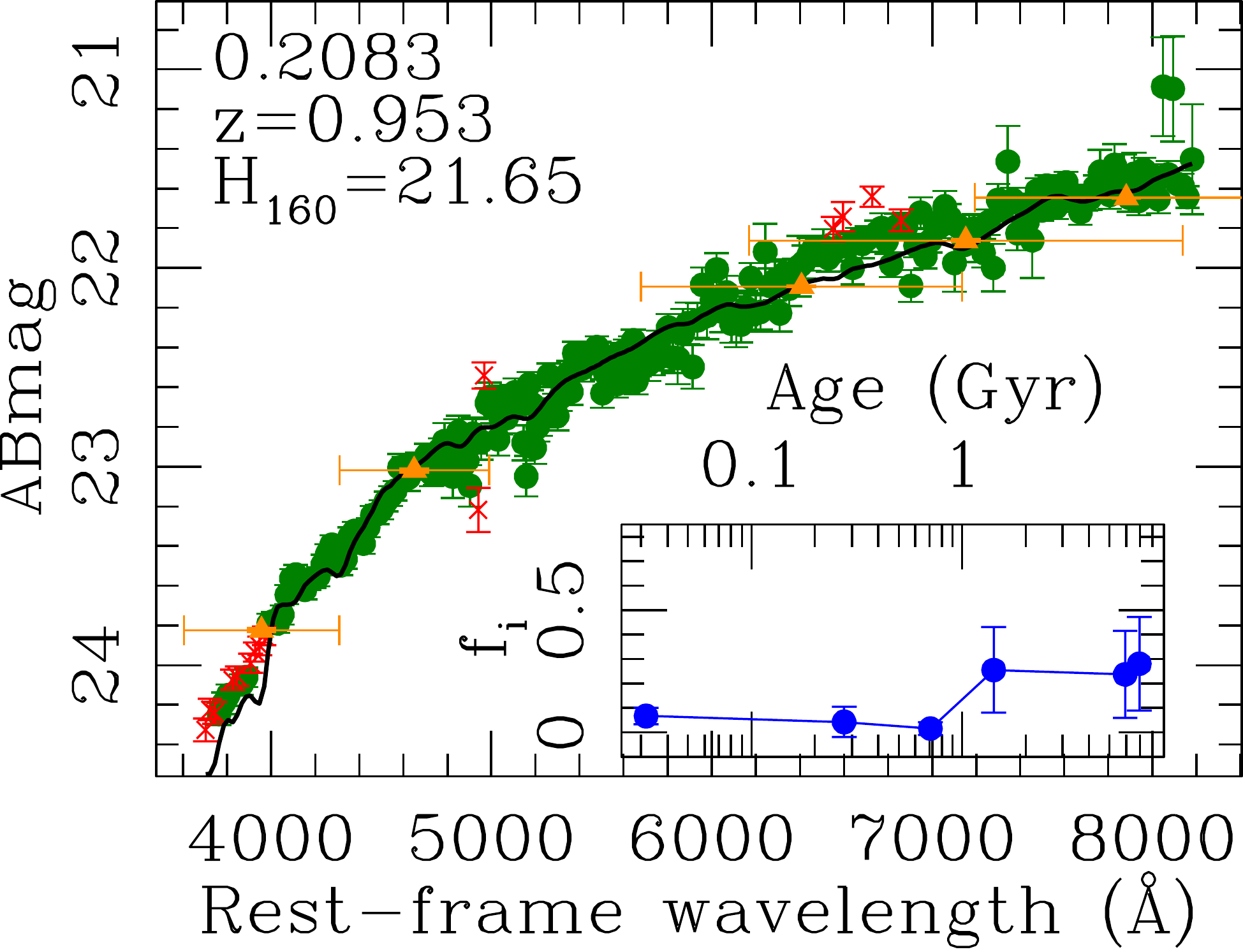}
\includegraphics[width=42mm]{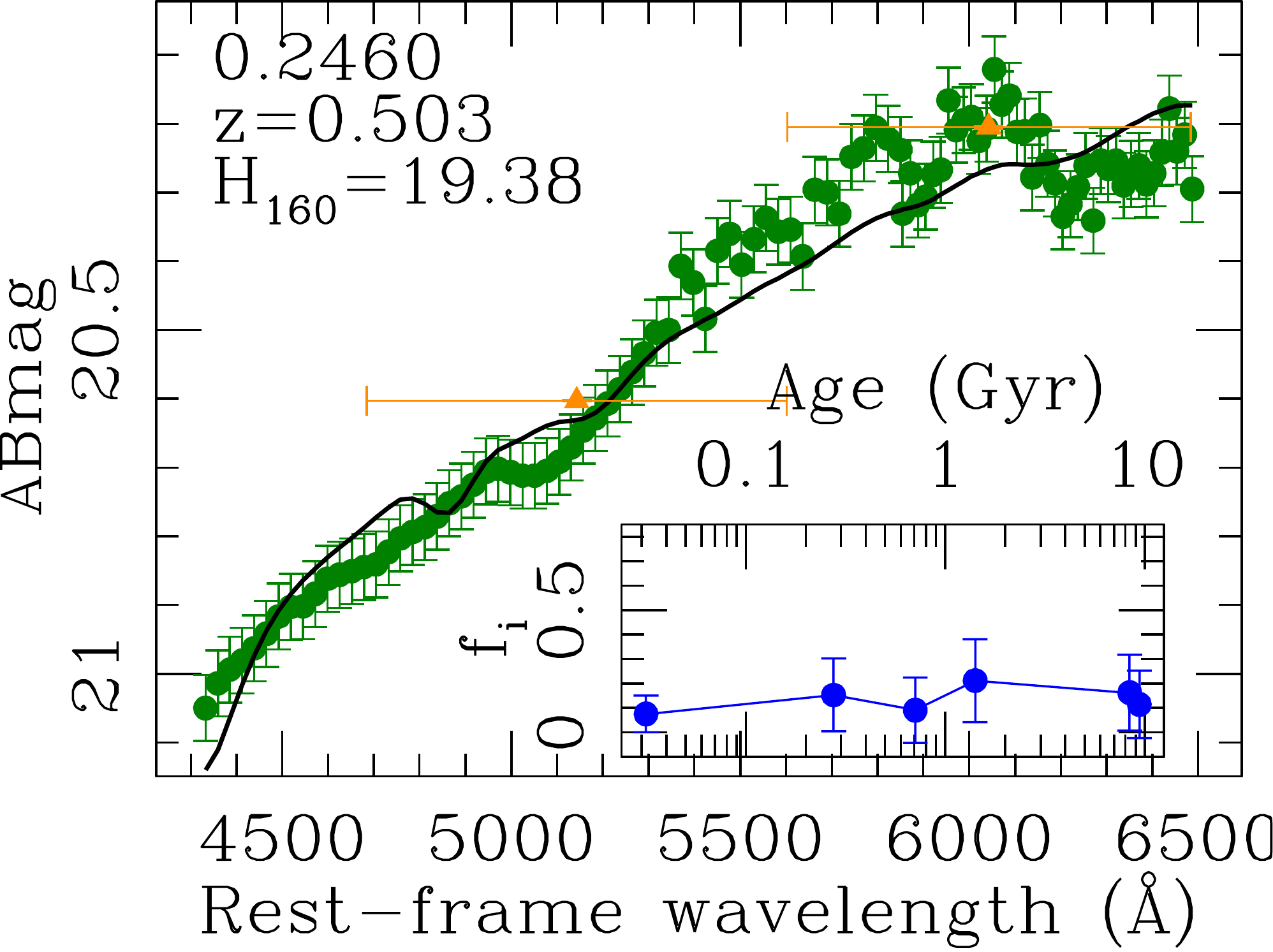}
\includegraphics[width=42mm]{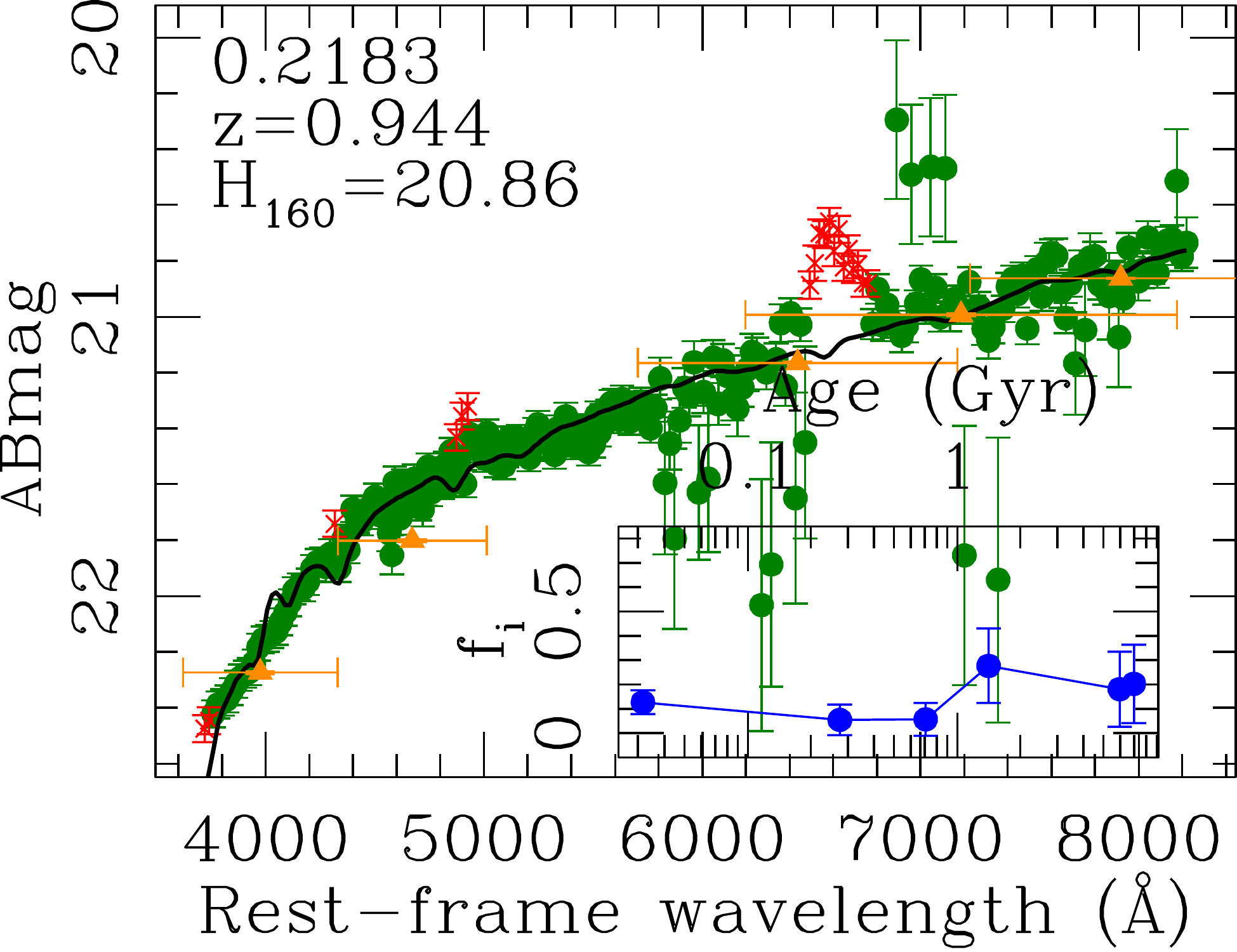}
\includegraphics[width=42mm]{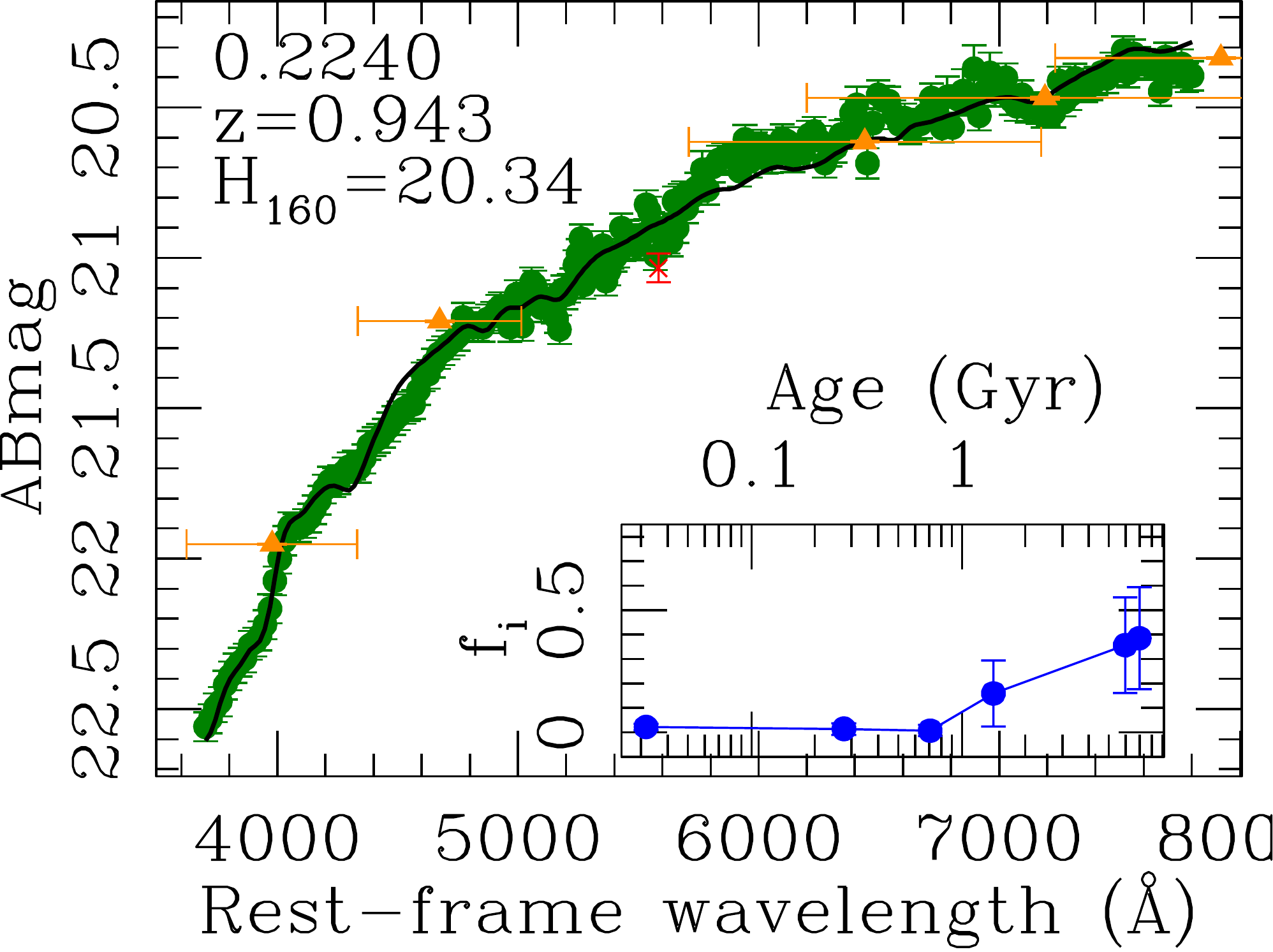}\\
\includegraphics[width=42mm]{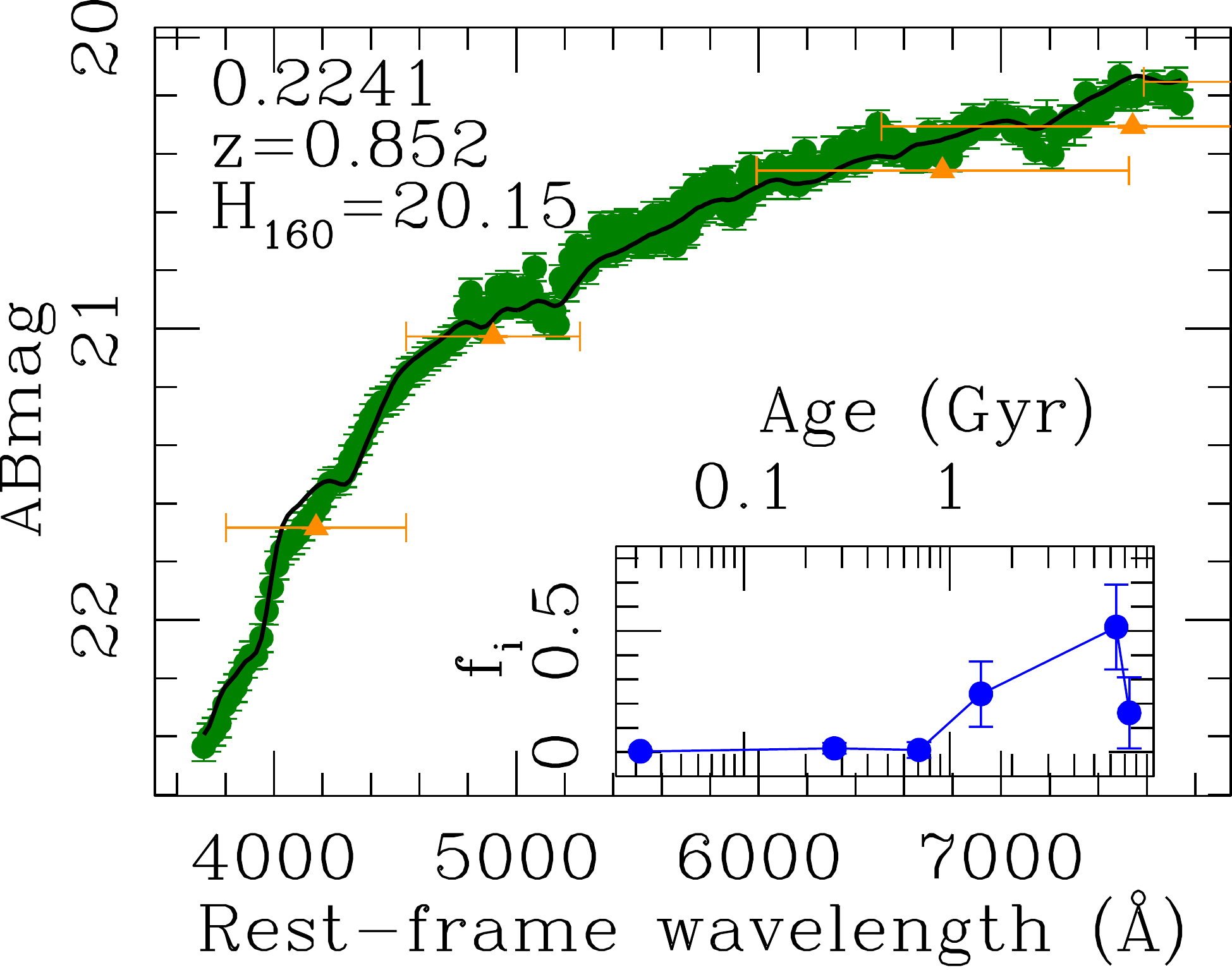}
\includegraphics[width=42mm]{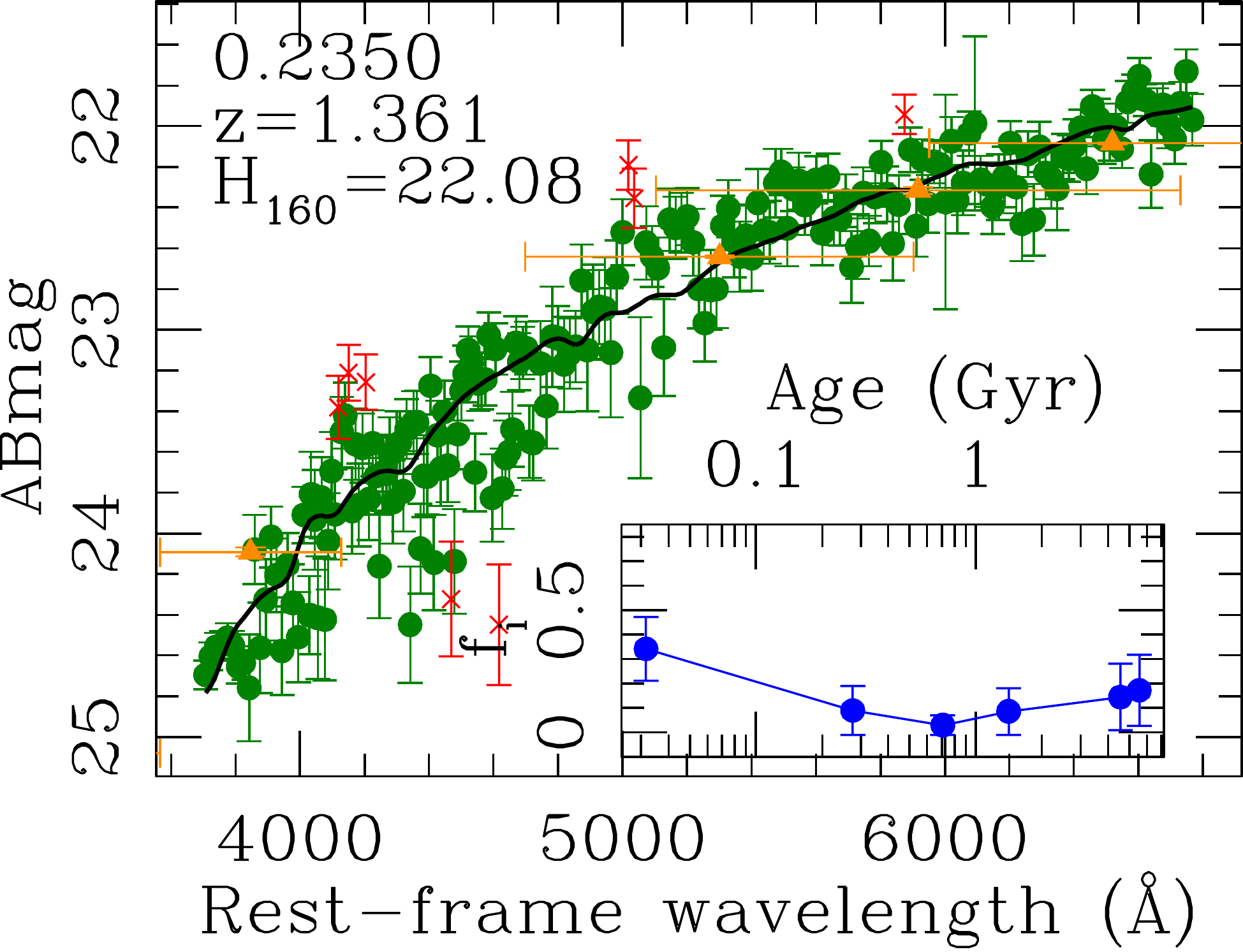}
\includegraphics[width=42mm]{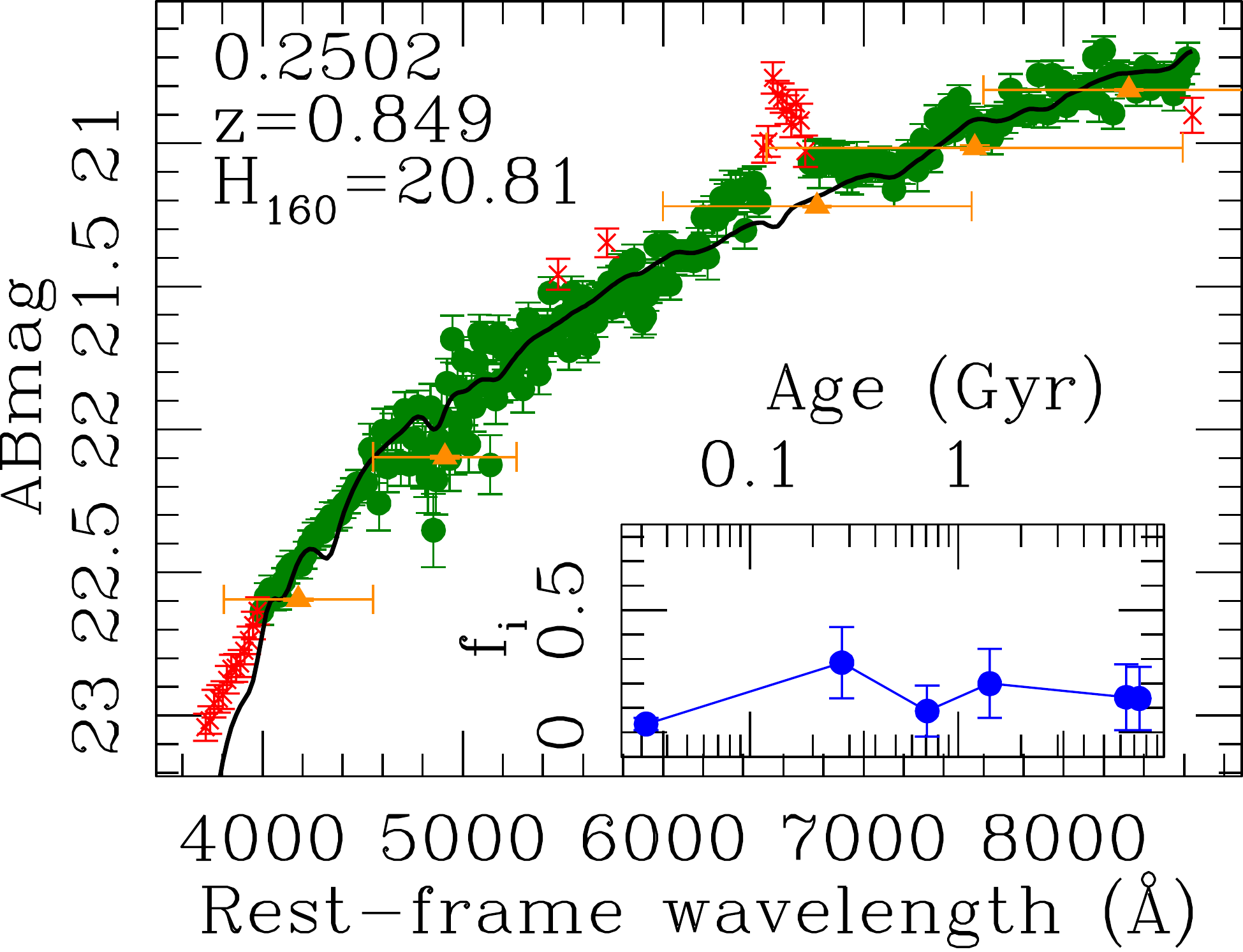}
\includegraphics[width=42mm]{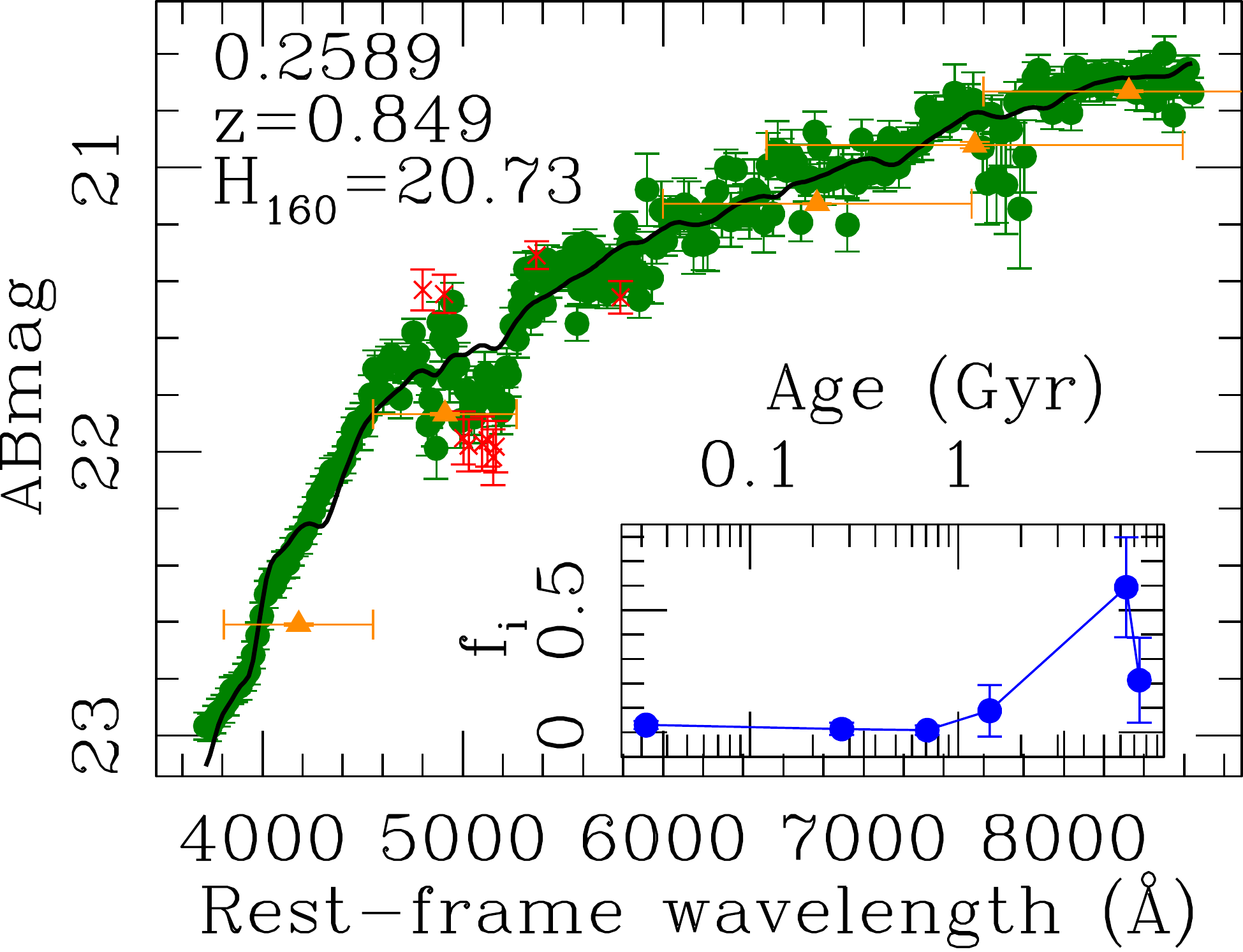}\\
\includegraphics[width=42mm]{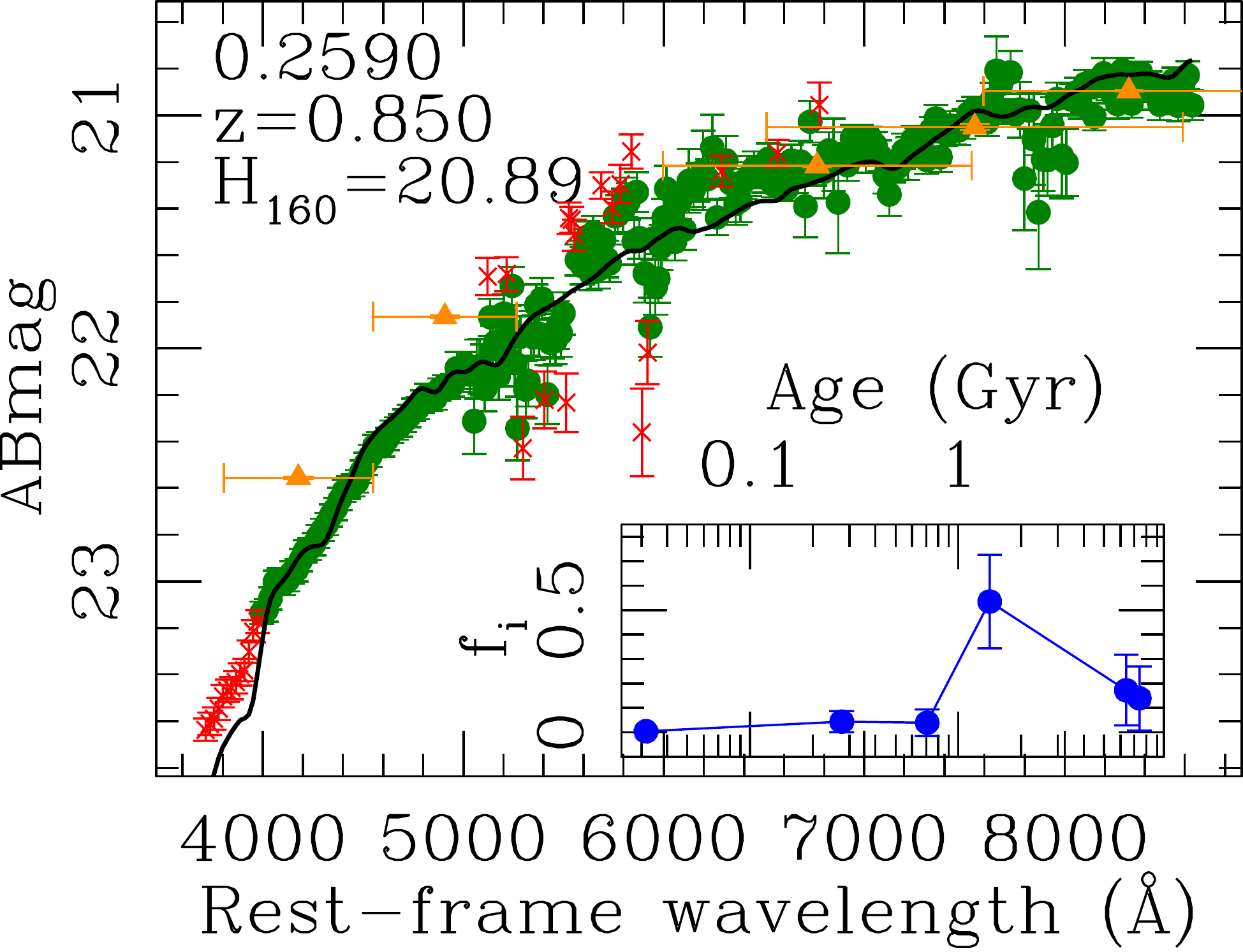}
\includegraphics[width=42mm]{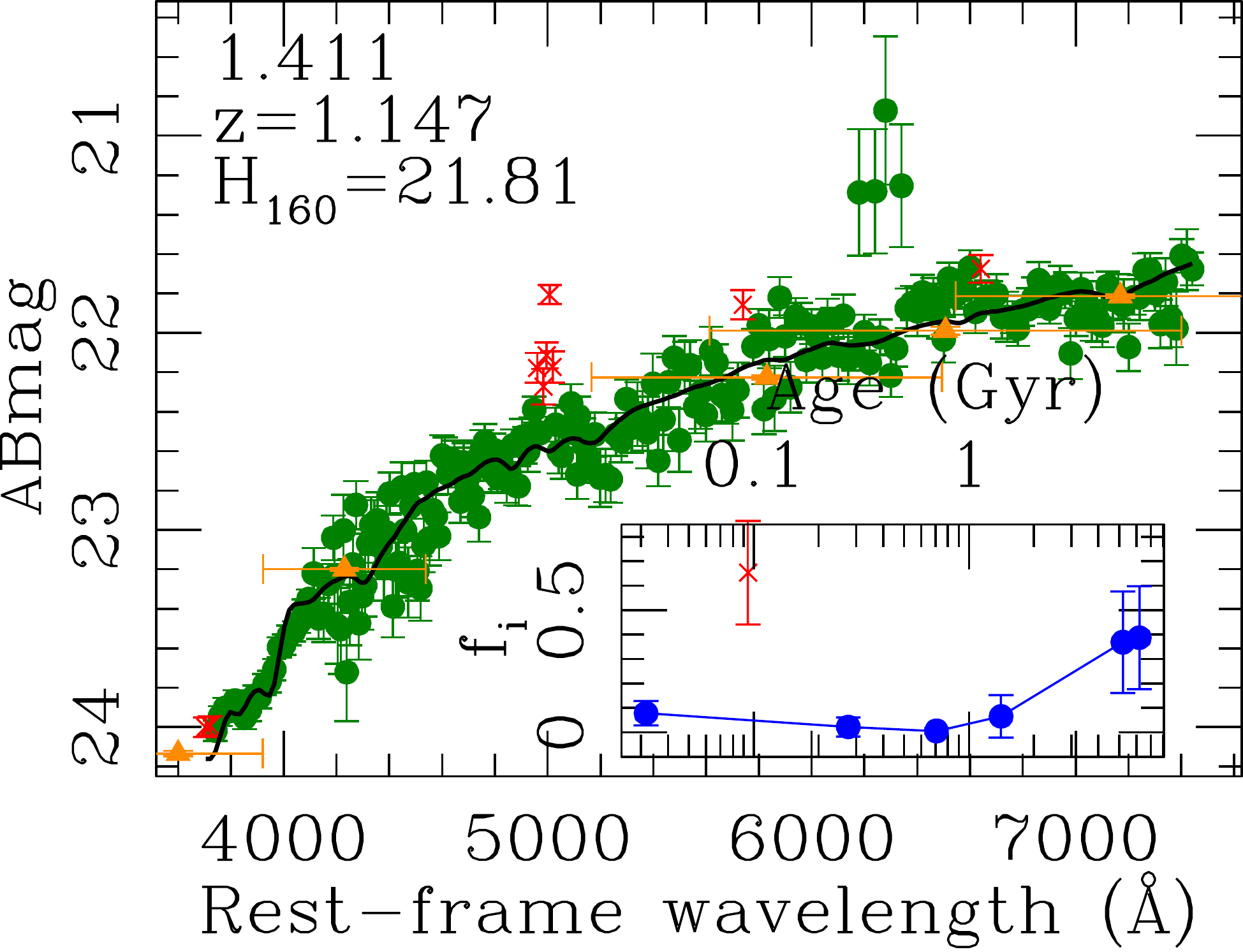}
\includegraphics[width=42mm]{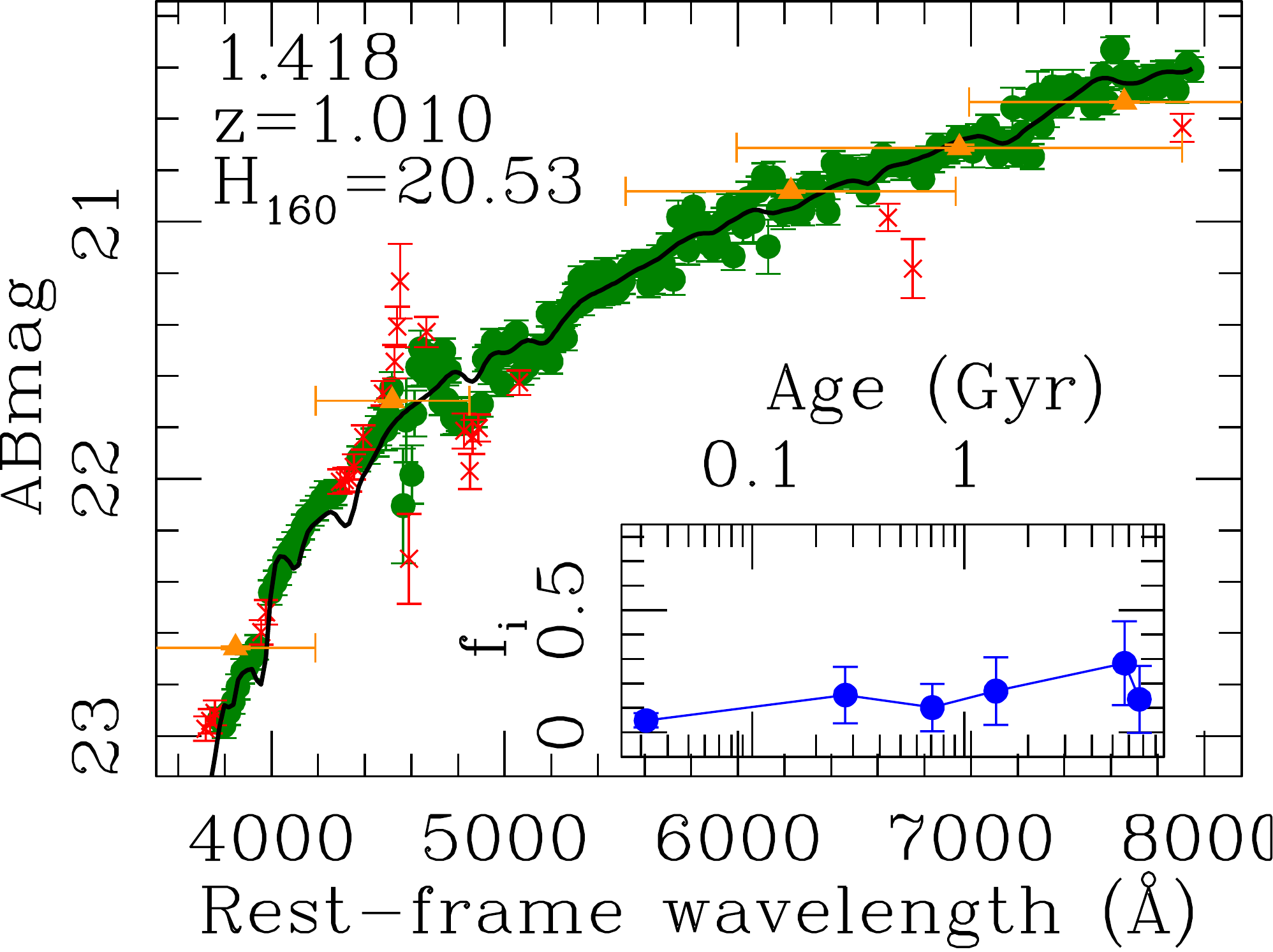}
\includegraphics[width=42mm]{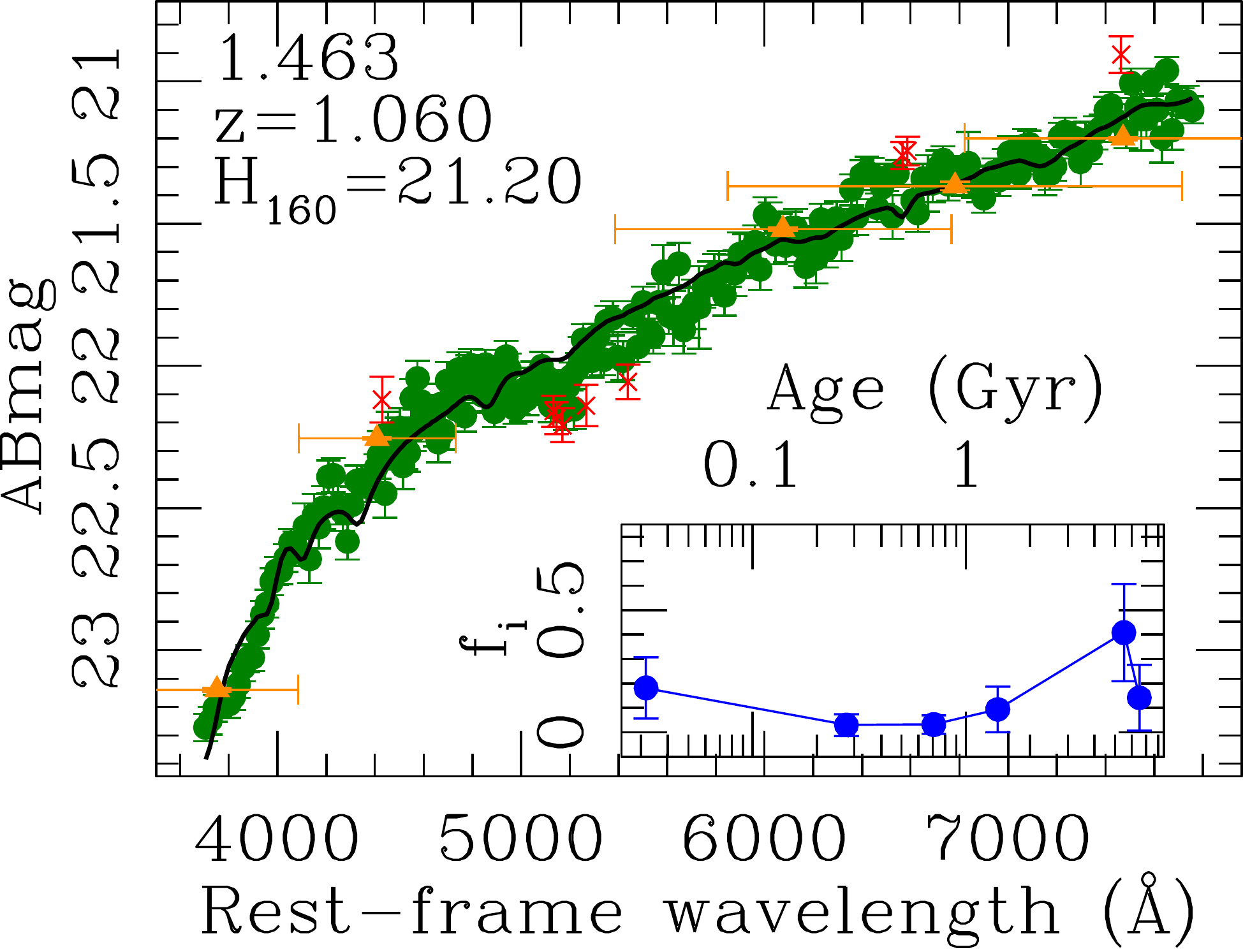}\\
\includegraphics[width=42mm]{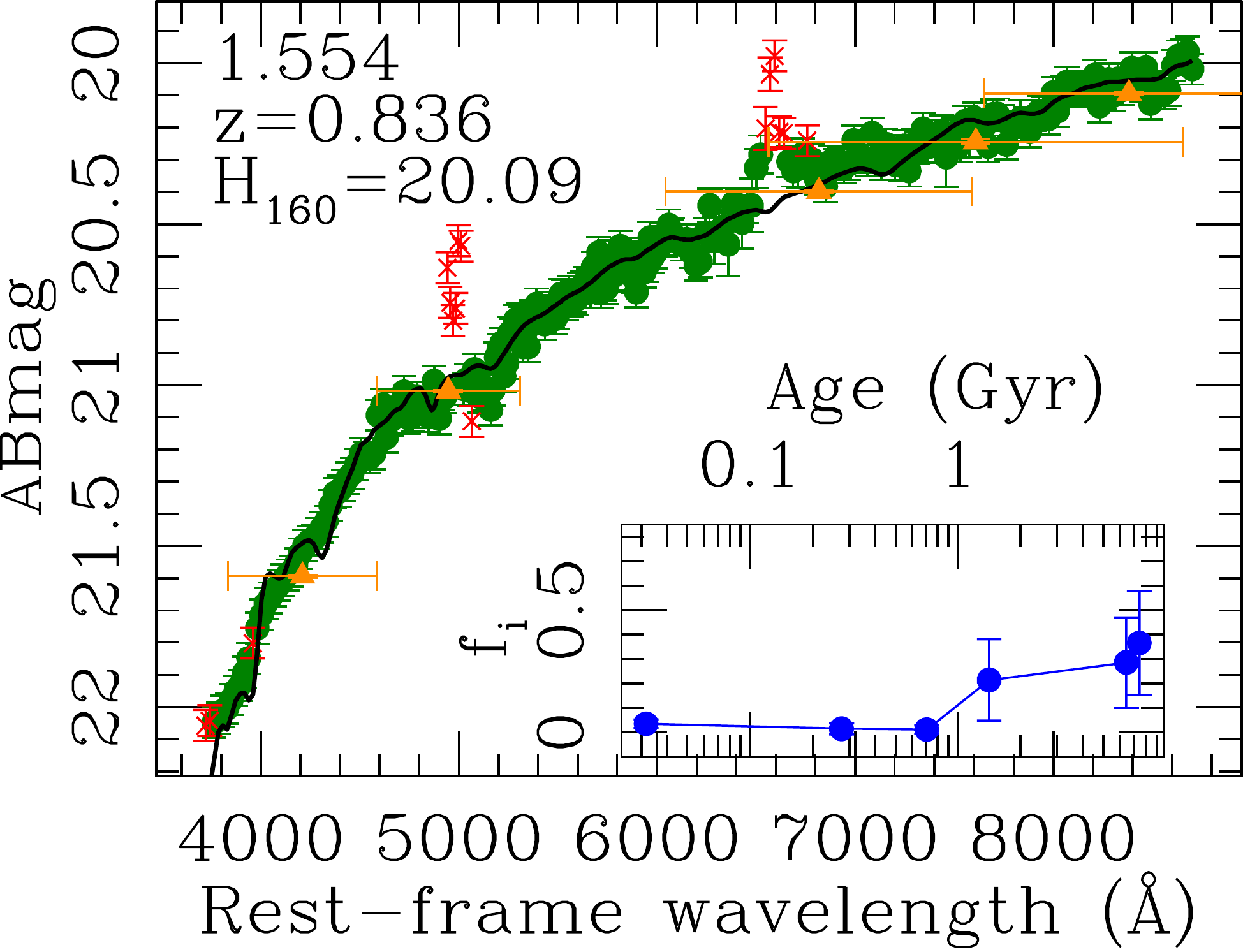}
\includegraphics[width=42mm]{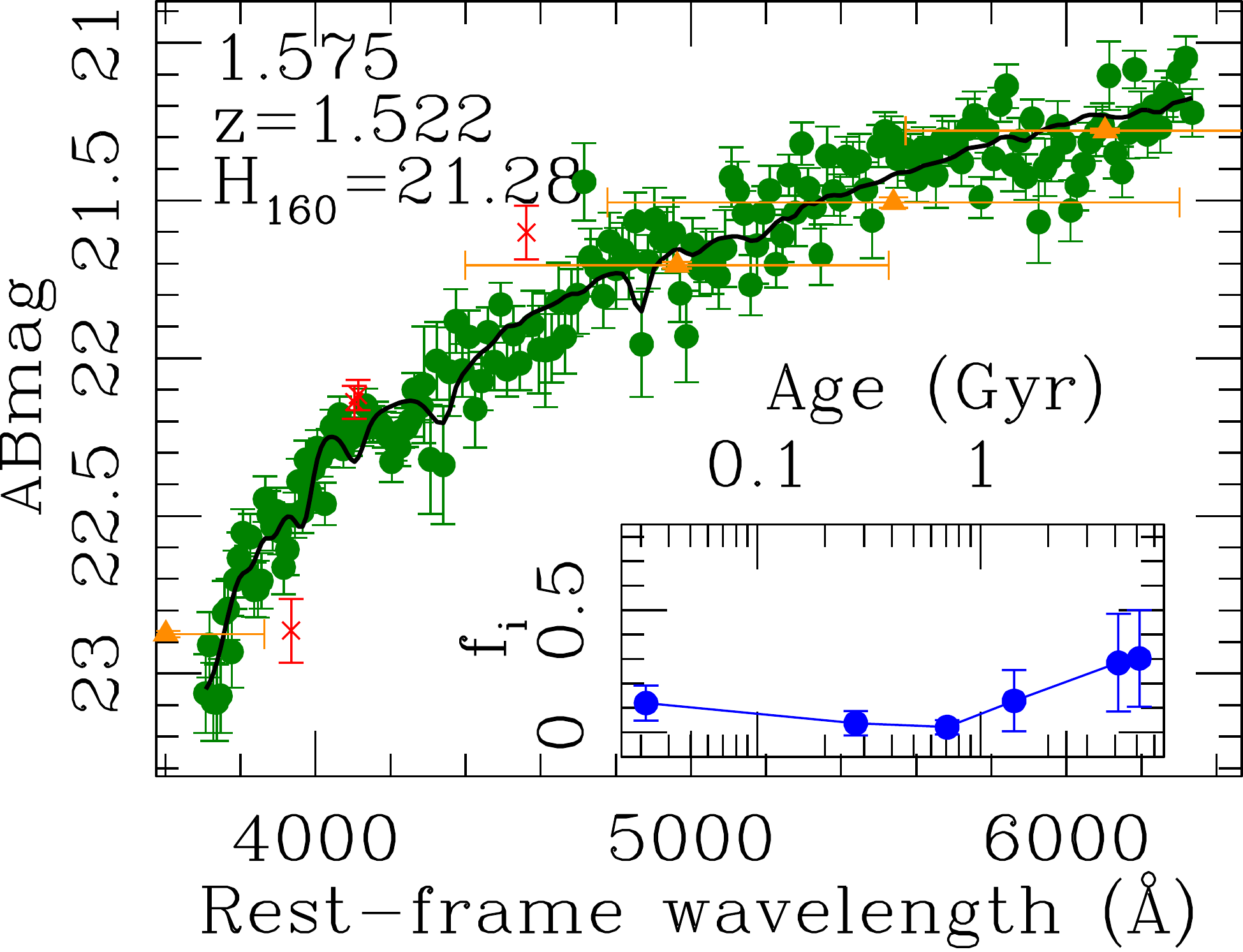}
\includegraphics[width=42mm]{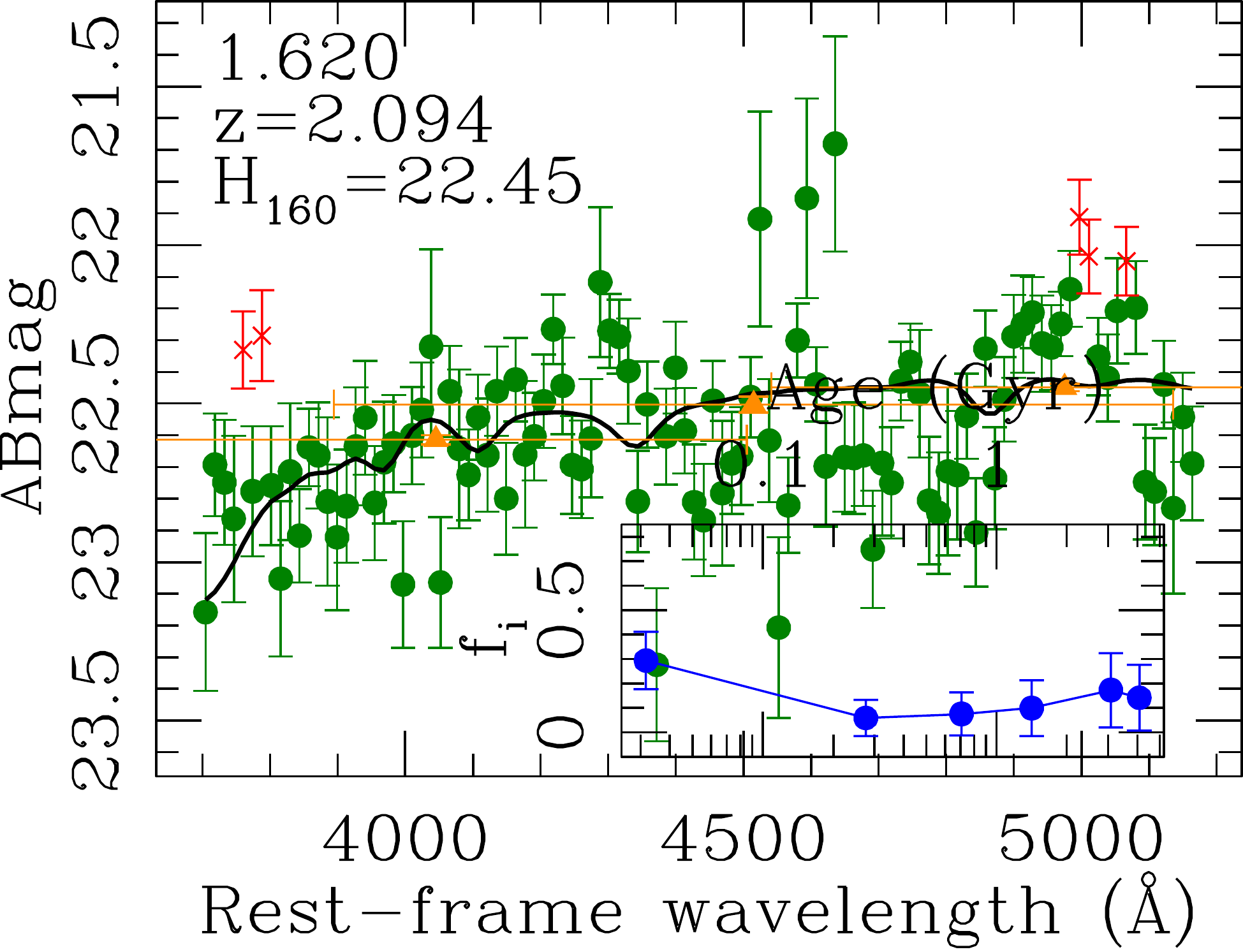}
\includegraphics[width=42mm]{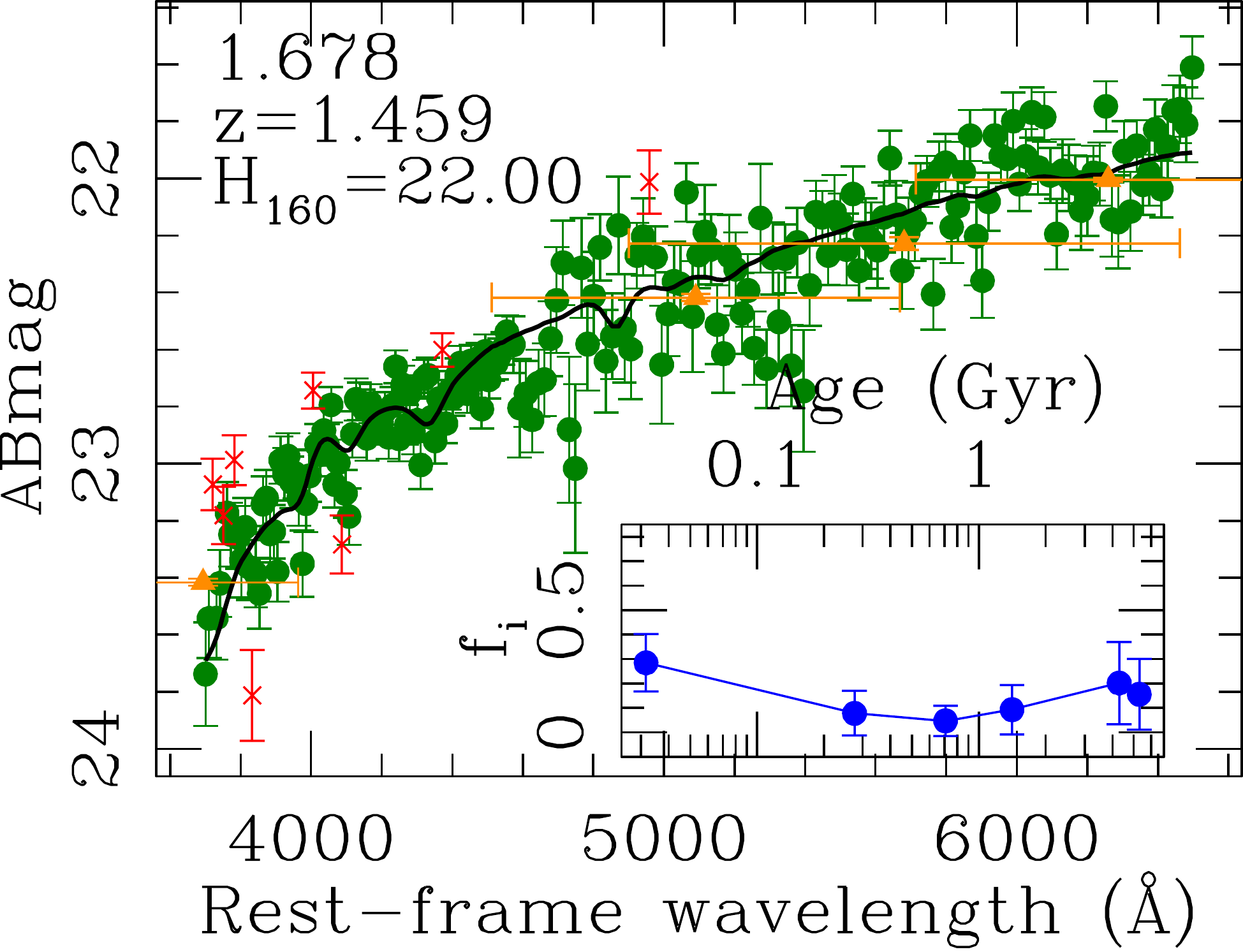}\\
\includegraphics[width=42mm]{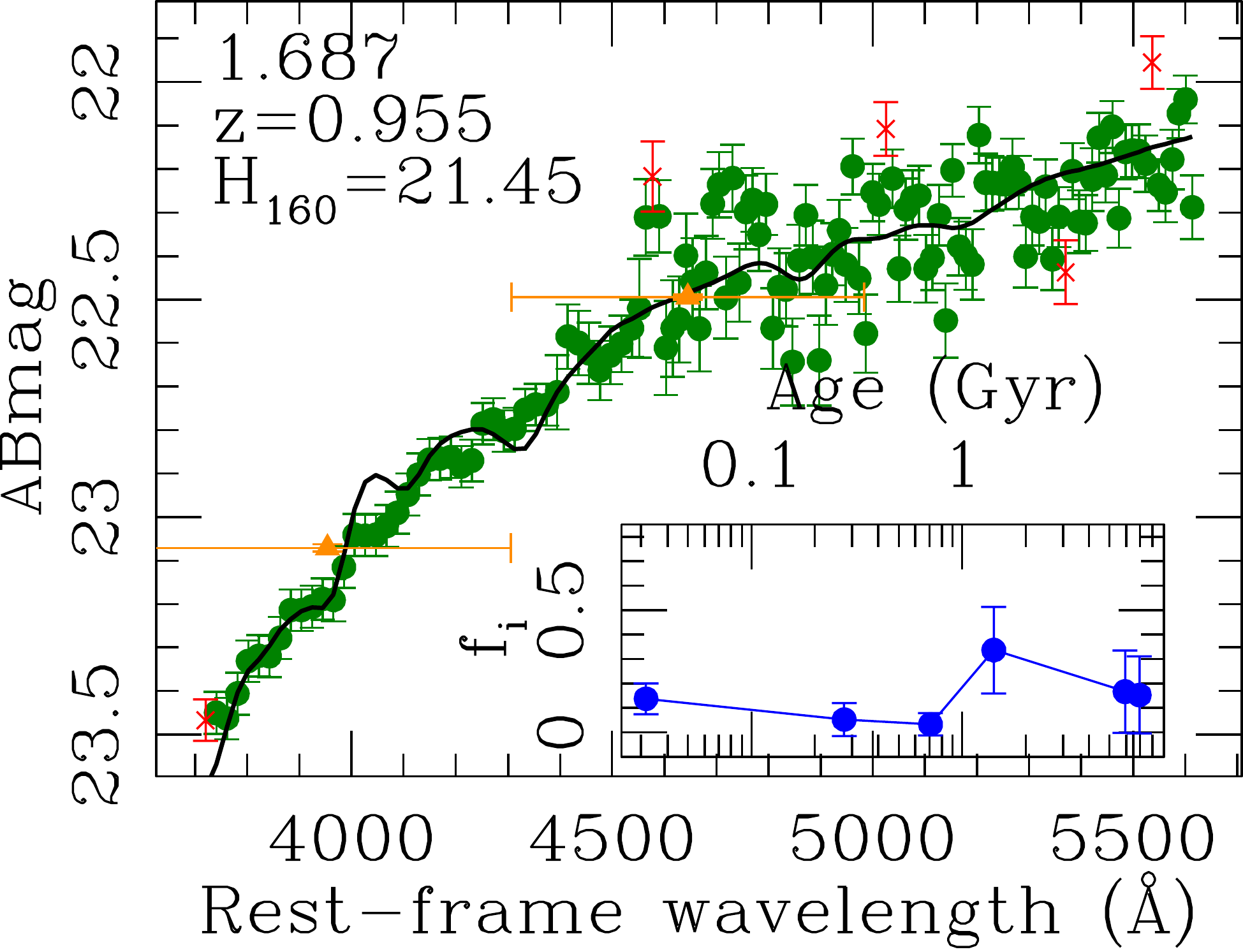}
\includegraphics[width=42mm]{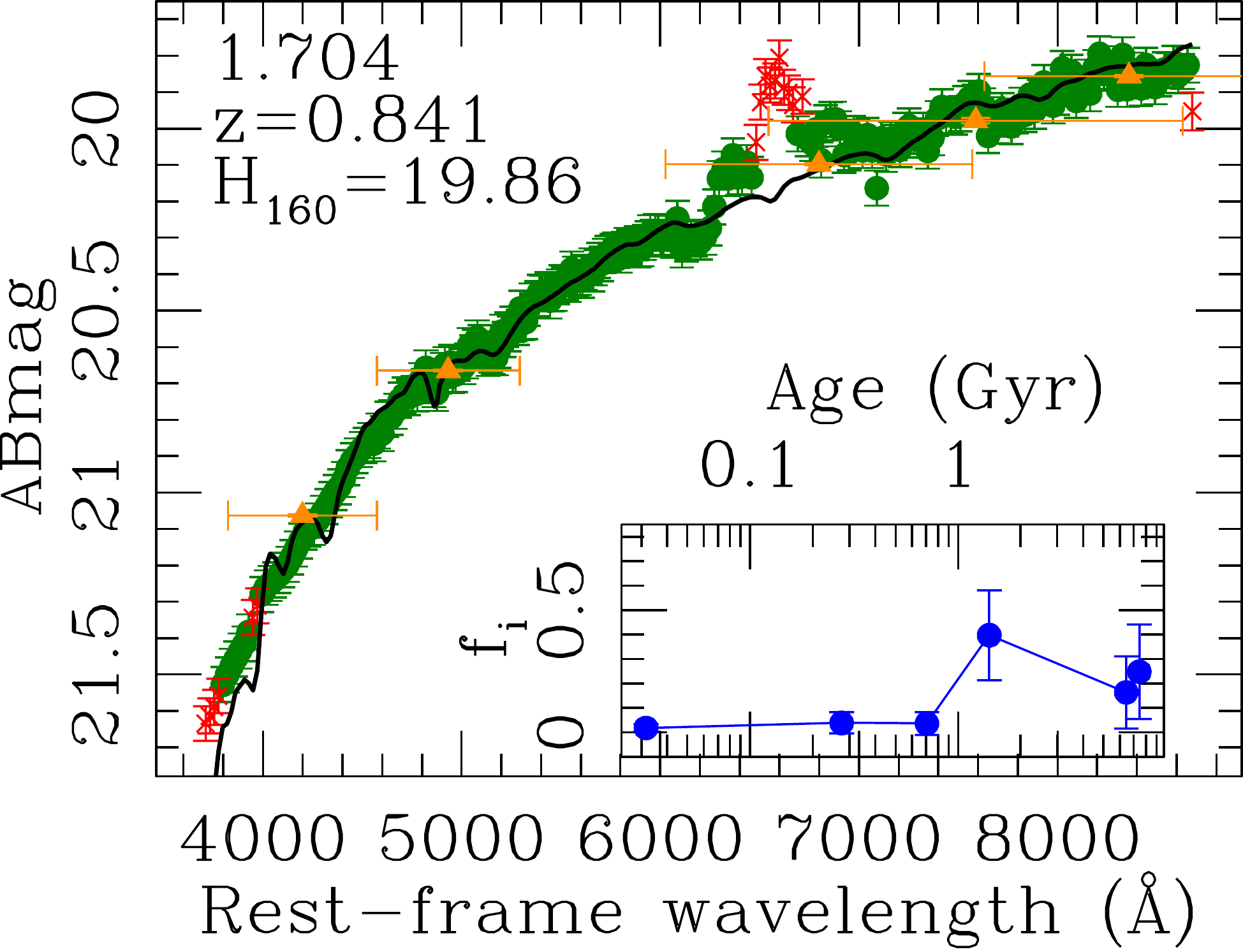}
\includegraphics[width=42mm]{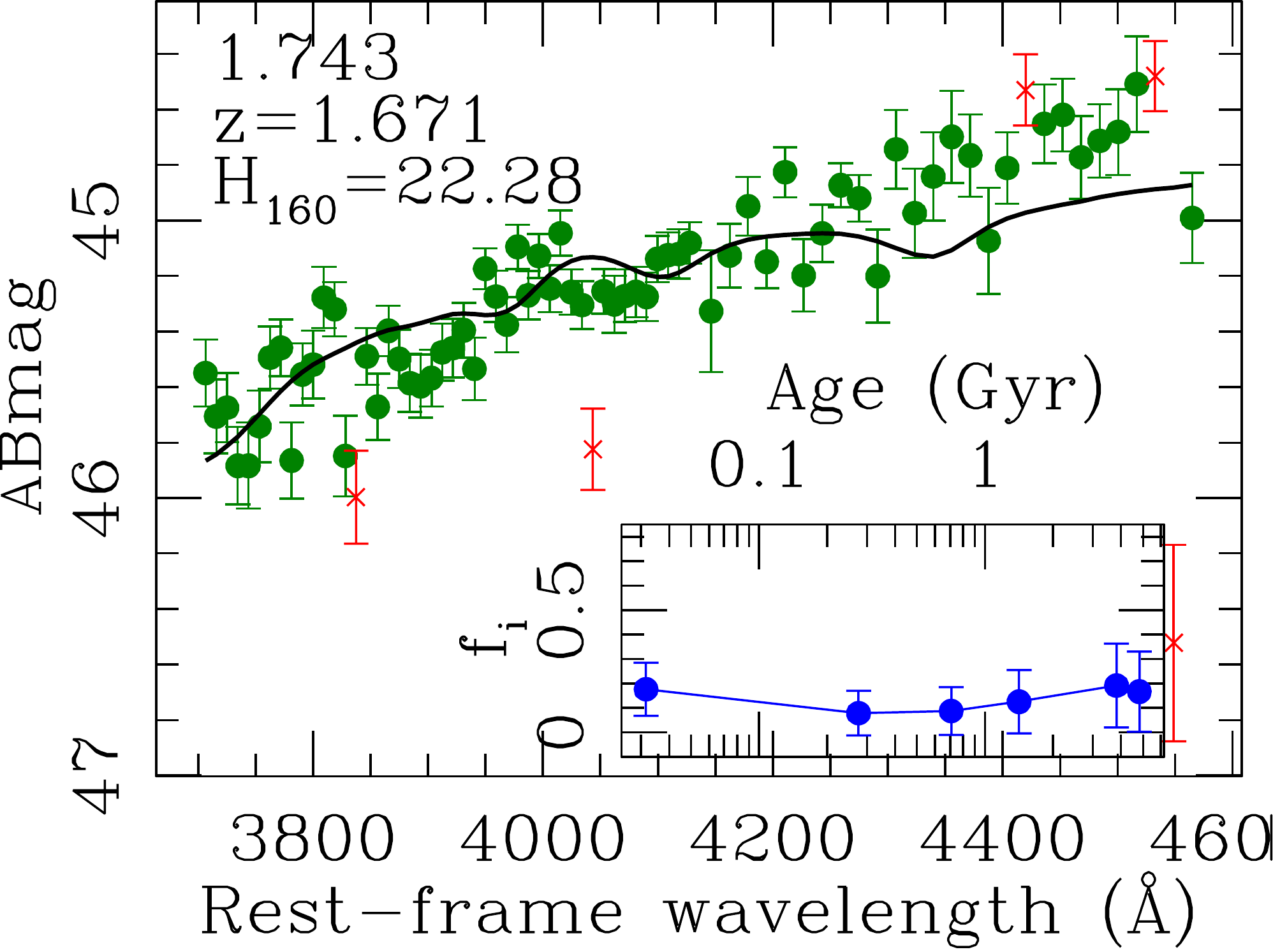}
\includegraphics[width=42mm]{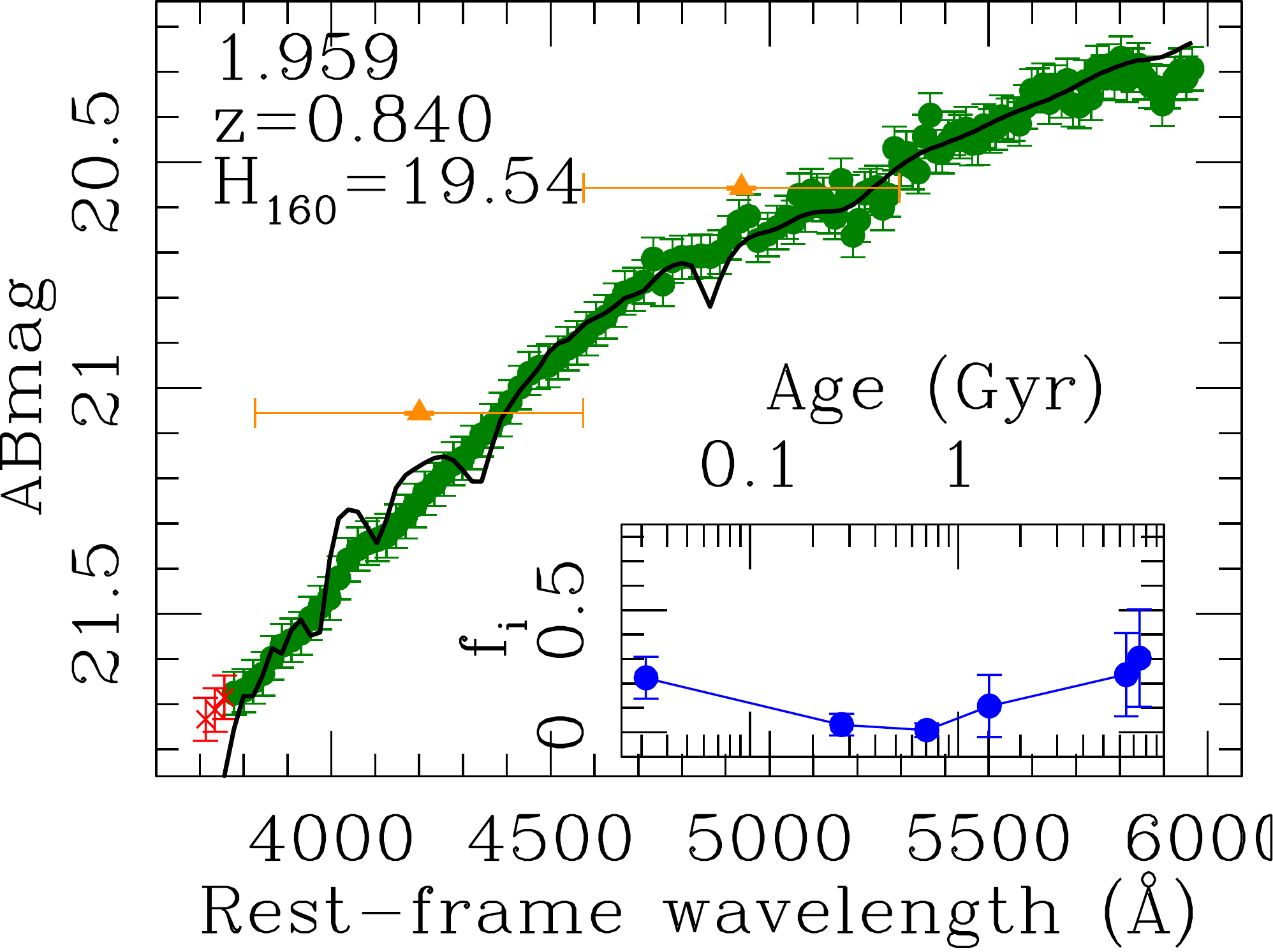}\\
\includegraphics[width=42mm]{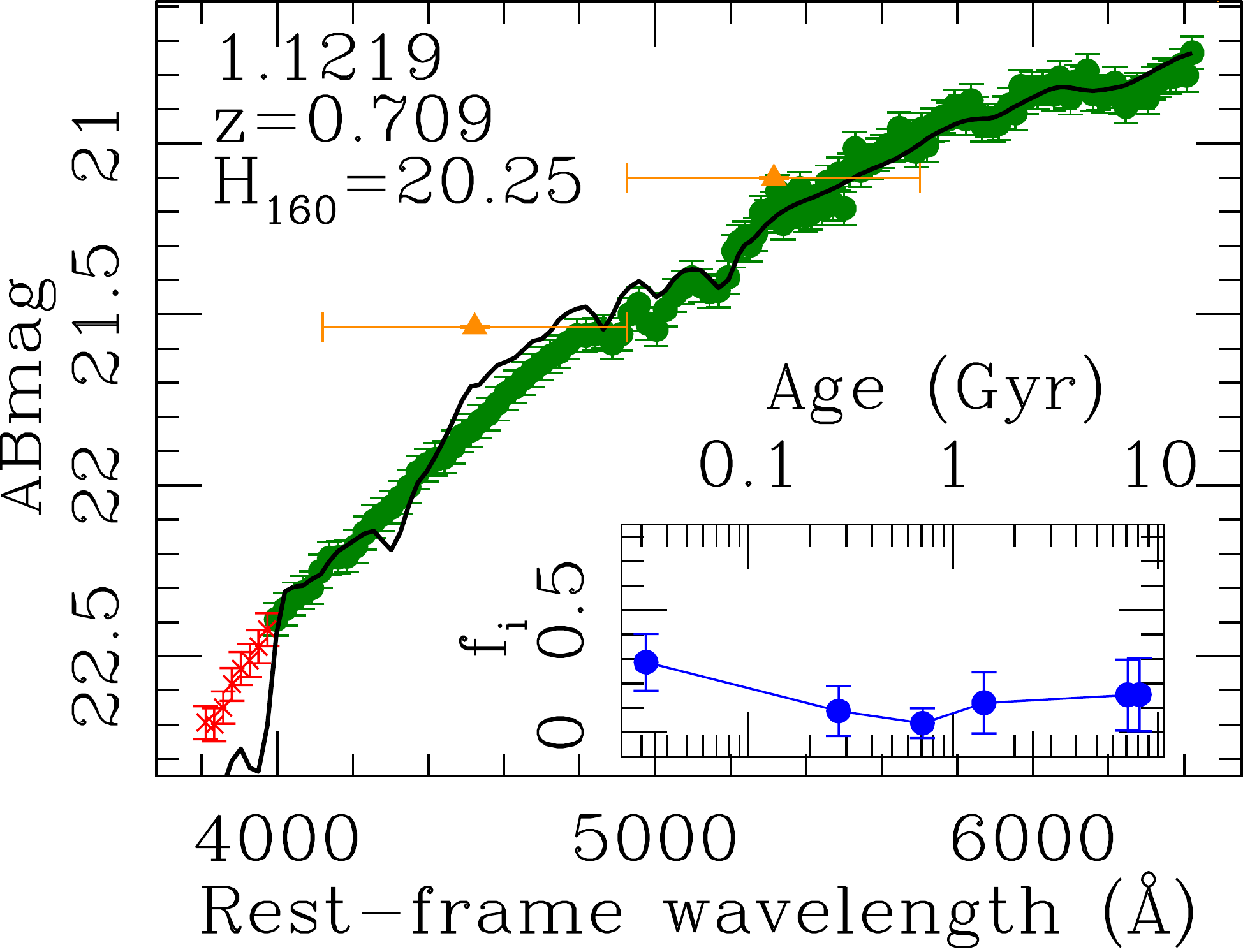}
\includegraphics[width=42mm]{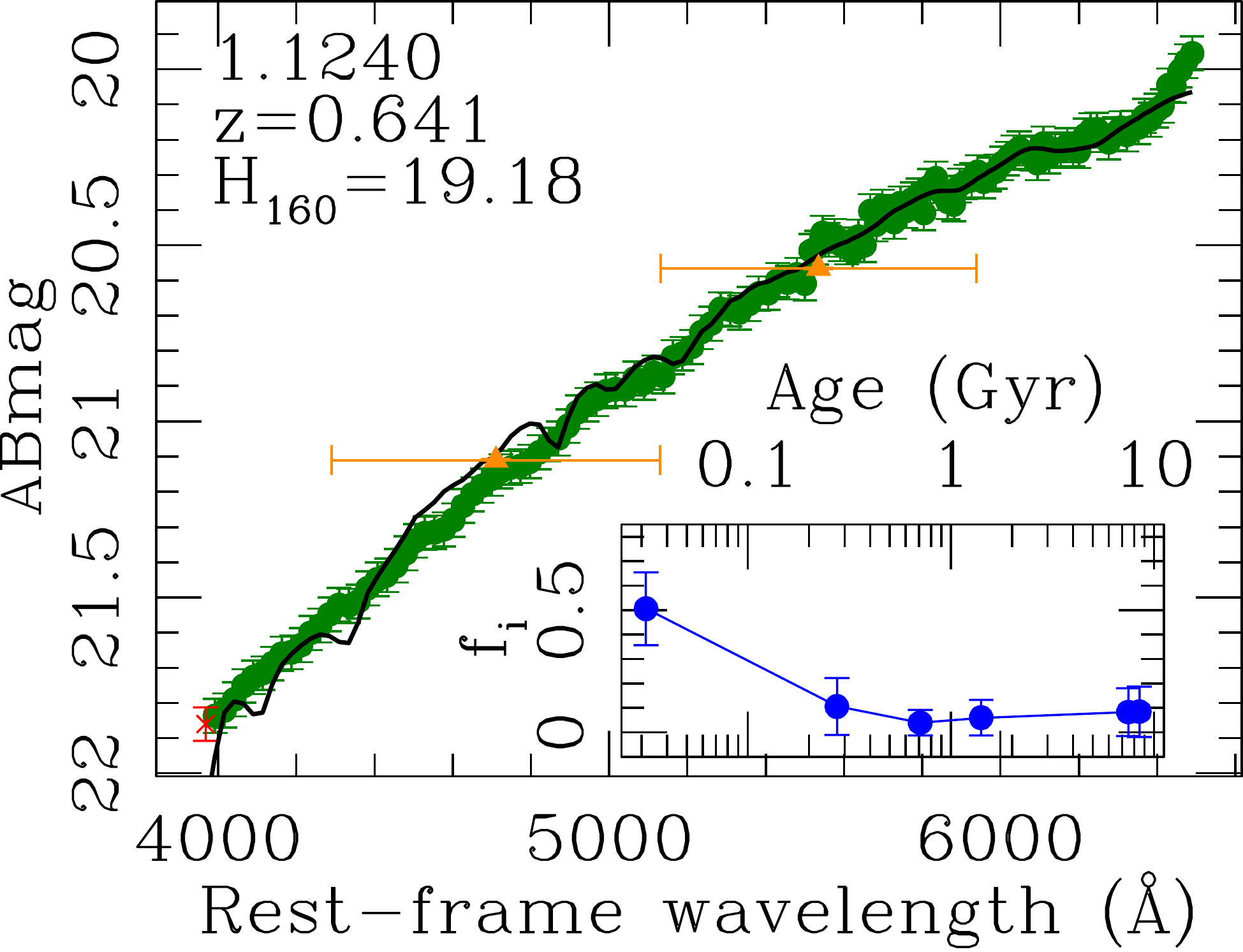}
\includegraphics[width=42mm]{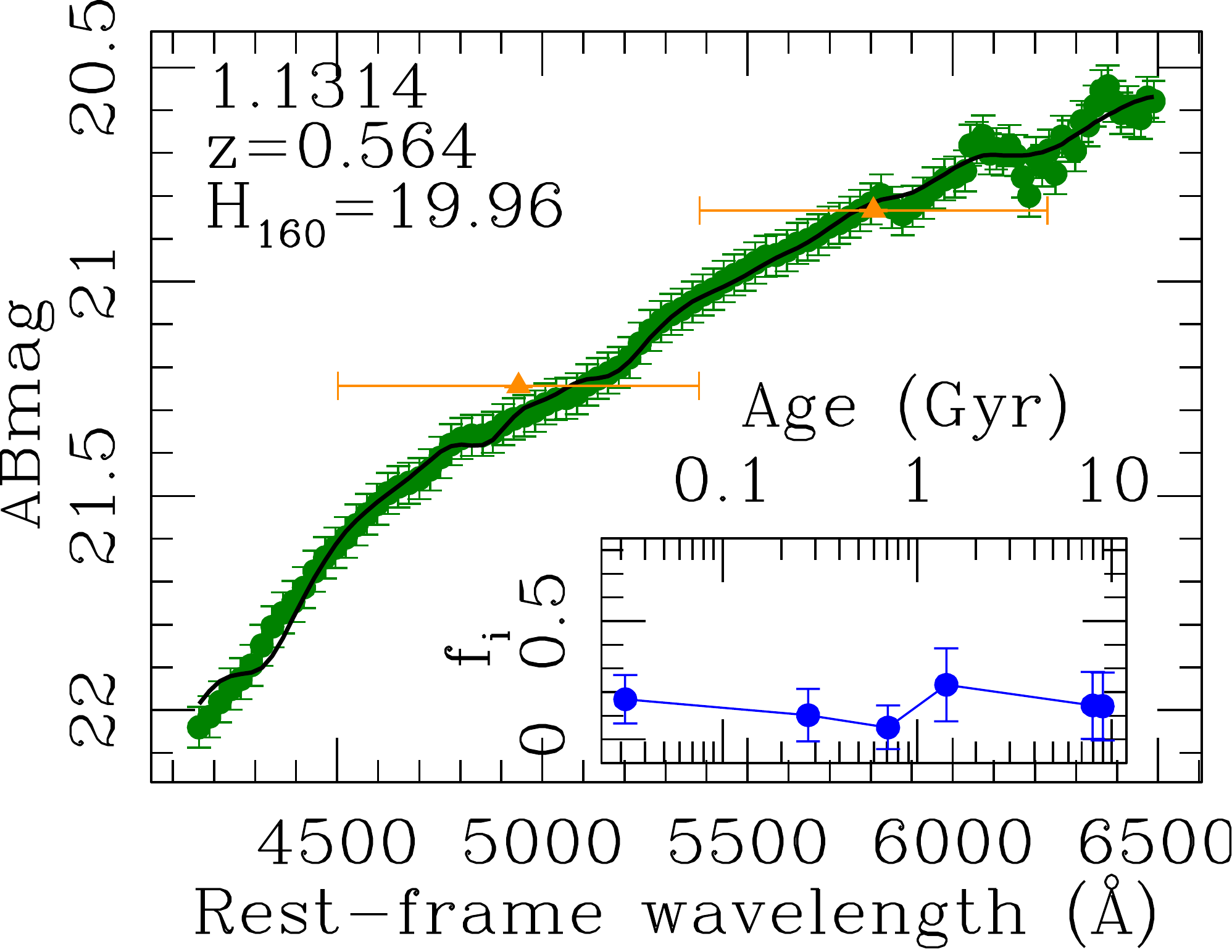}
\includegraphics[width=42mm]{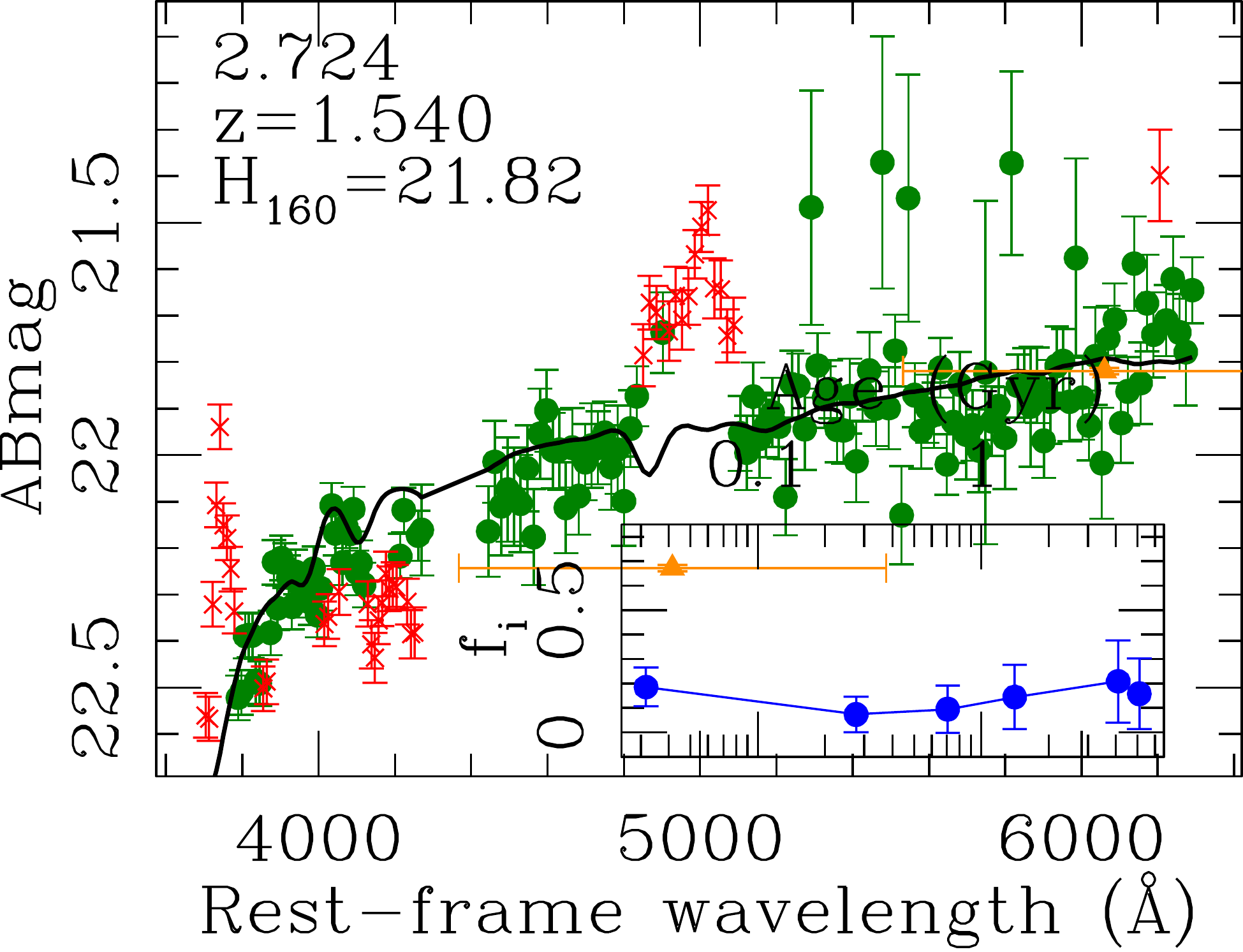}\\
\includegraphics[width=42mm]{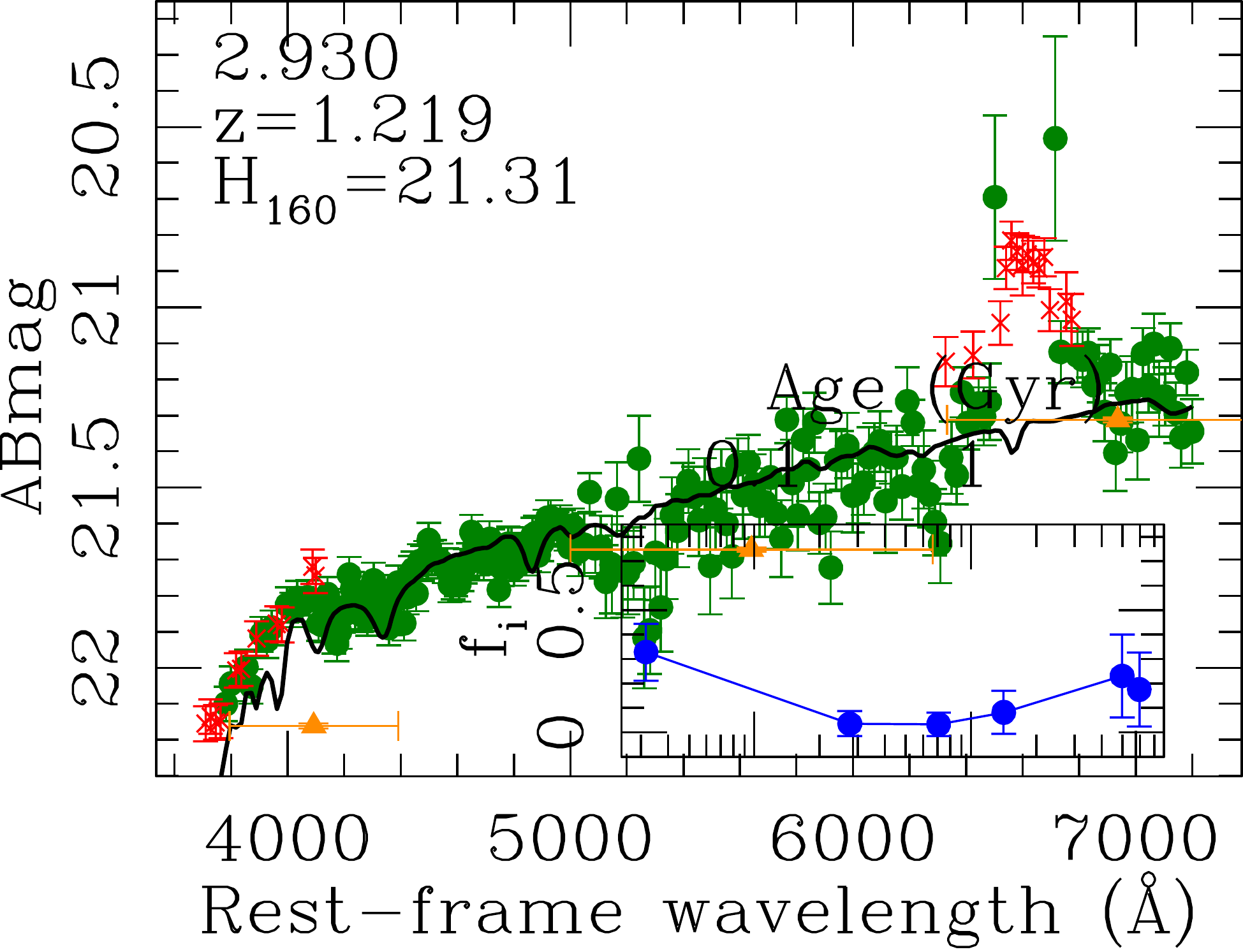}
\includegraphics[width=42mm]{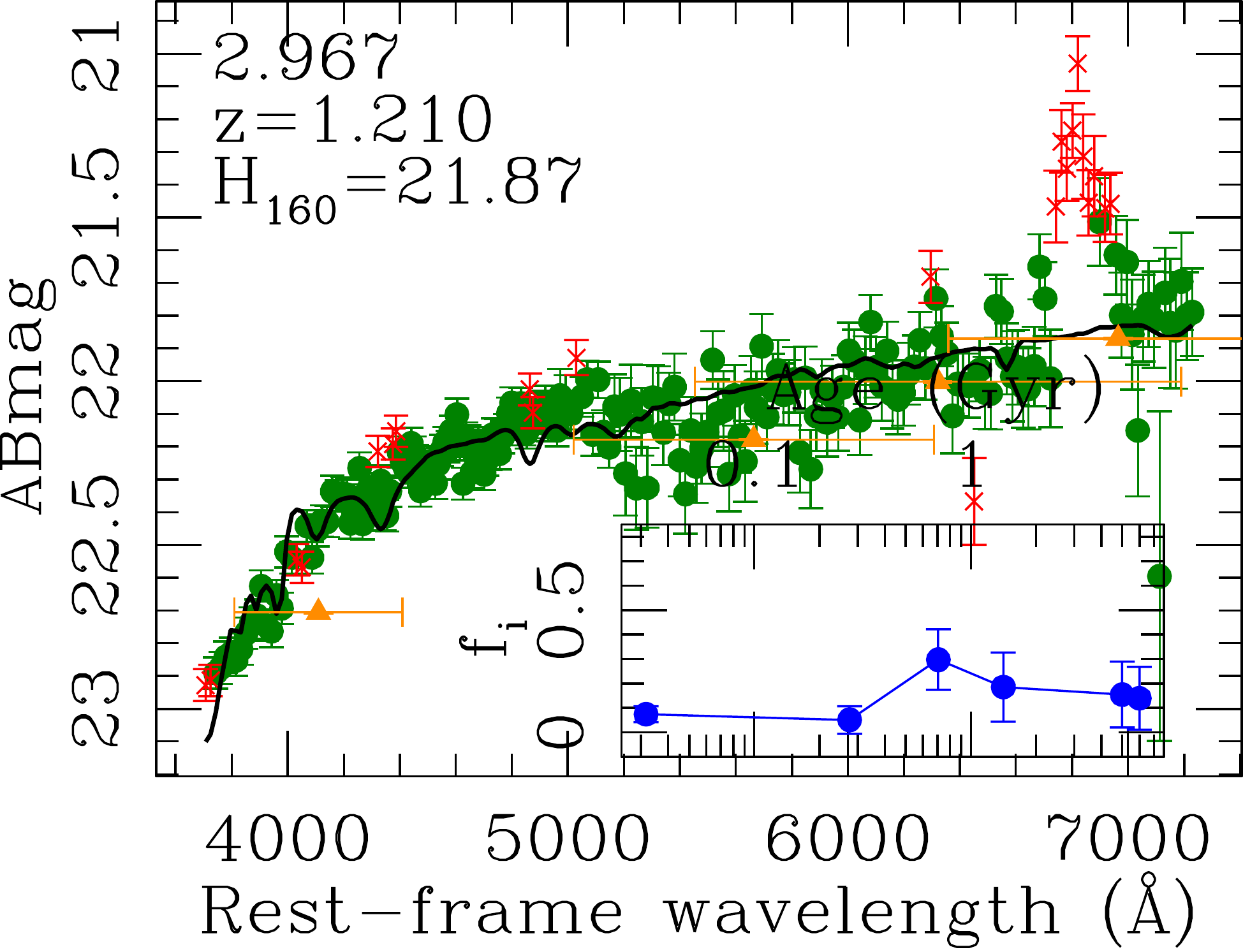}
\includegraphics[width=42mm]{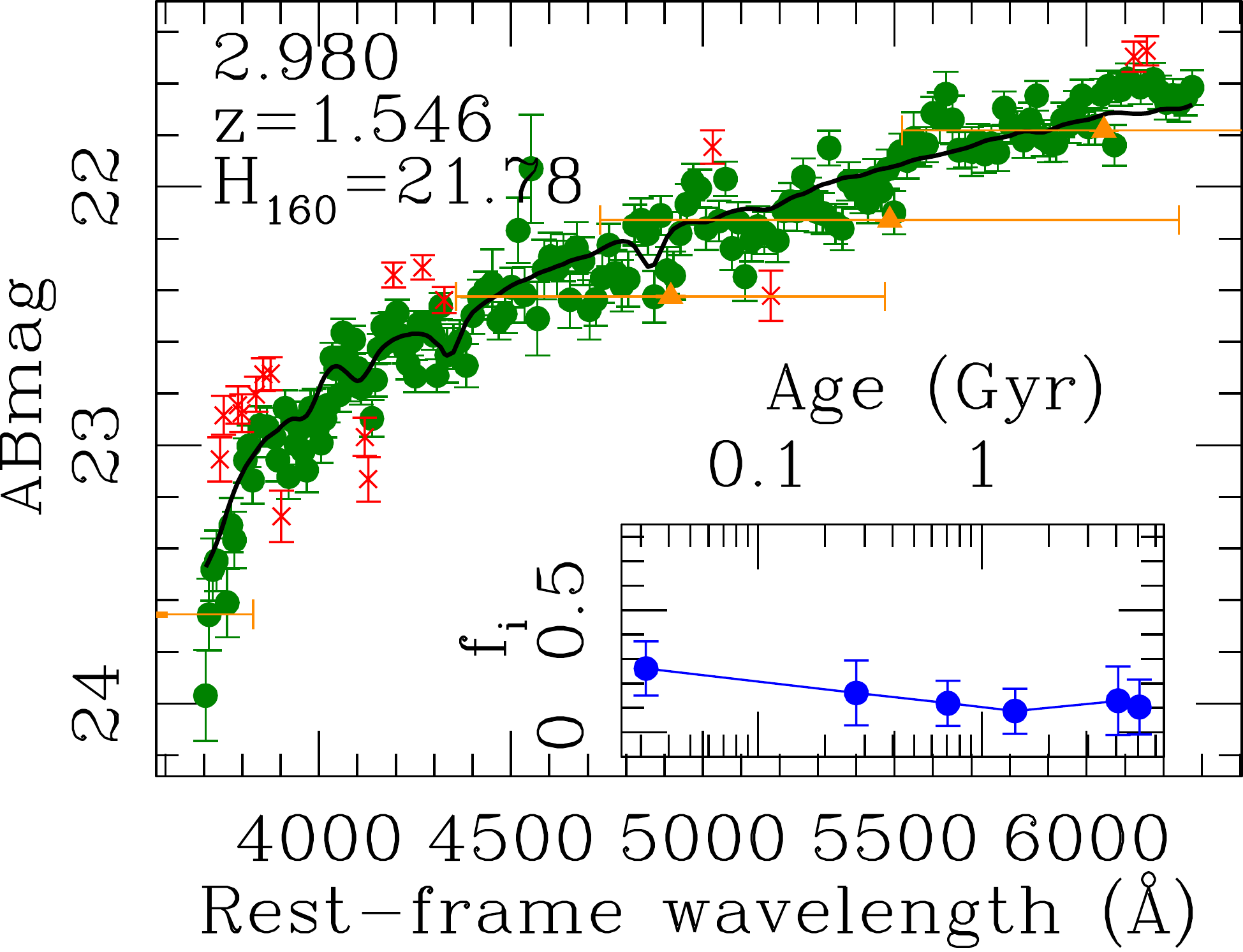}
\includegraphics[width=42mm]{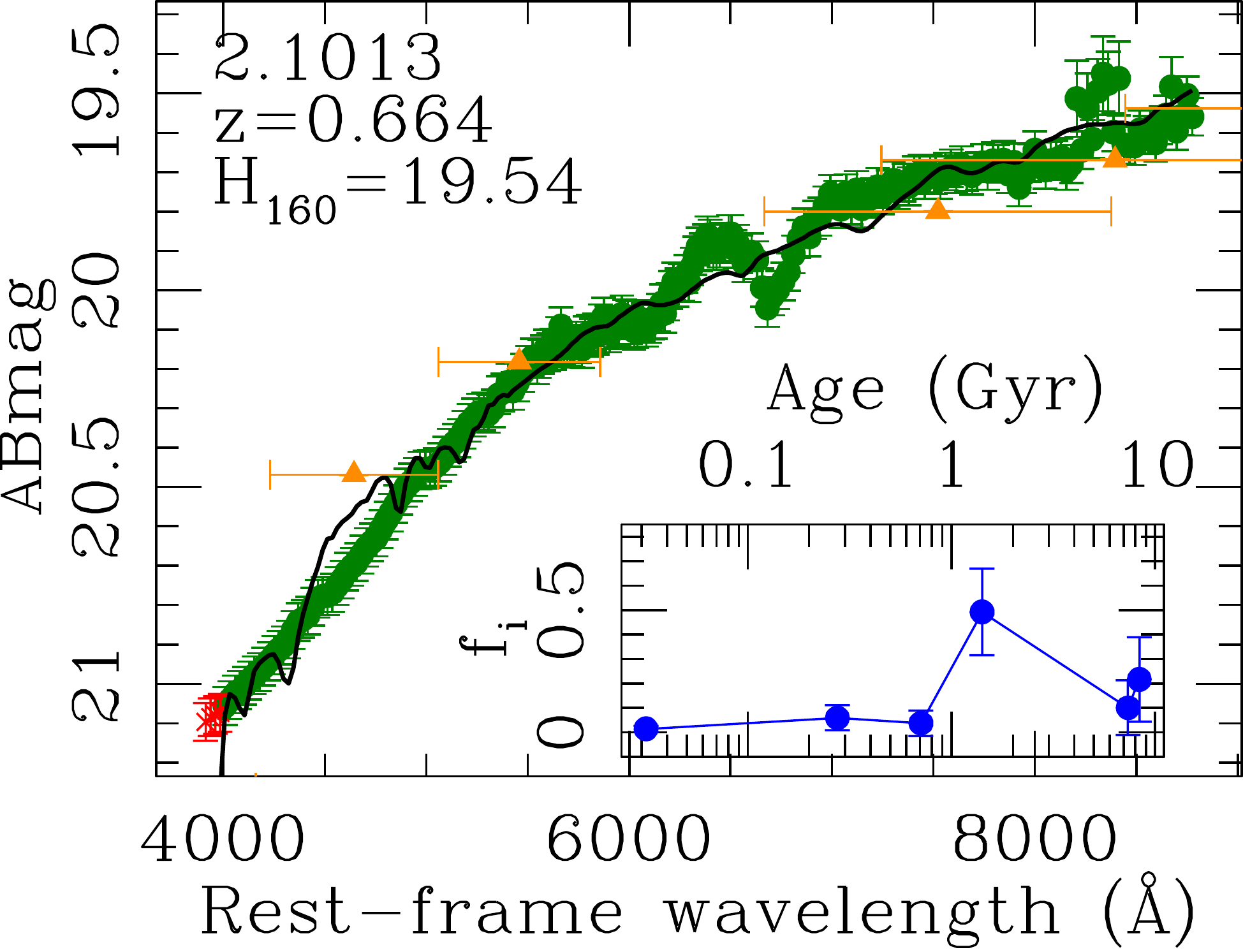}
\end{center}
\caption{Spectral fits of the complete sample.
The notation follows that of Fig.~\ref{fig:sed}.
Note that Base Model 6 (that has
the same age as BM5) is displaced by $+$1\,Gyr.
}
\label{fig:appFits}
\end{figure*}
%%%%%%%%%%%%%%%%%%%%%%%%%%%%%%%%%%%%%%%%%%%%%%%%

\begin{figure*}
\begin{center}
\includegraphics[width=42mm]{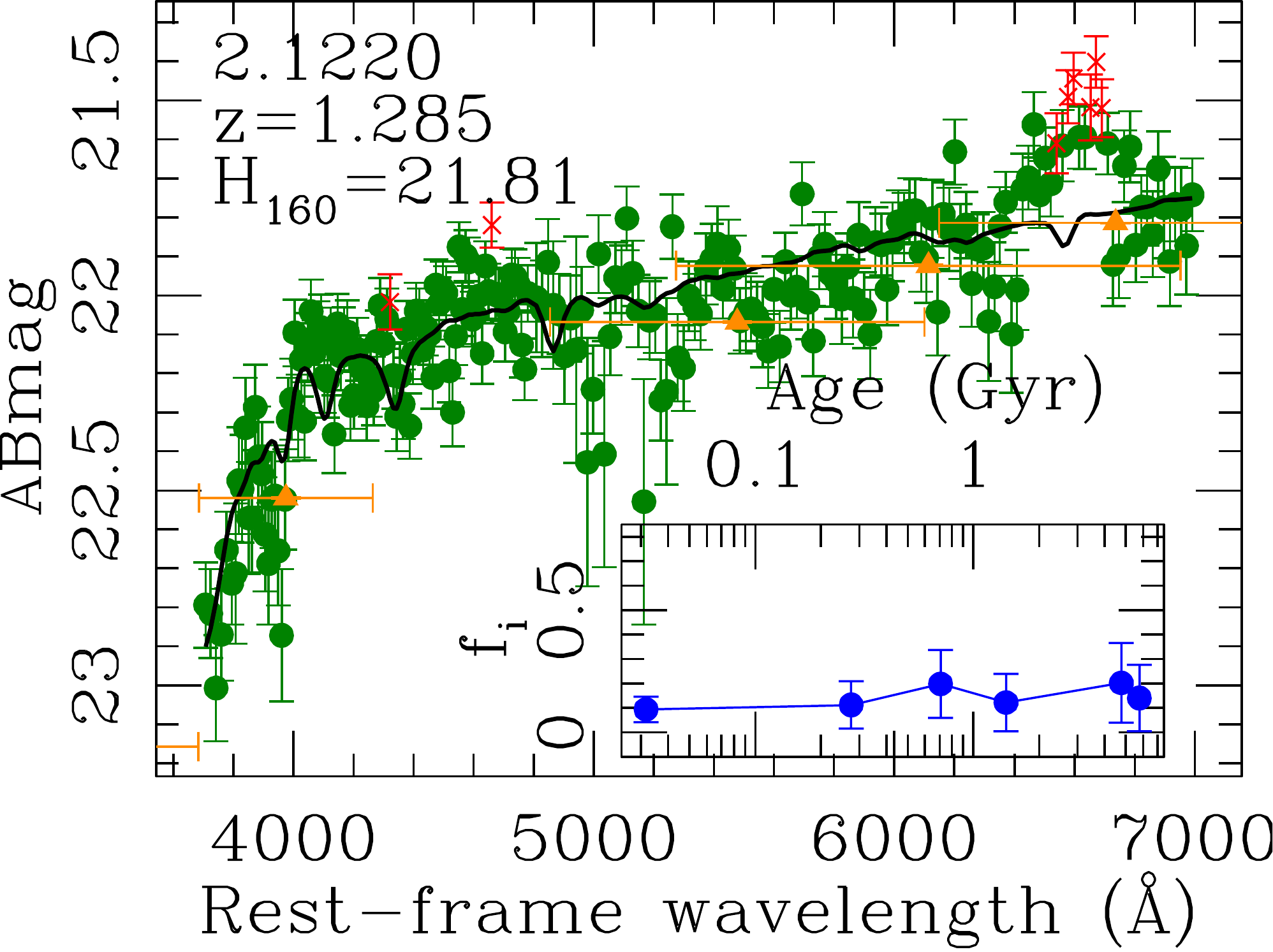}
\includegraphics[width=42mm]{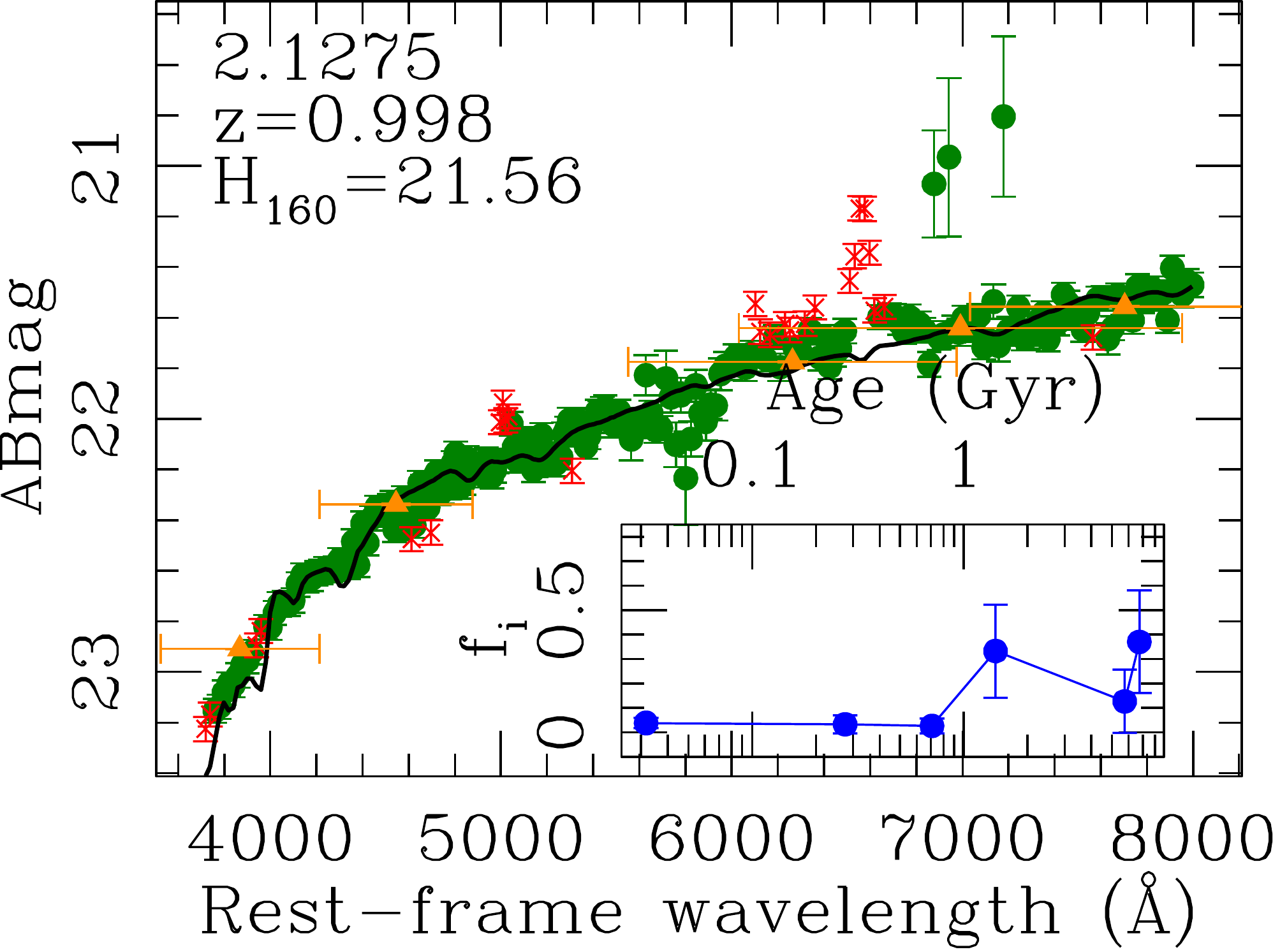}
\includegraphics[width=42mm]{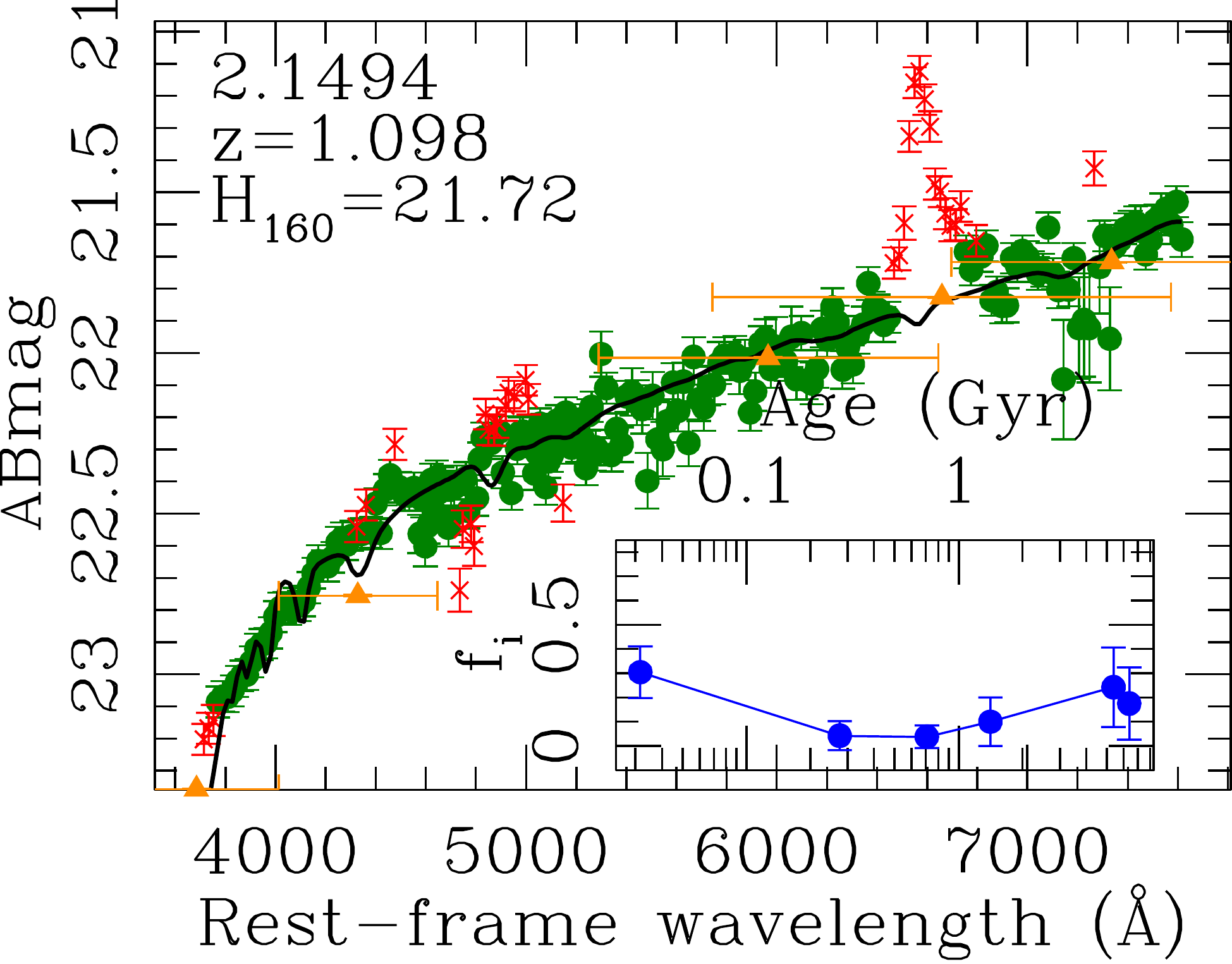}
\includegraphics[width=42mm]{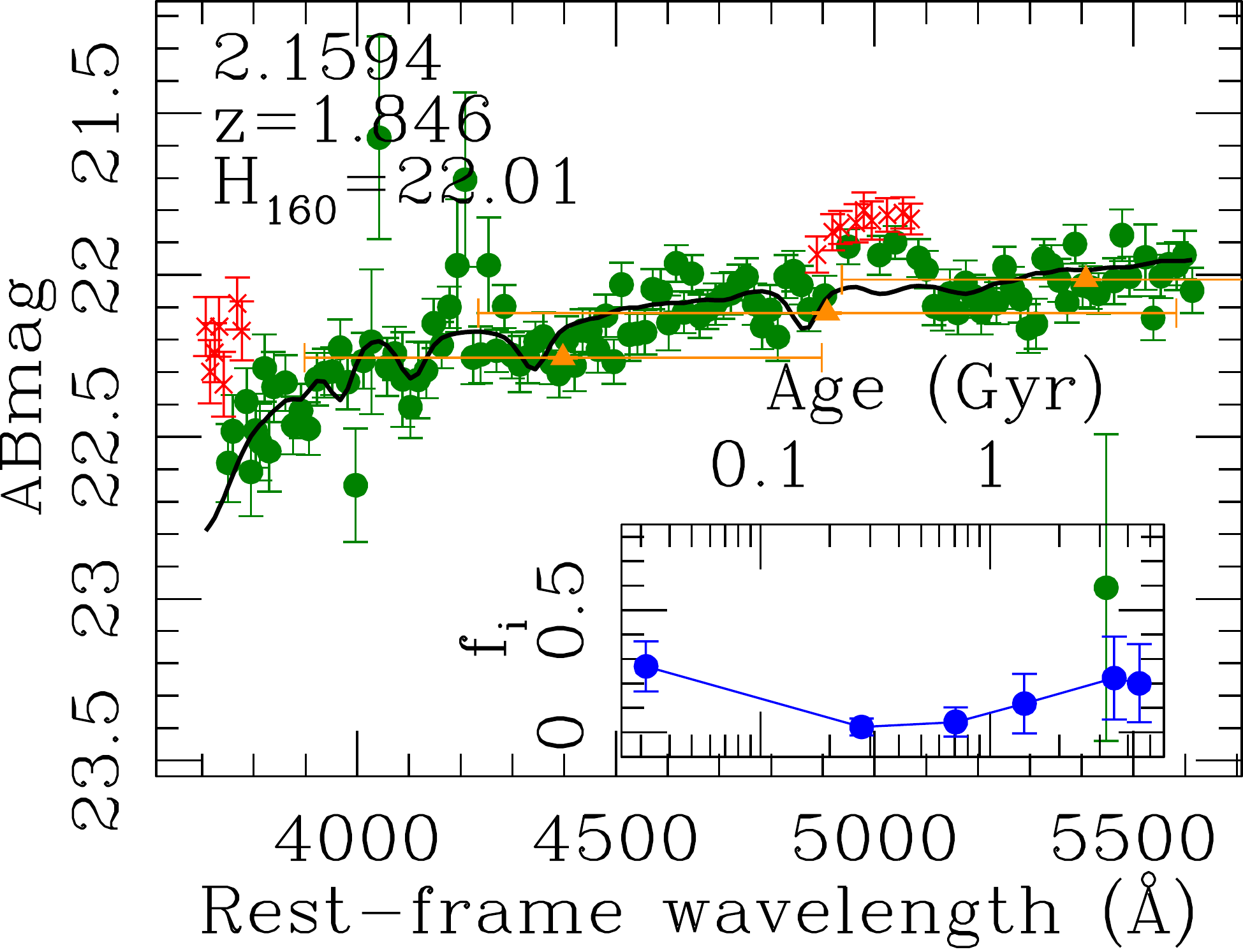}\\
\includegraphics[width=42mm]{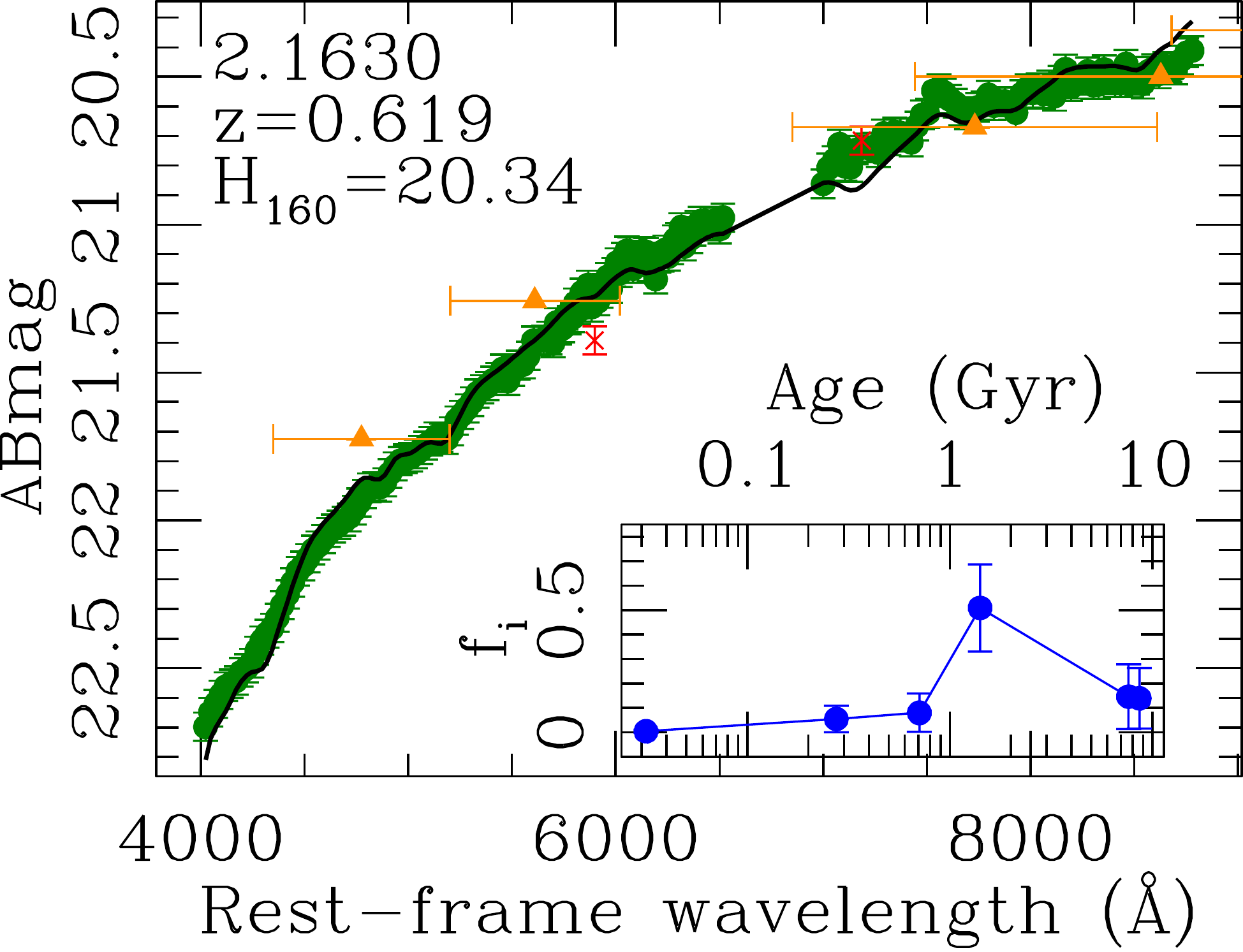}
\includegraphics[width=42mm]{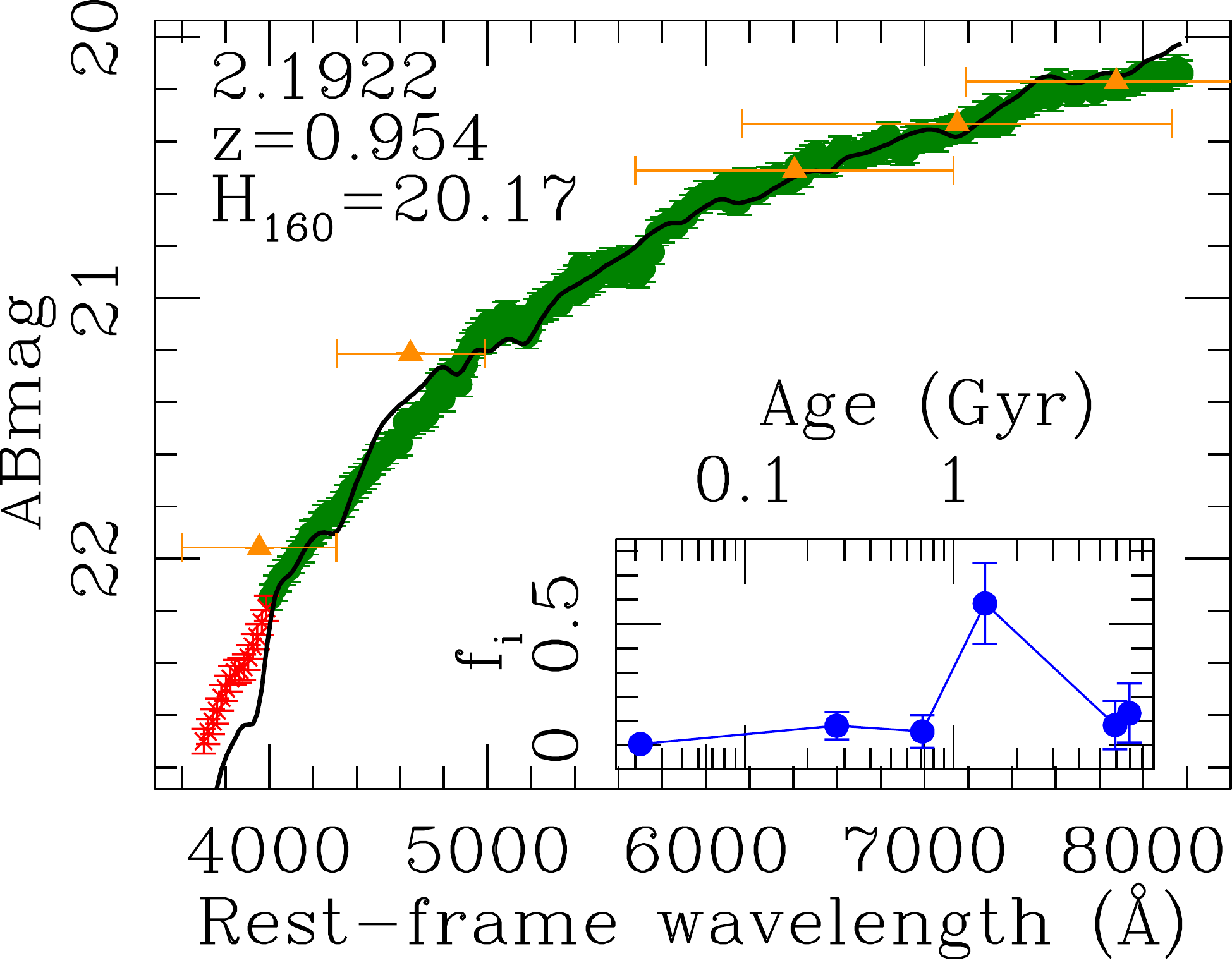}
\includegraphics[width=42mm]{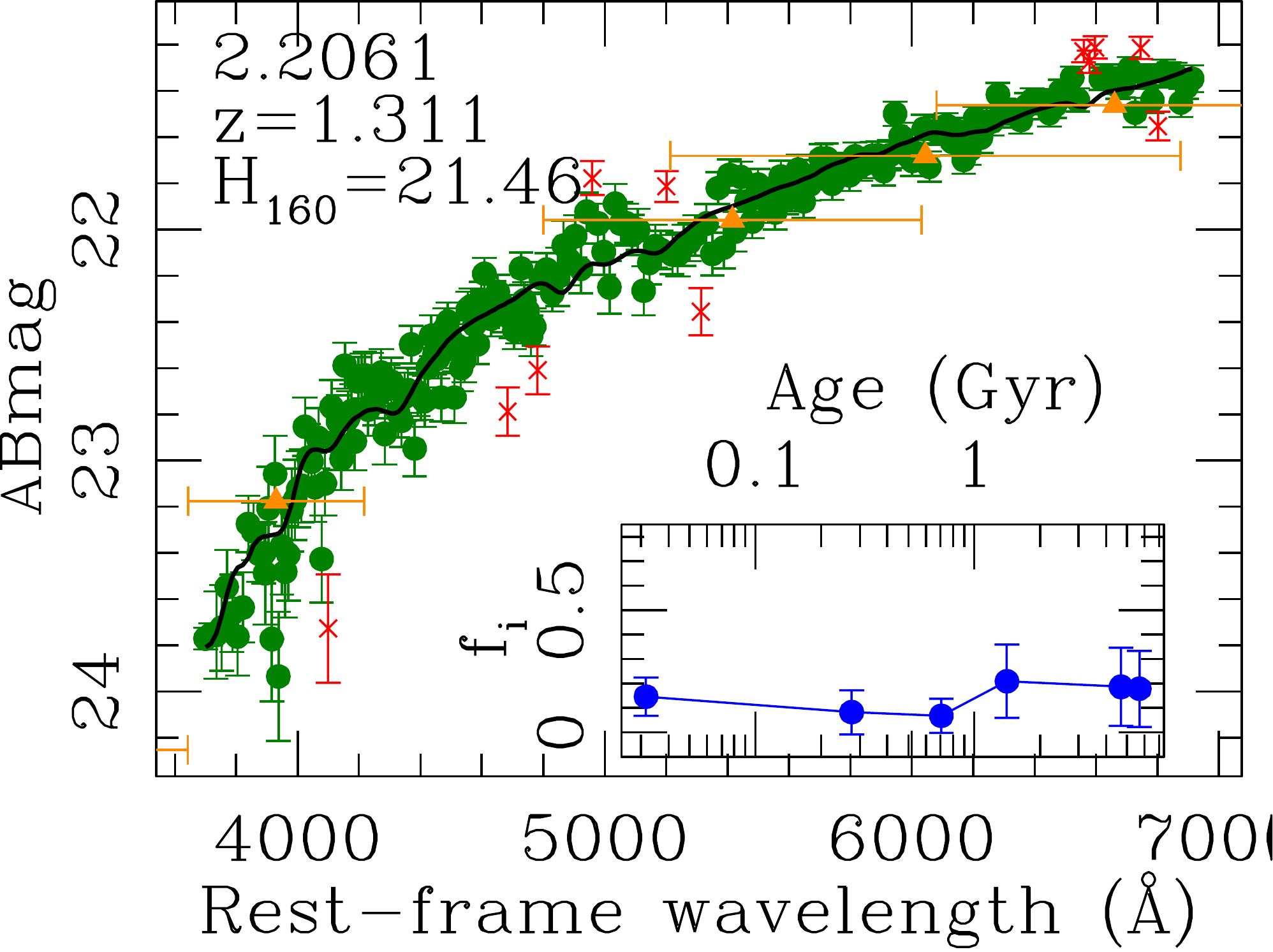}
\includegraphics[width=42mm]{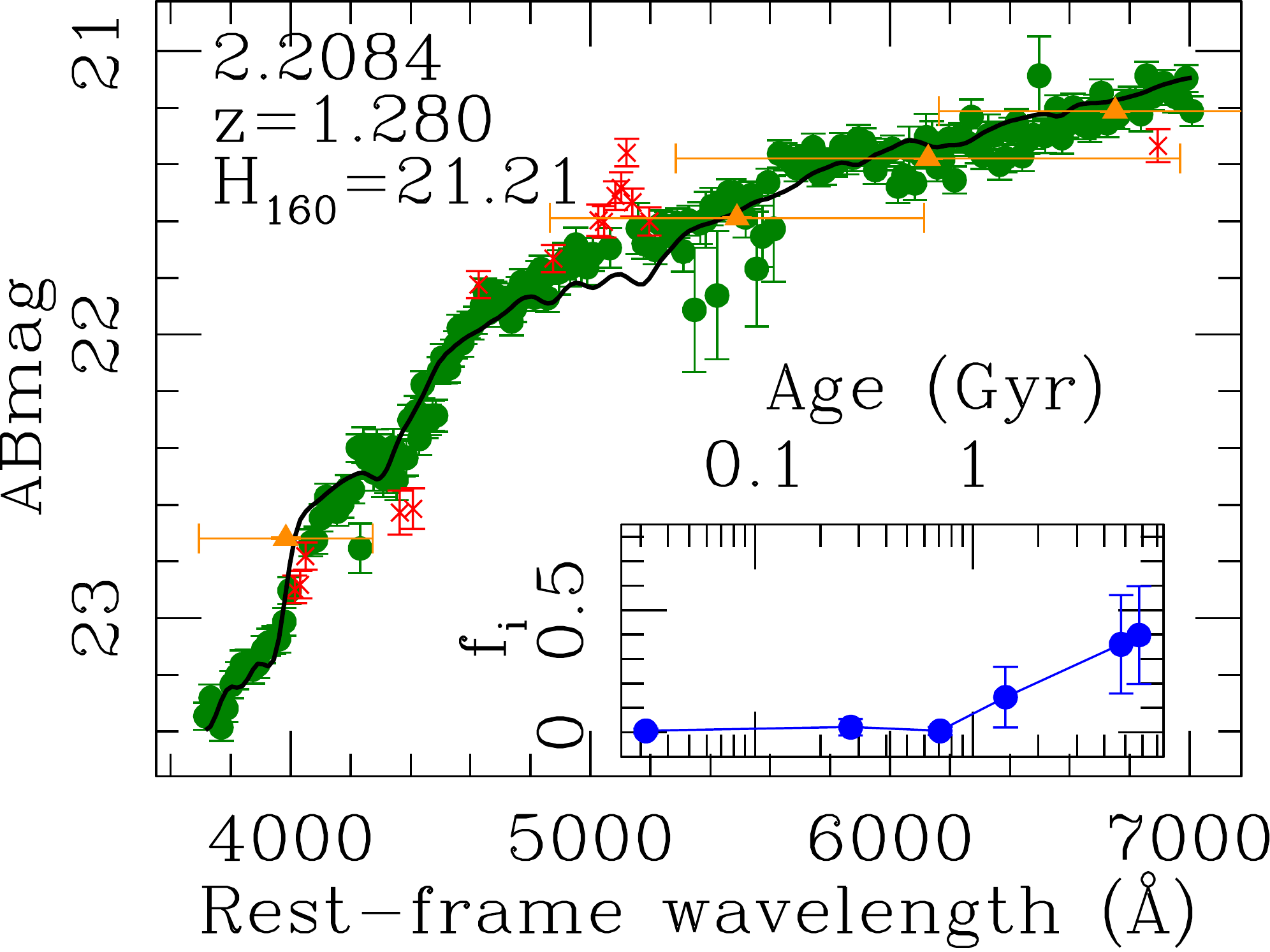}\\
\includegraphics[width=42mm]{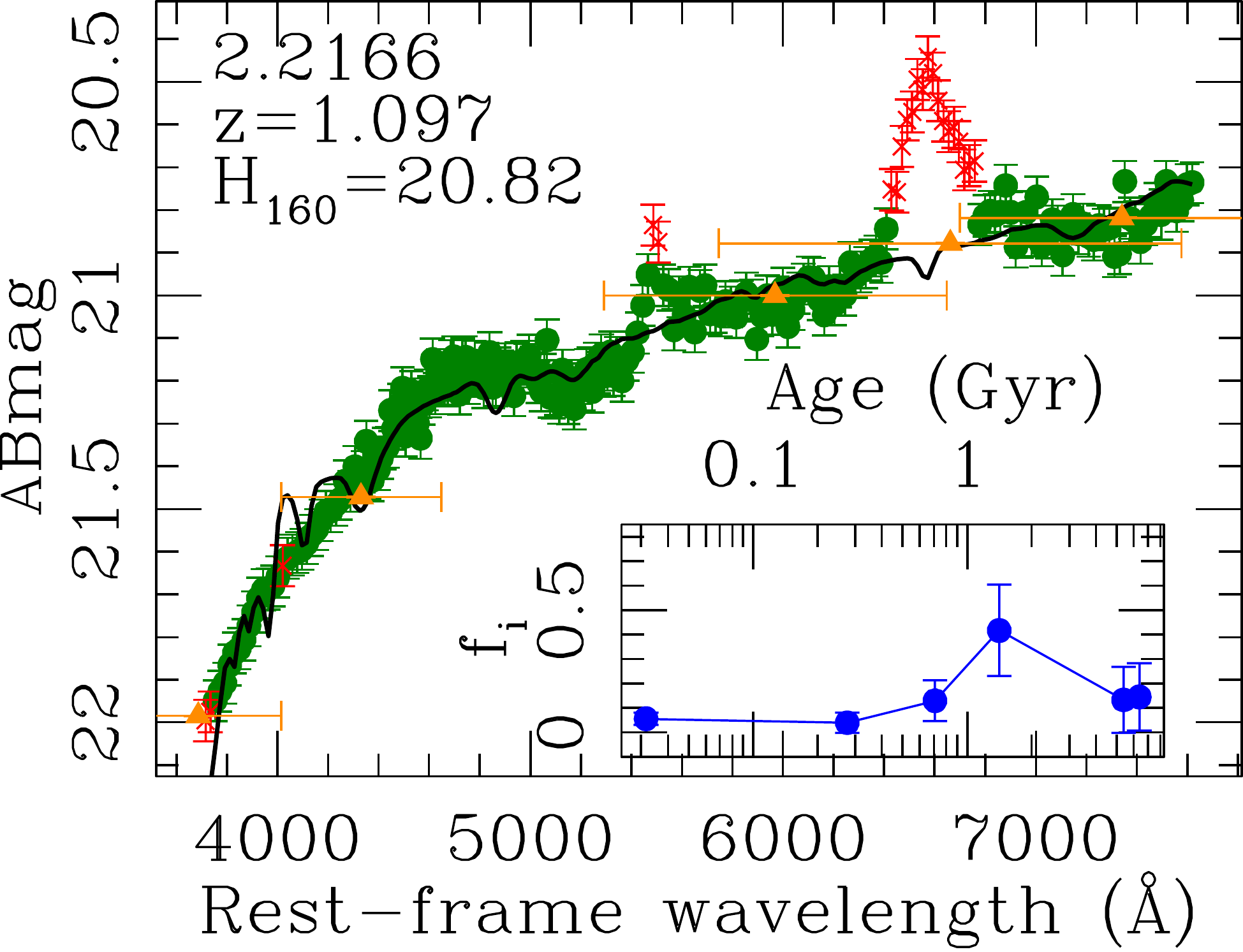}
\includegraphics[width=42mm]{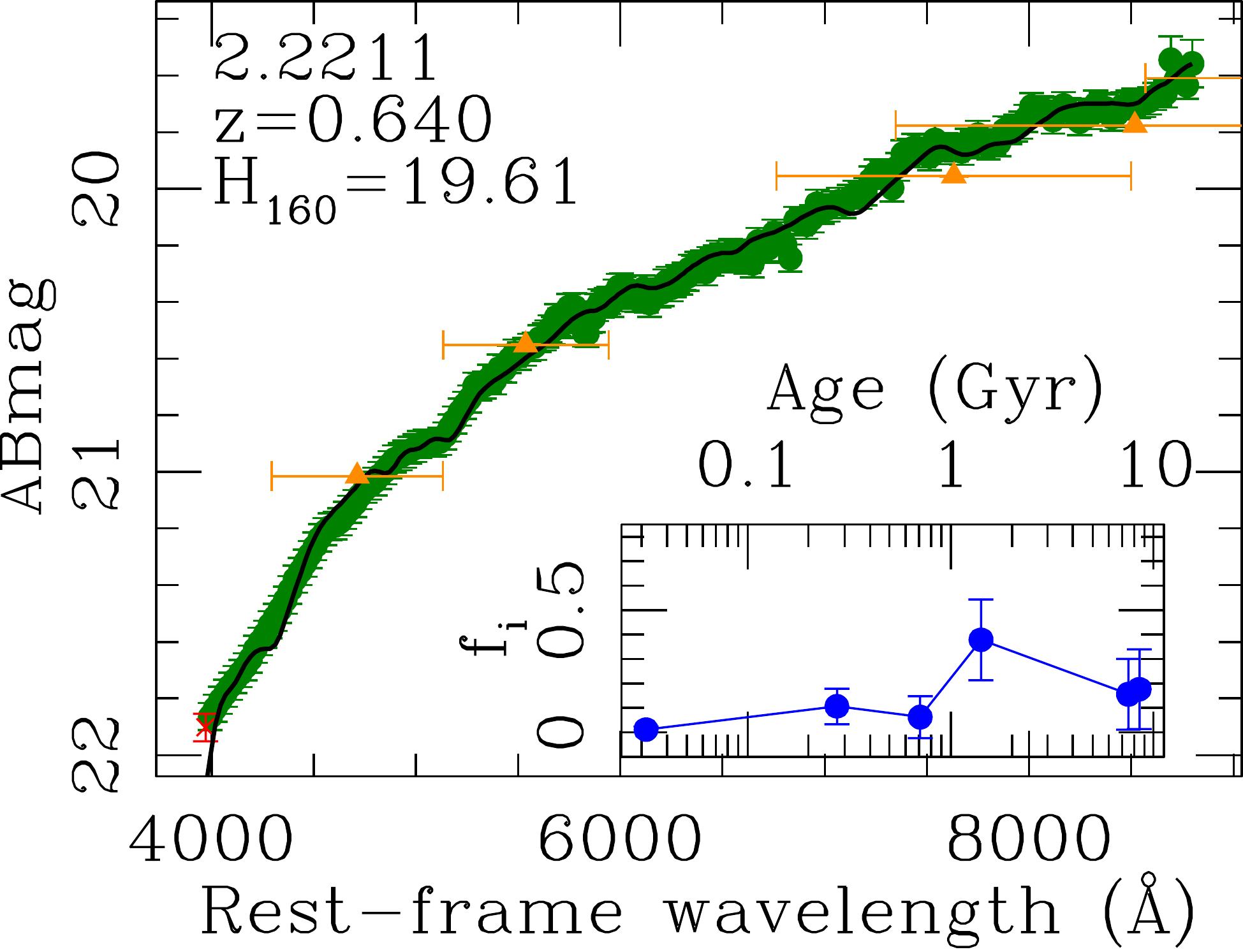}
\includegraphics[width=42mm]{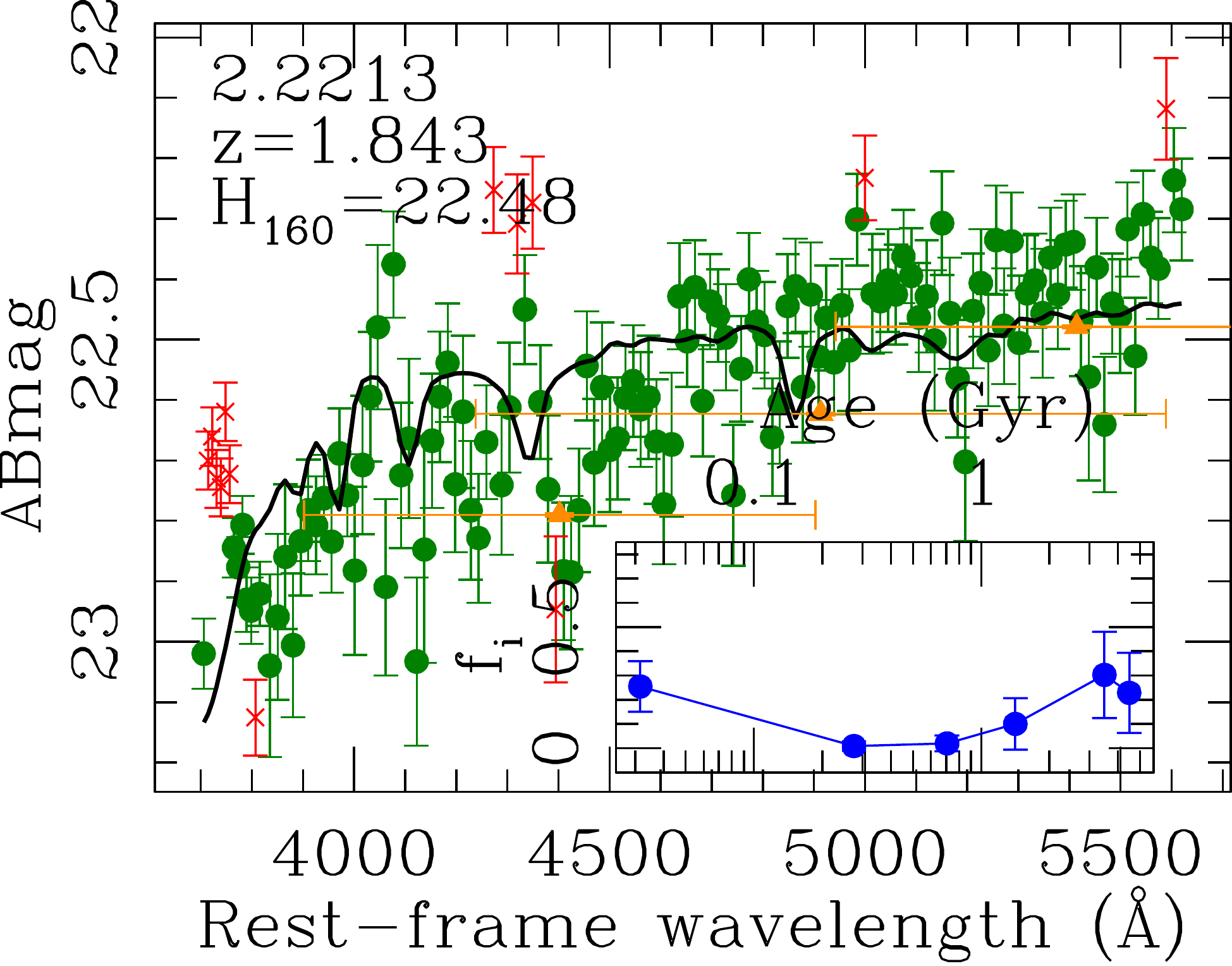}
\includegraphics[width=42mm]{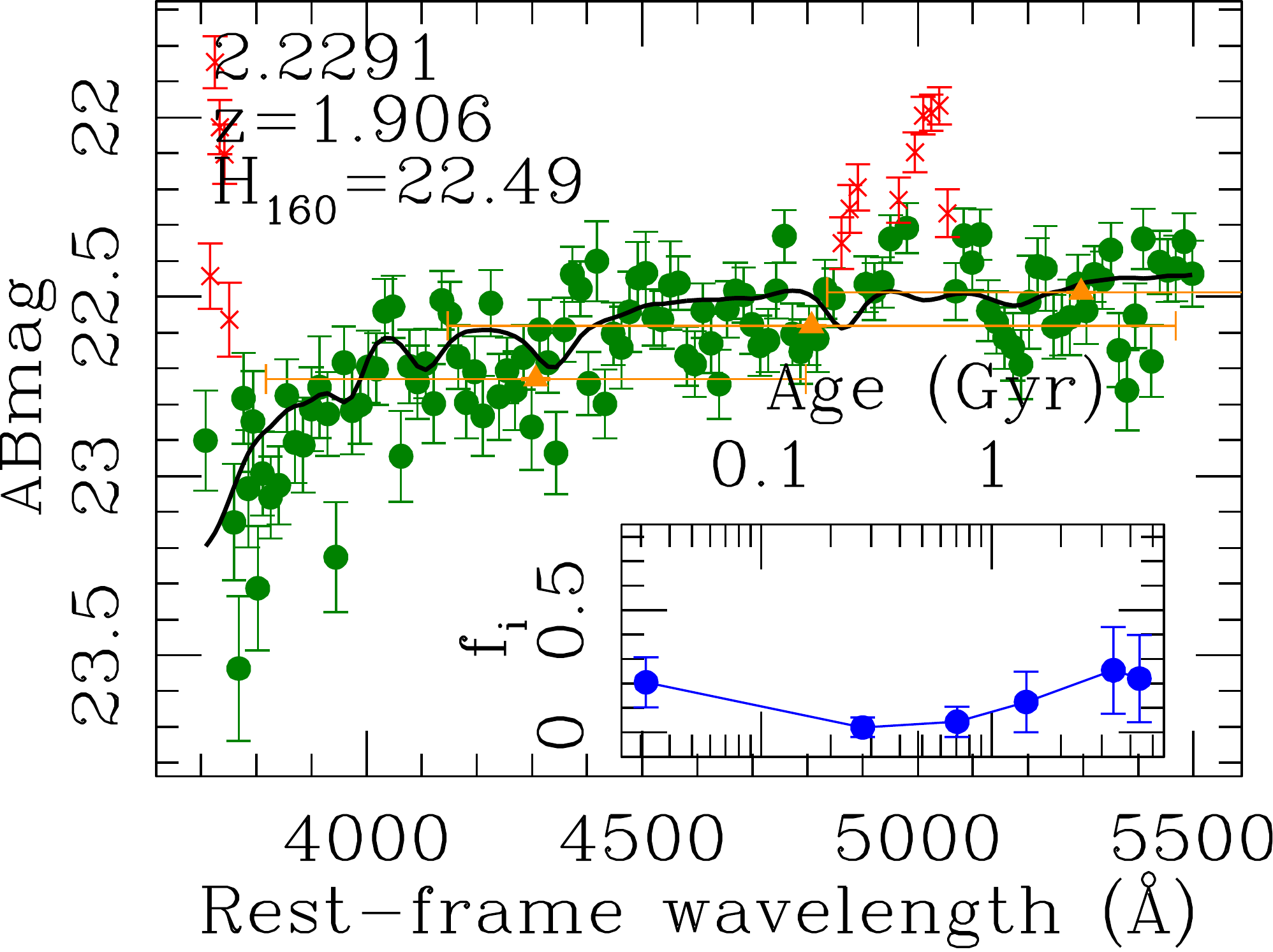}\\
\includegraphics[width=42mm]{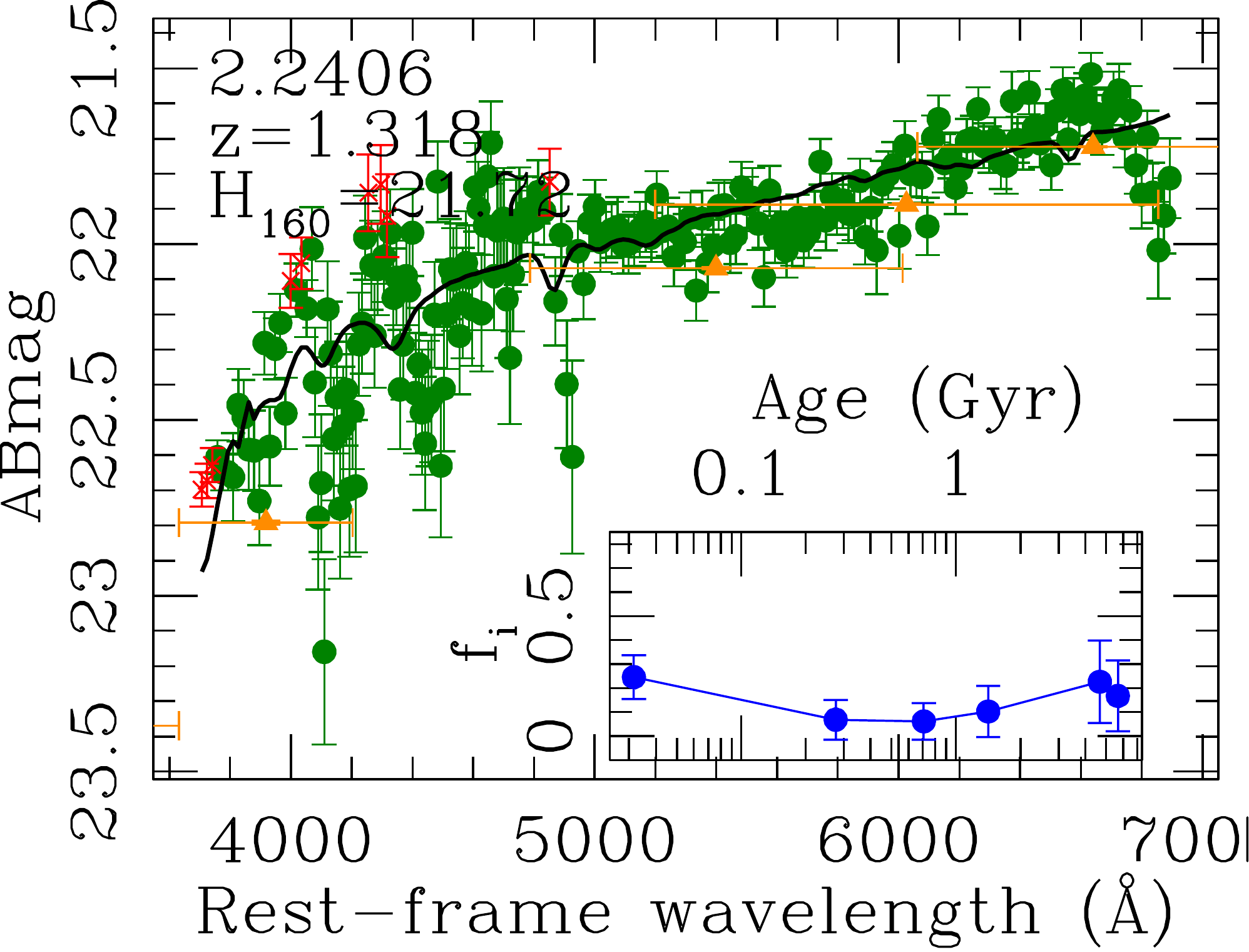}
\includegraphics[width=42mm]{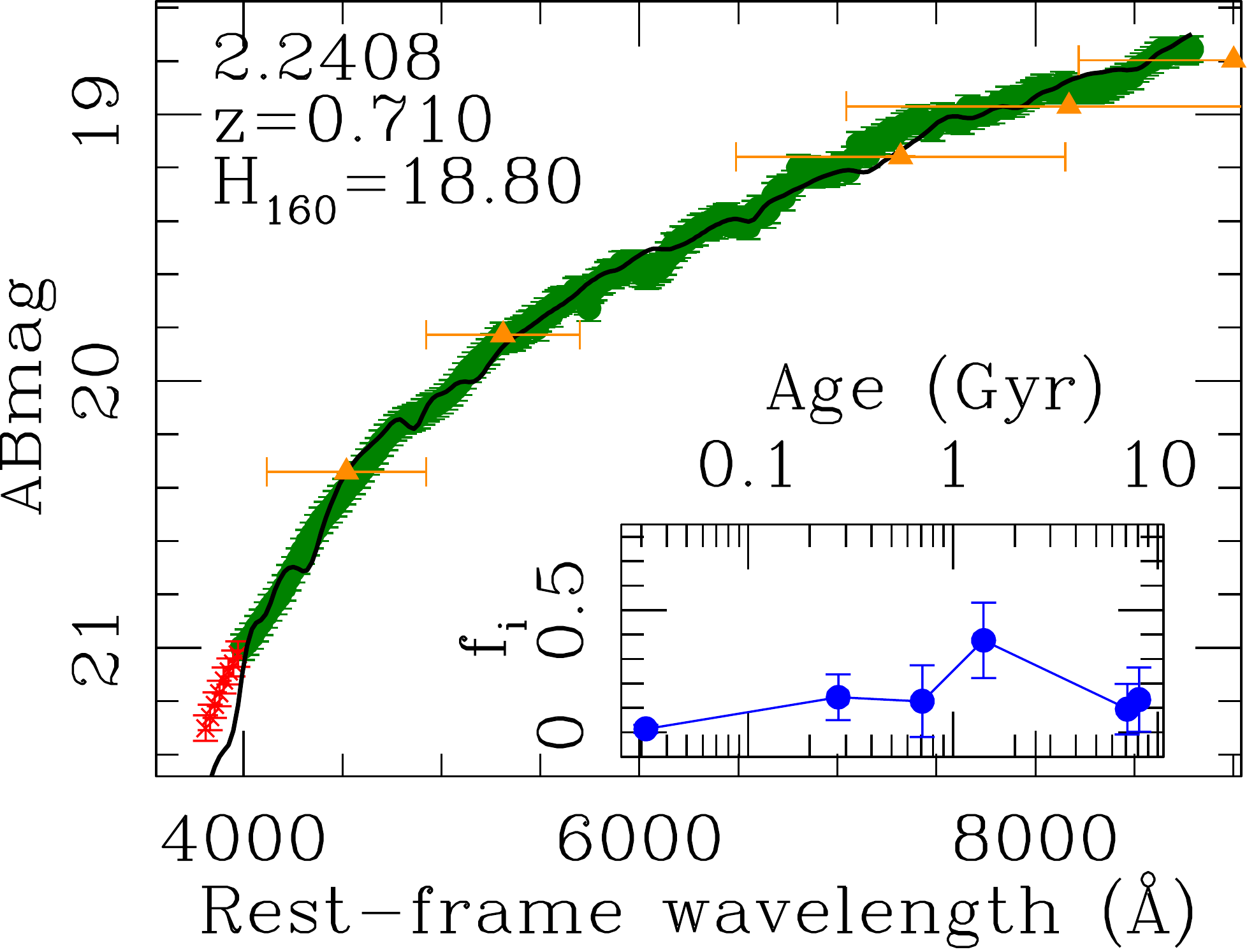}
\includegraphics[width=42mm]{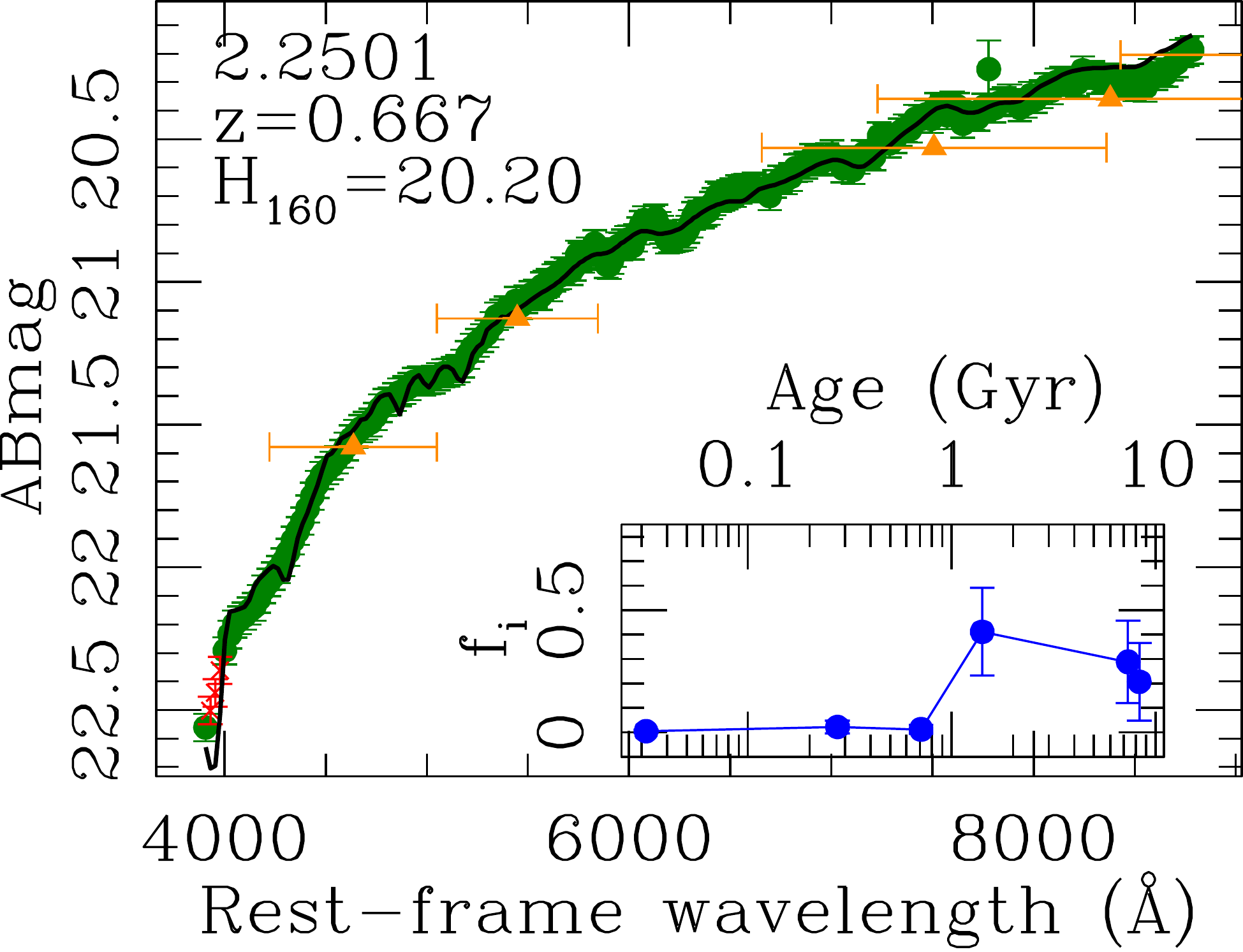}
\includegraphics[width=42mm]{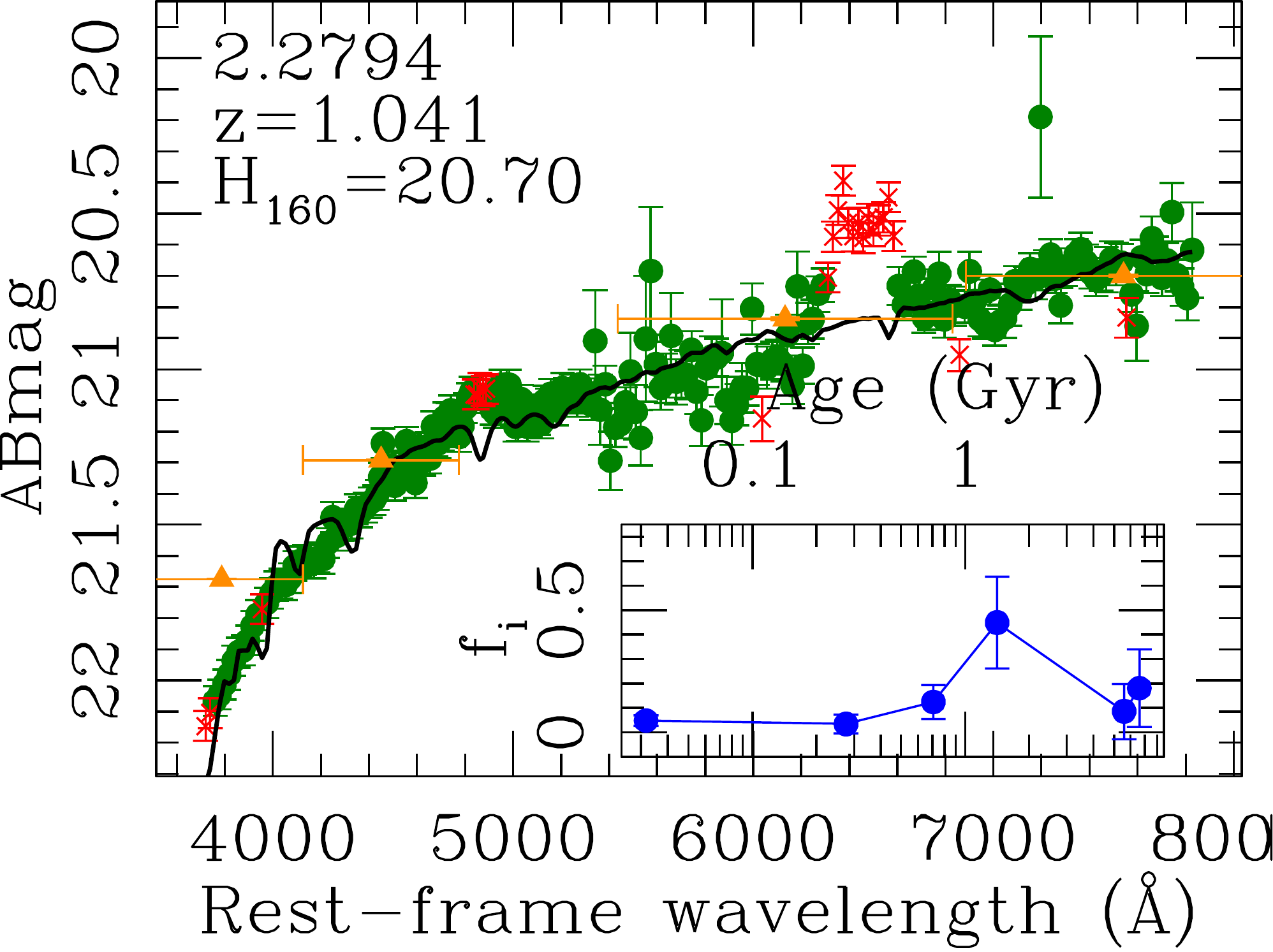}\\
\includegraphics[width=42mm]{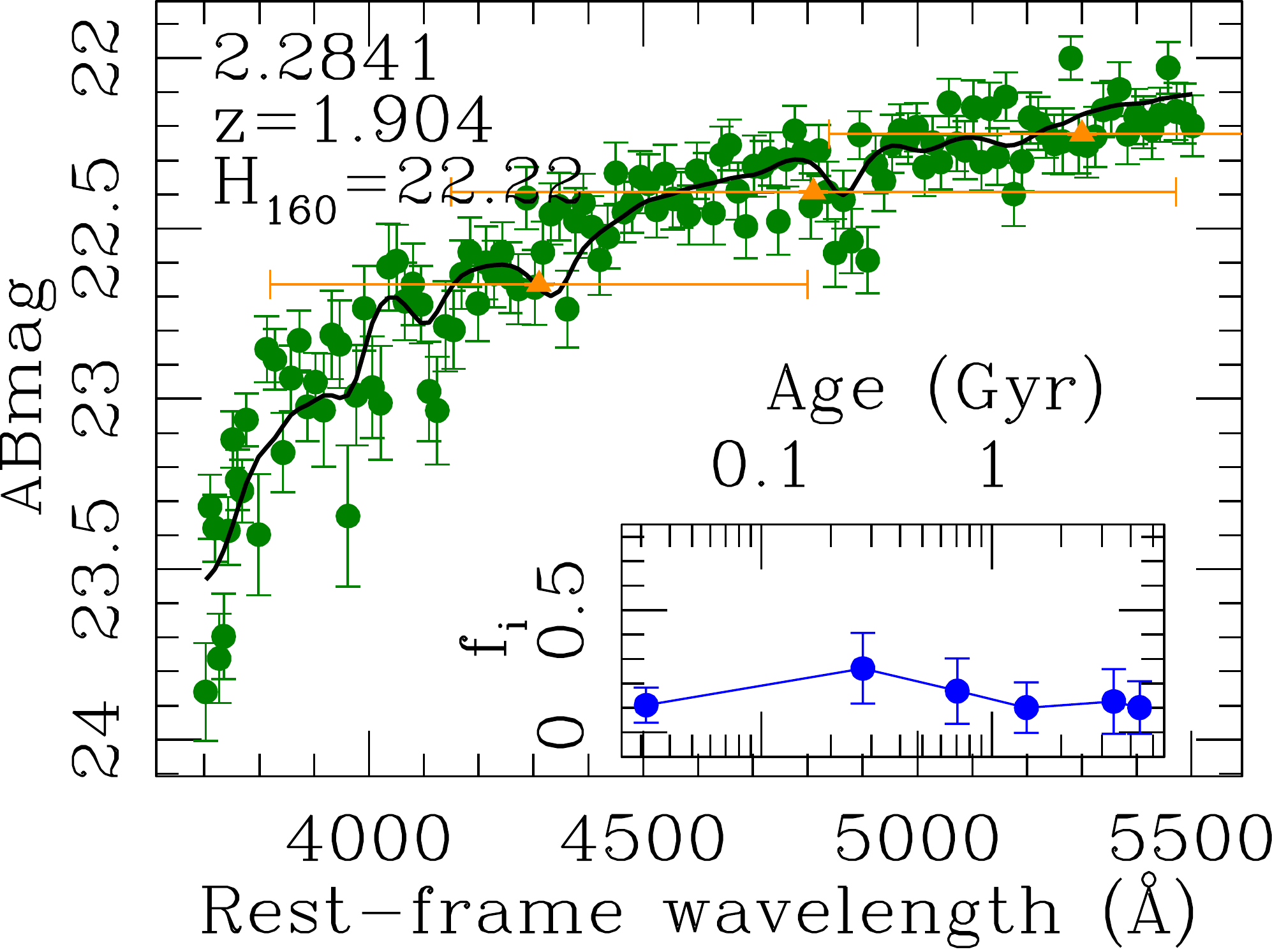}
\includegraphics[width=42mm]{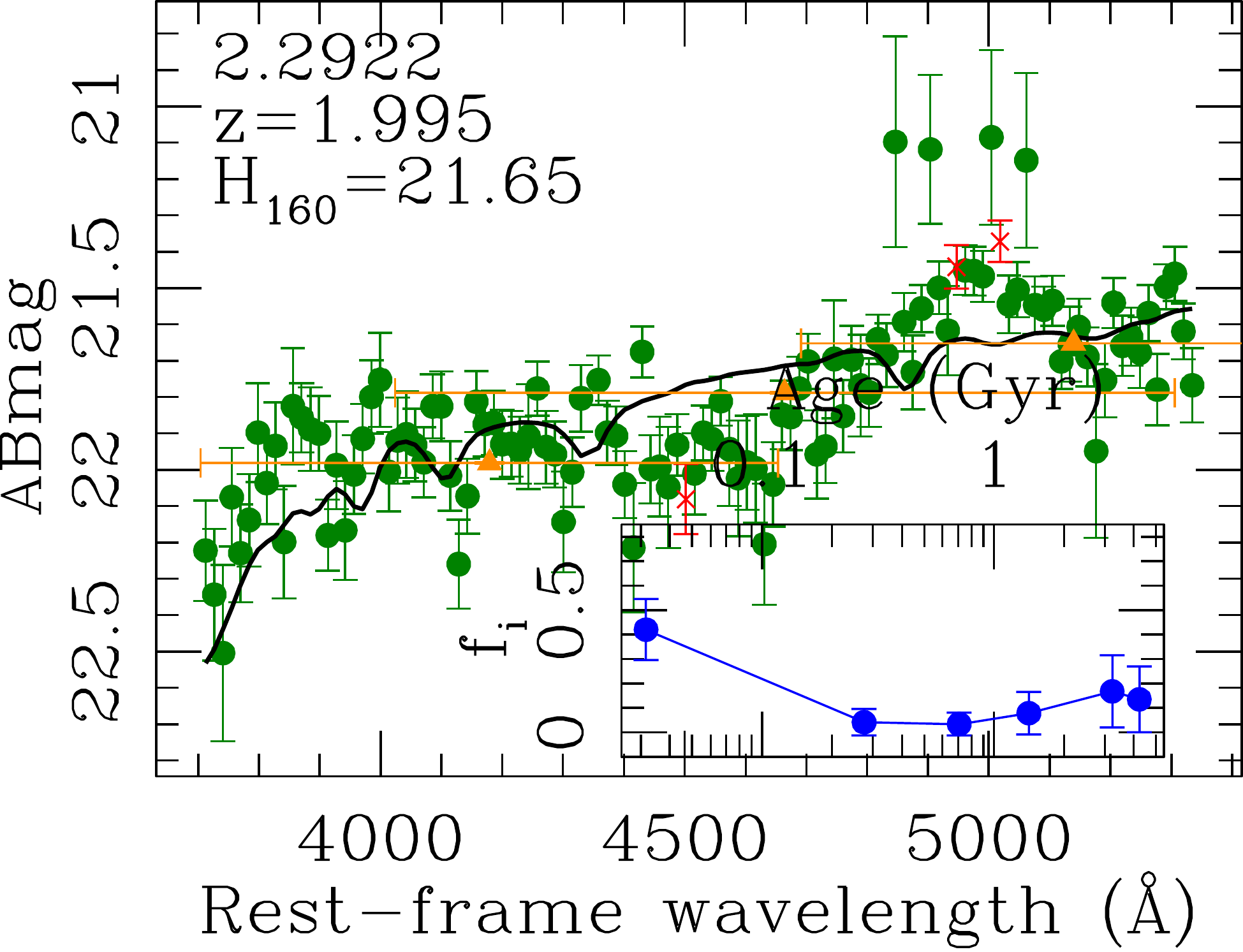}
\includegraphics[width=42mm]{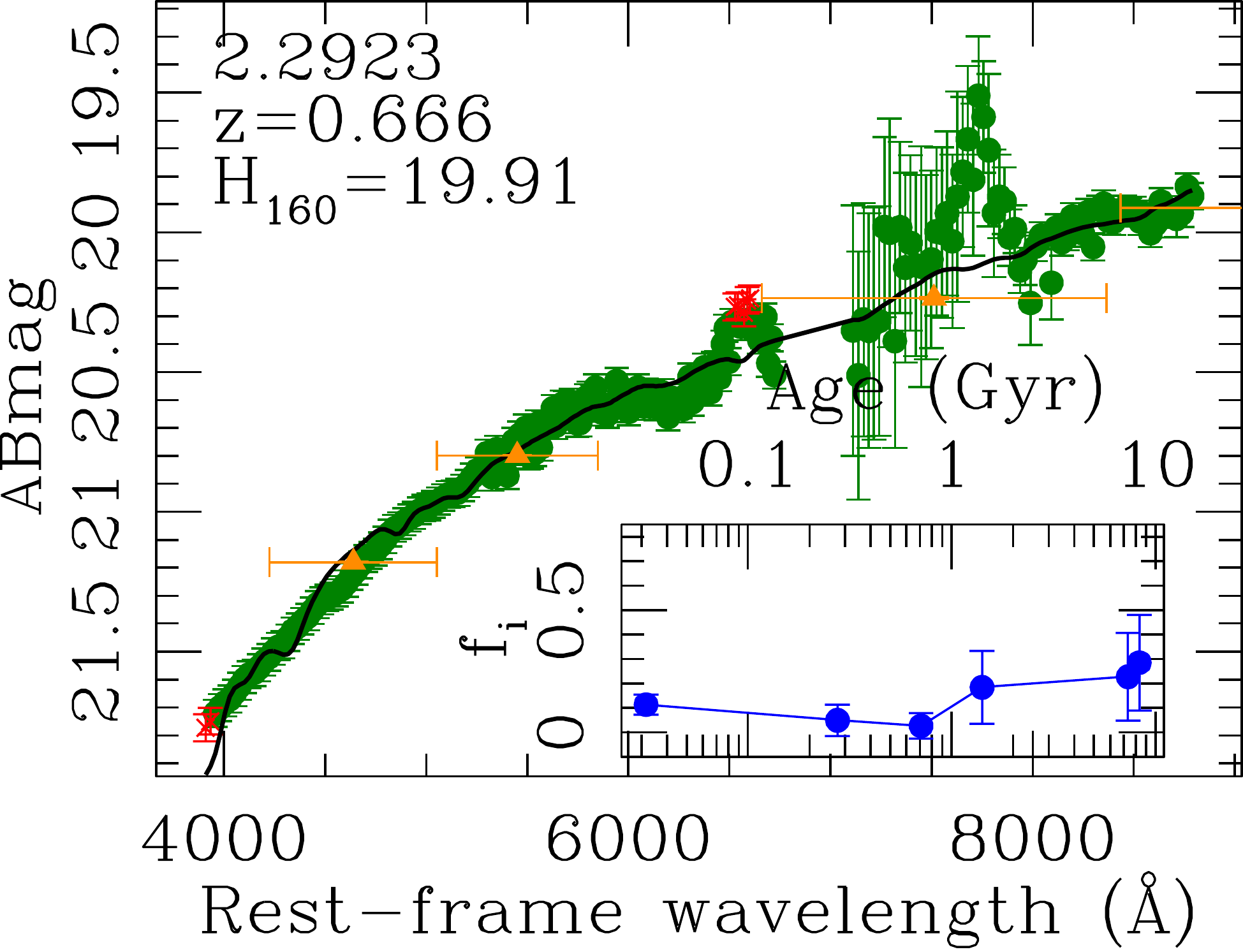}
\includegraphics[width=42mm]{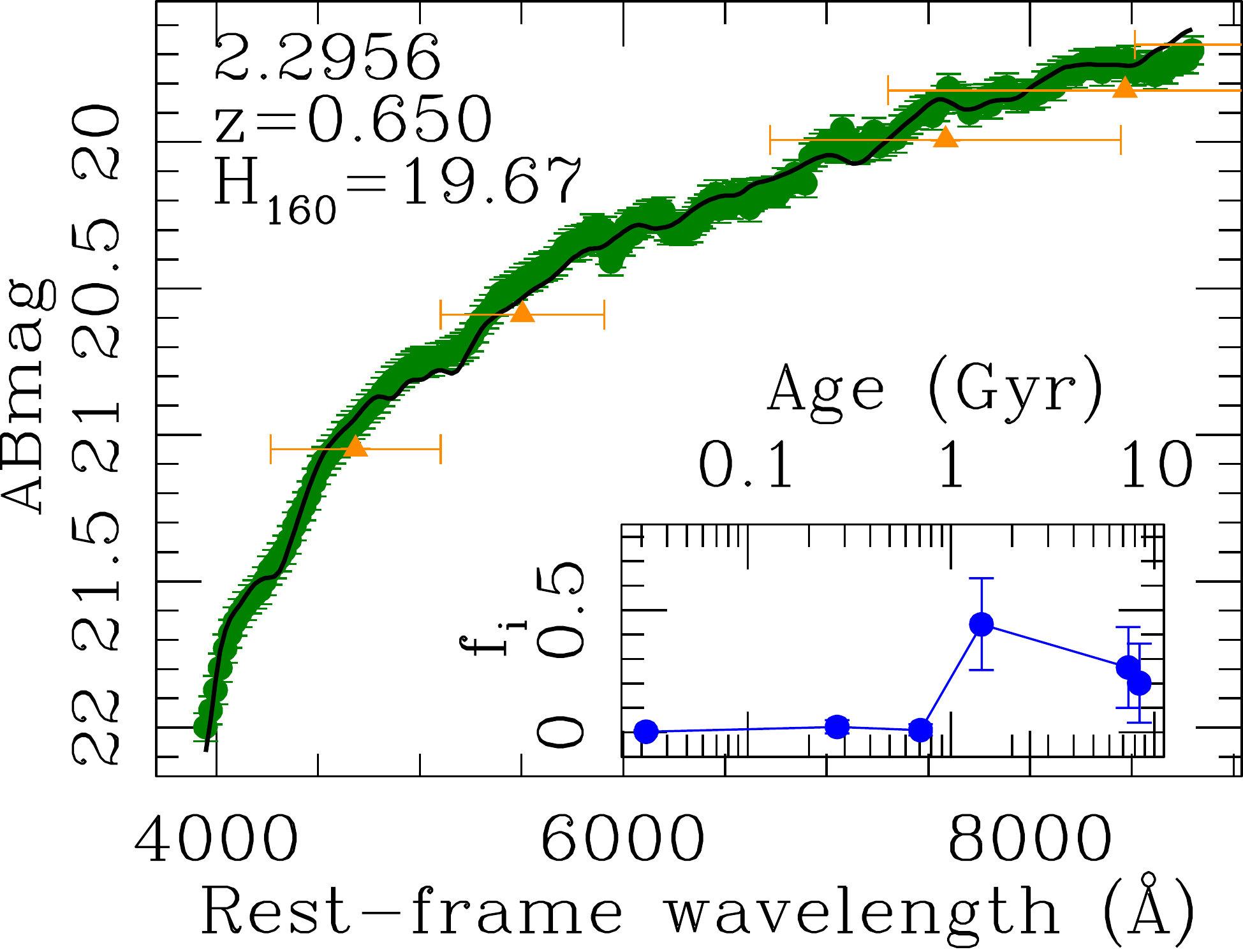}\\
\includegraphics[width=42mm]{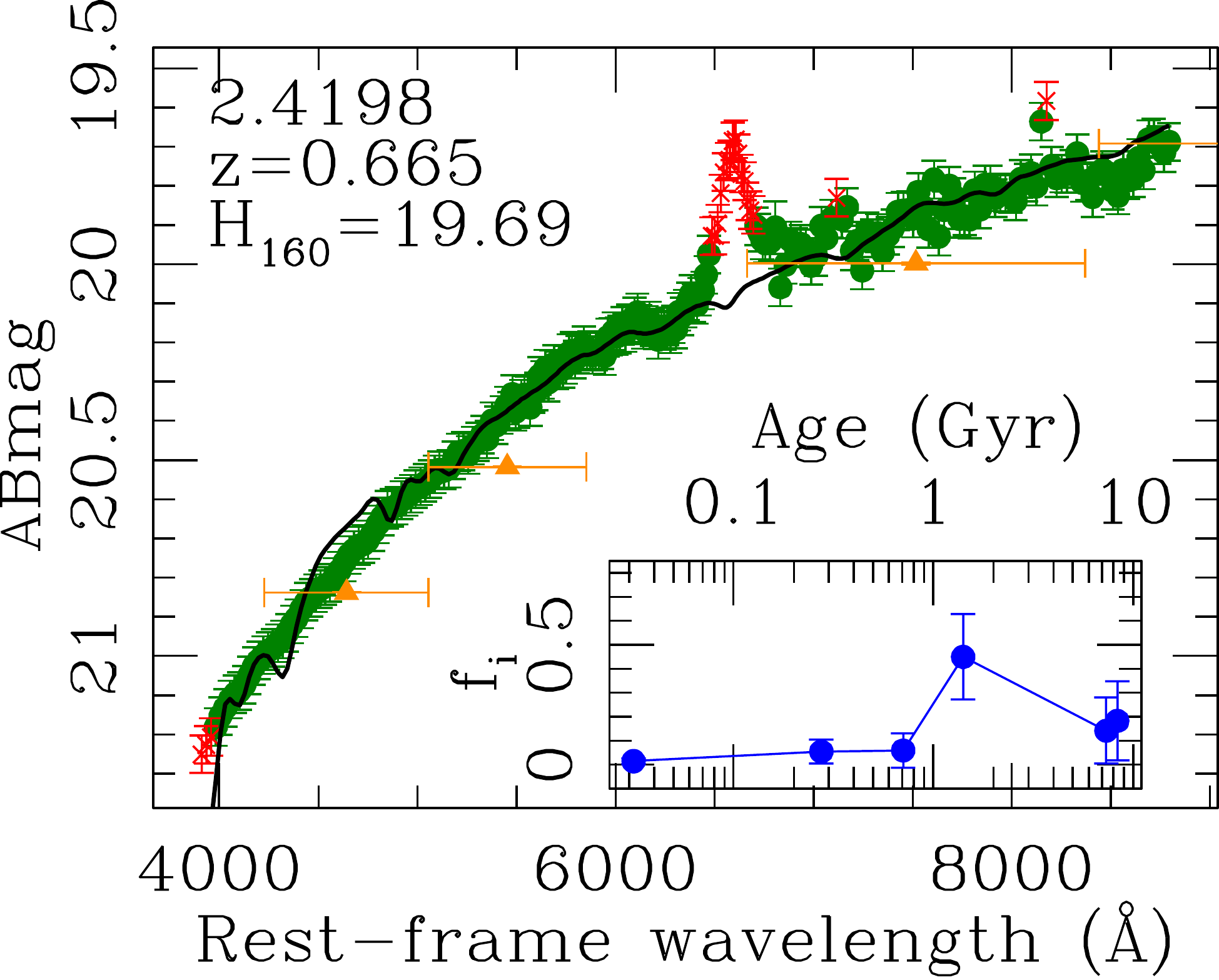}
\includegraphics[width=42mm]{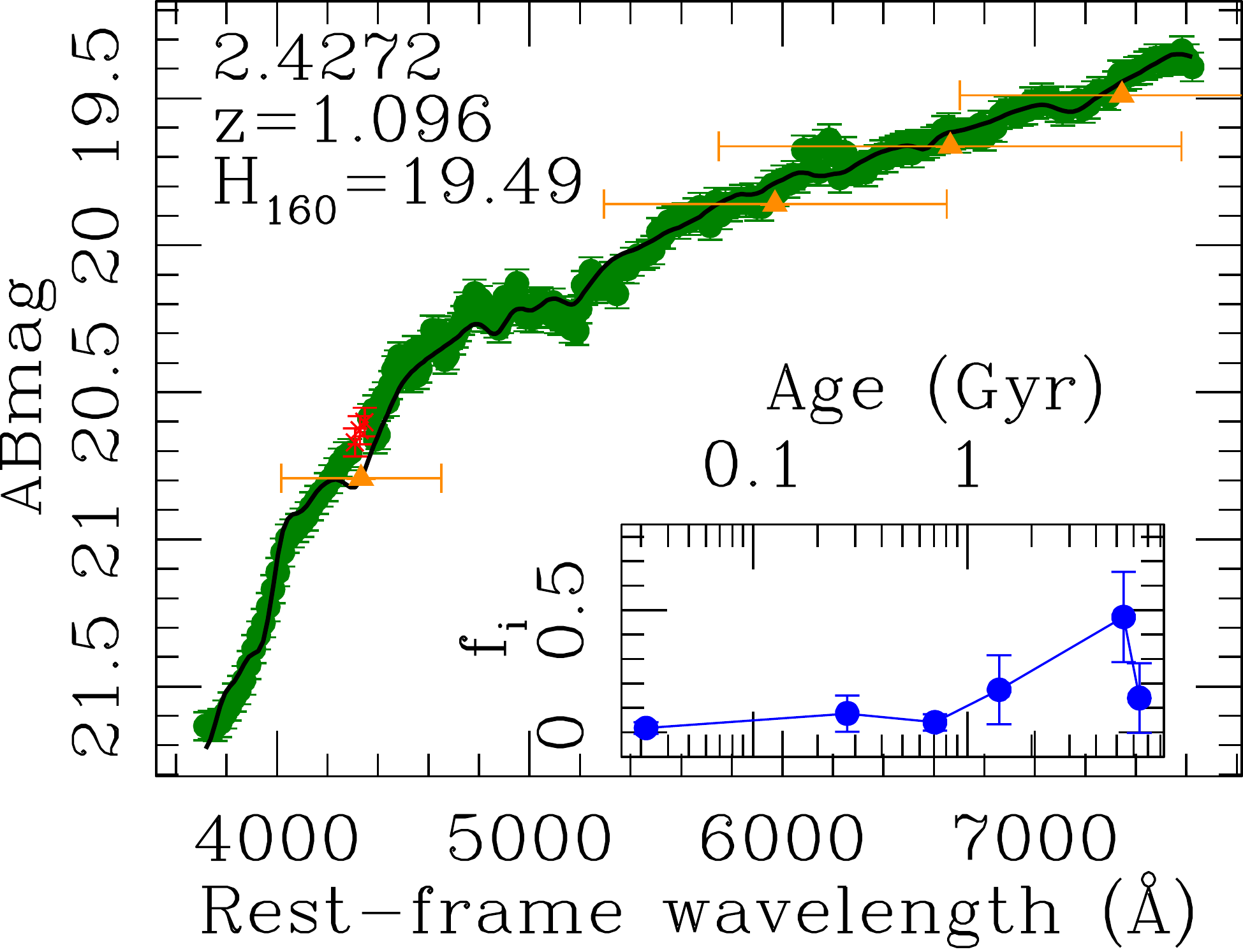}
\end{center}
\caption{Continuation of Fig.~\ref{fig:appFits}.
The notation follows that of Fig.~\ref{fig:sed}.
Note that Base Model 6 (that has
the same age as BM5) is displaced by $+$1\,Gyr.
}
\label{fig:appFits2}
\end{figure*}
%%%%%%%%%%%%%%%%%%%%%%%%%%%%%%%%%%%%%%%%%%%%%%%%

%\fi

%%%%%%%%%%%%%%%%%%%%%%%%%%%%%%%%%%%%%%%%%%%%%%%%
\section{Model tests}
\label{App:Tests}
\begin{figure*}
\begin{center}
\includegraphics[angle=90,width=55mm]{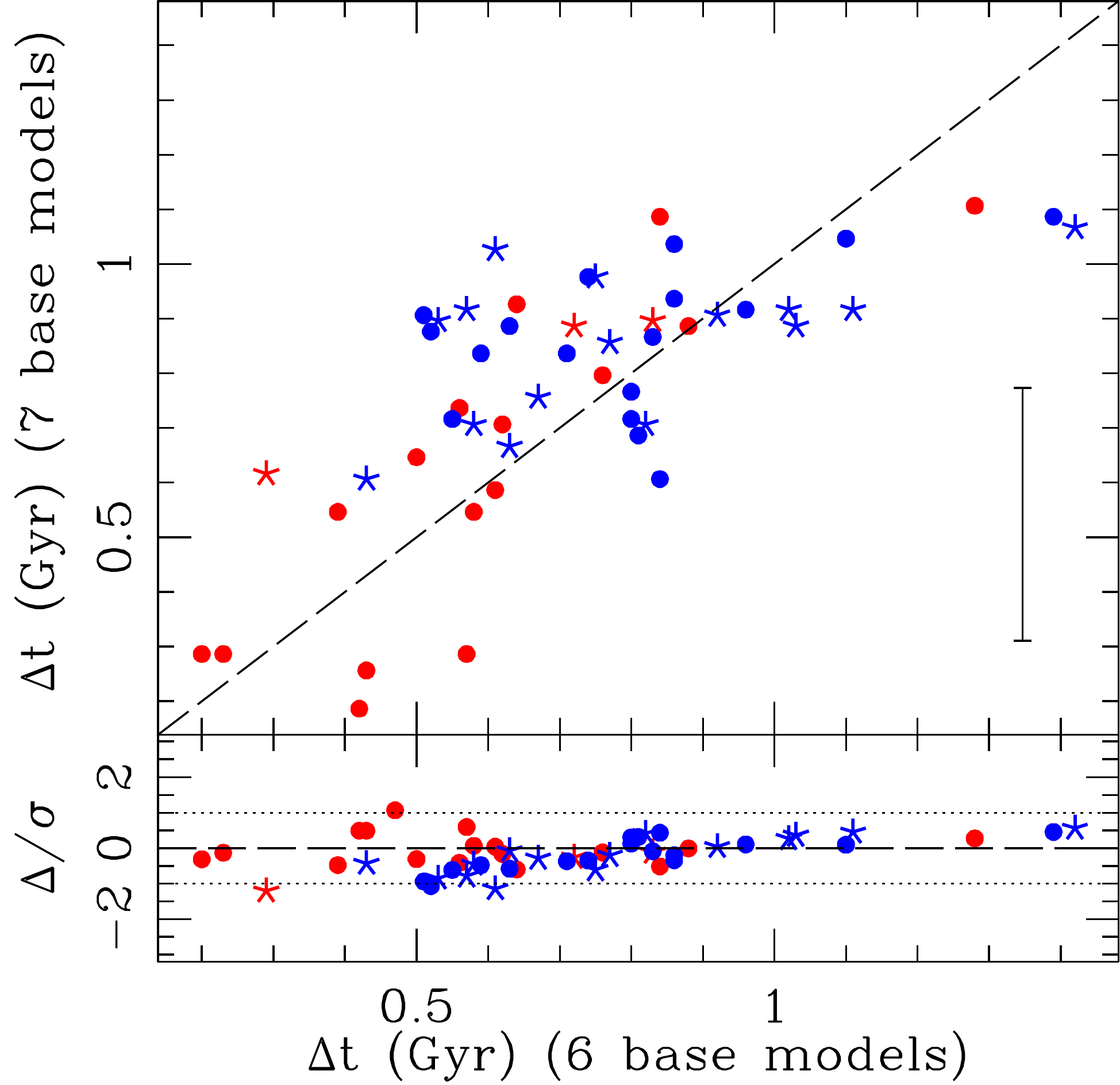}
\includegraphics[angle=90,width=55mm]{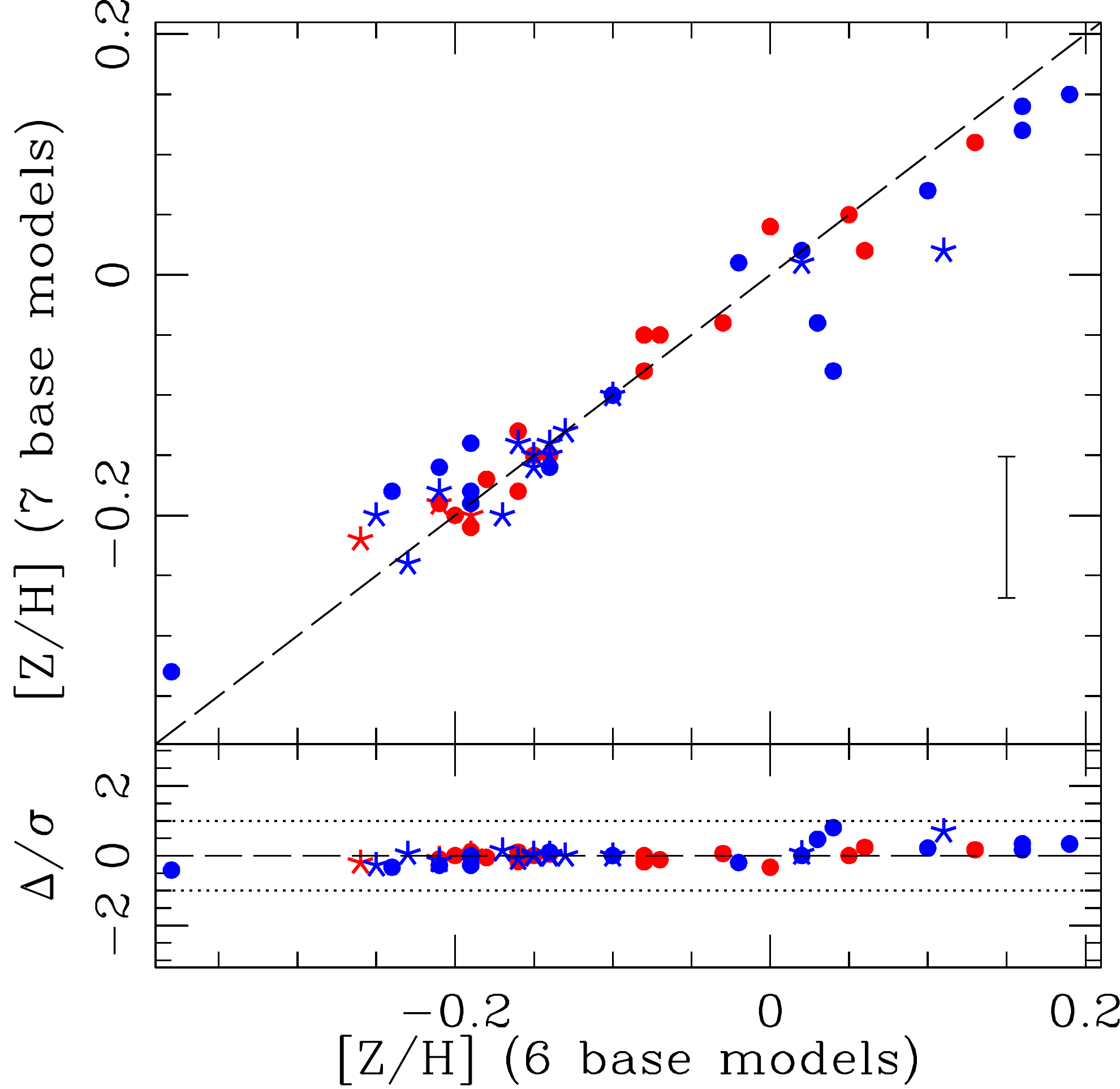}
\includegraphics[angle=90,width=55mm]{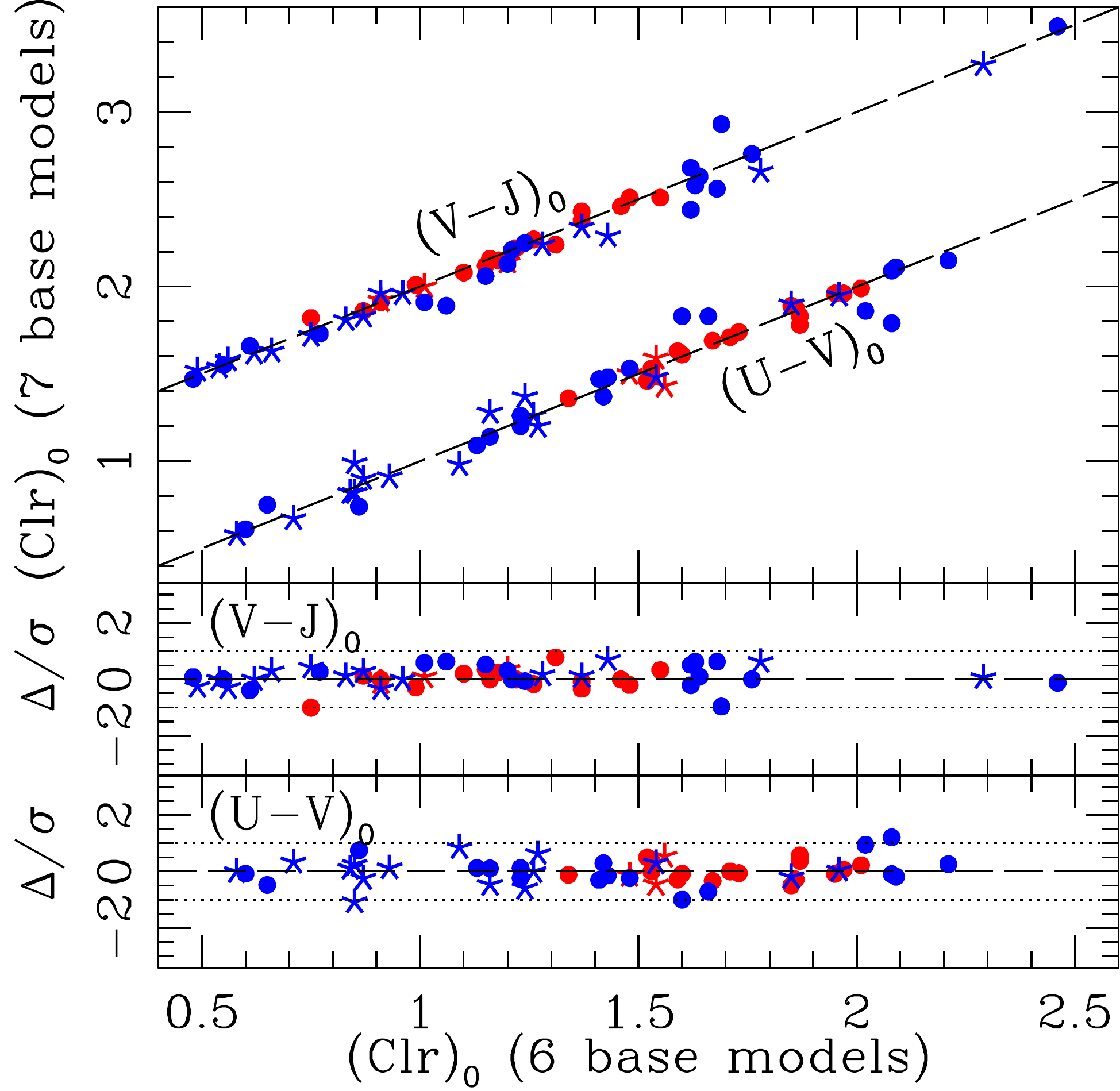}\\
\includegraphics[angle=90,width=55mm]{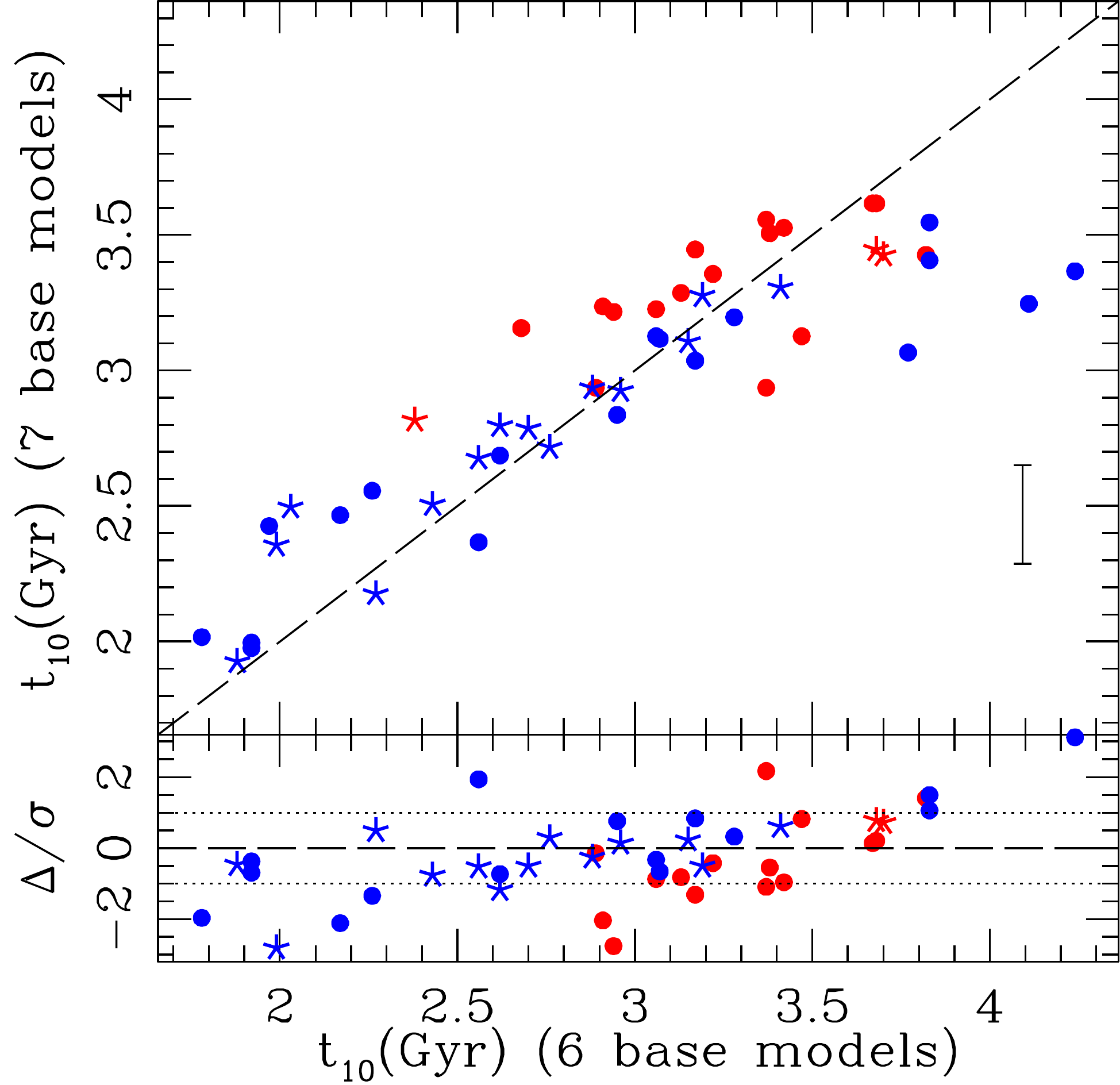}
\includegraphics[angle=90,width=55mm]{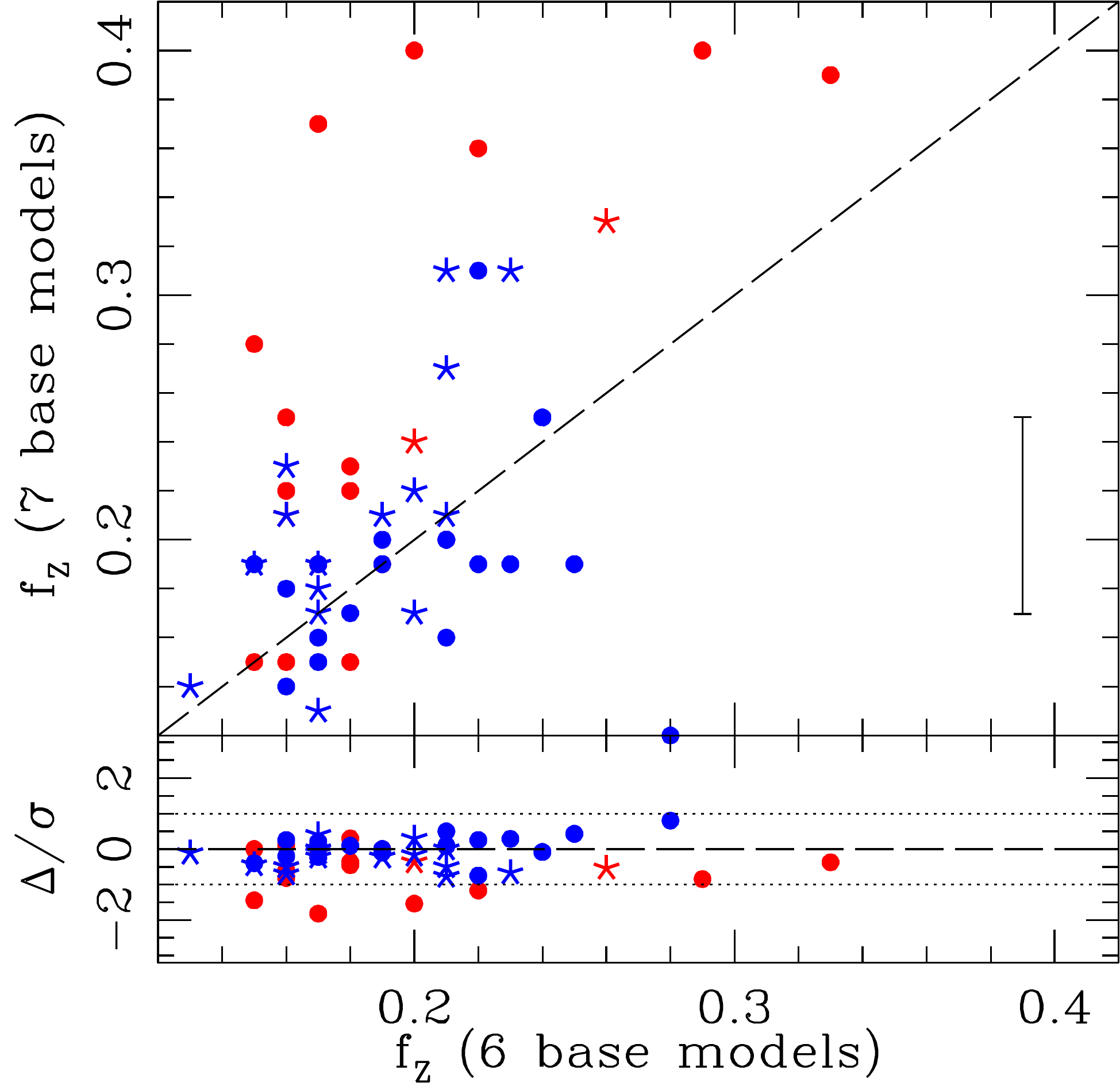}
\includegraphics[angle=90,width=55mm]{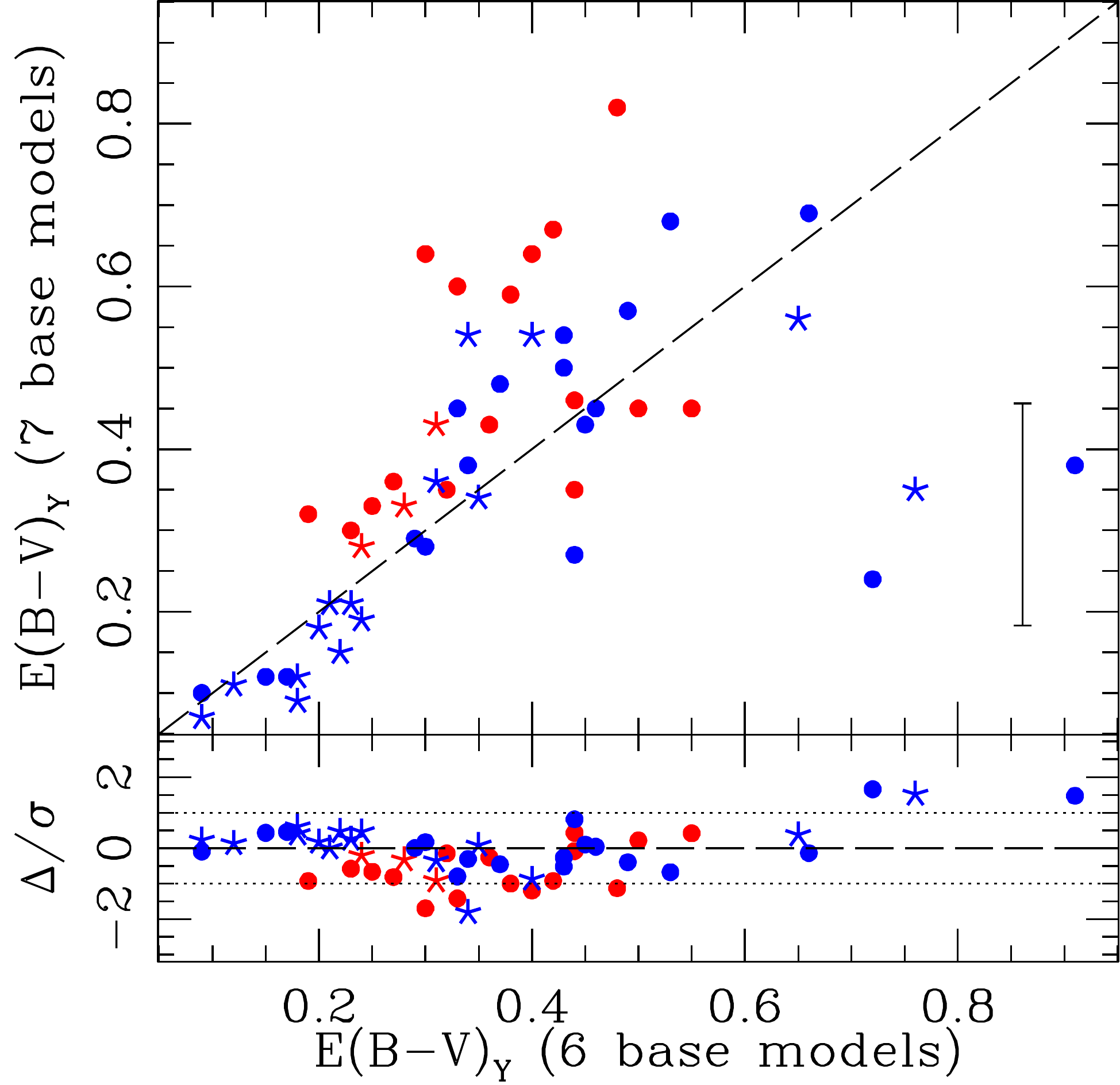}\\
\includegraphics[angle=90,width=55mm]{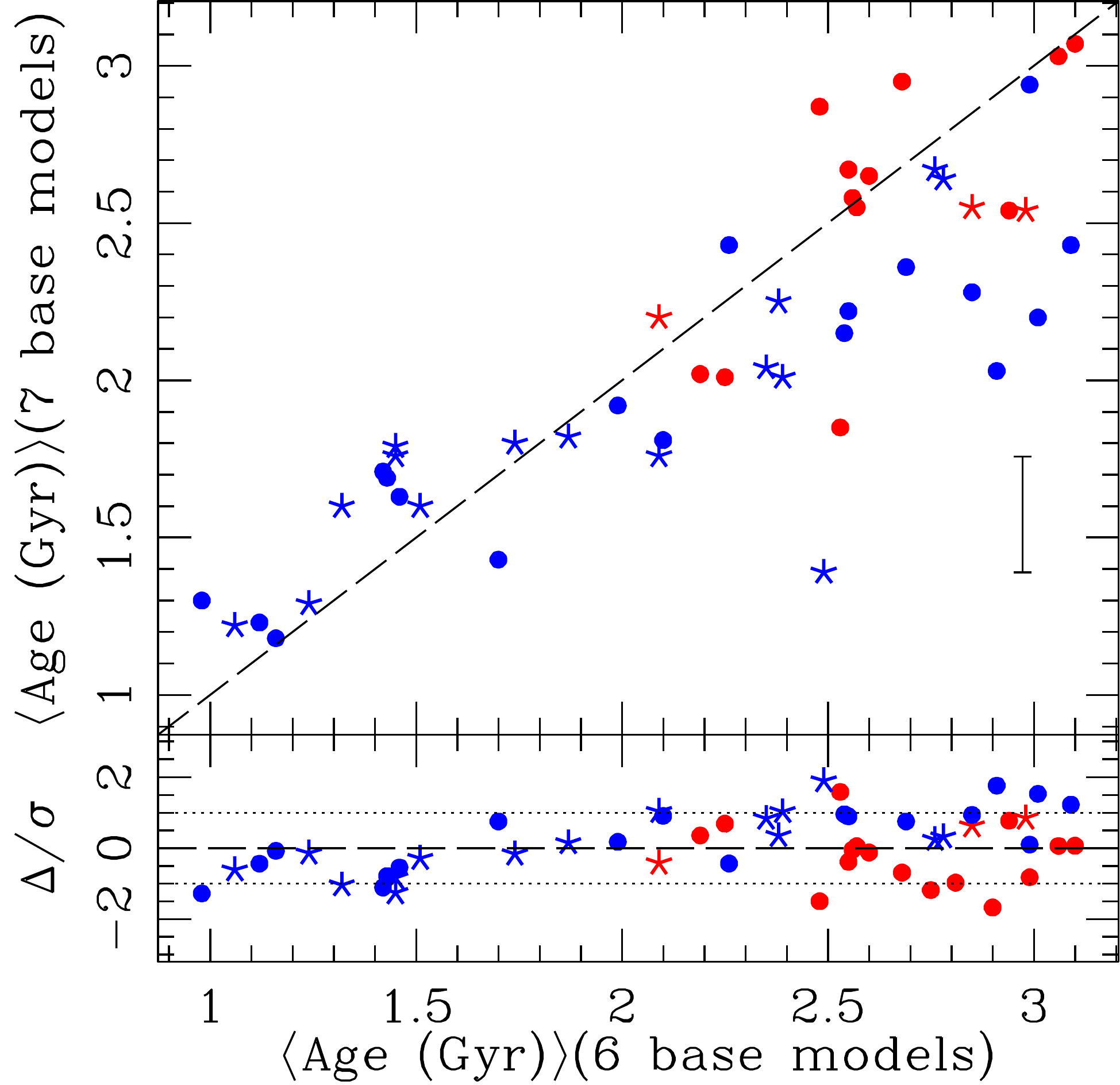}
\includegraphics[angle=90,width=55mm]{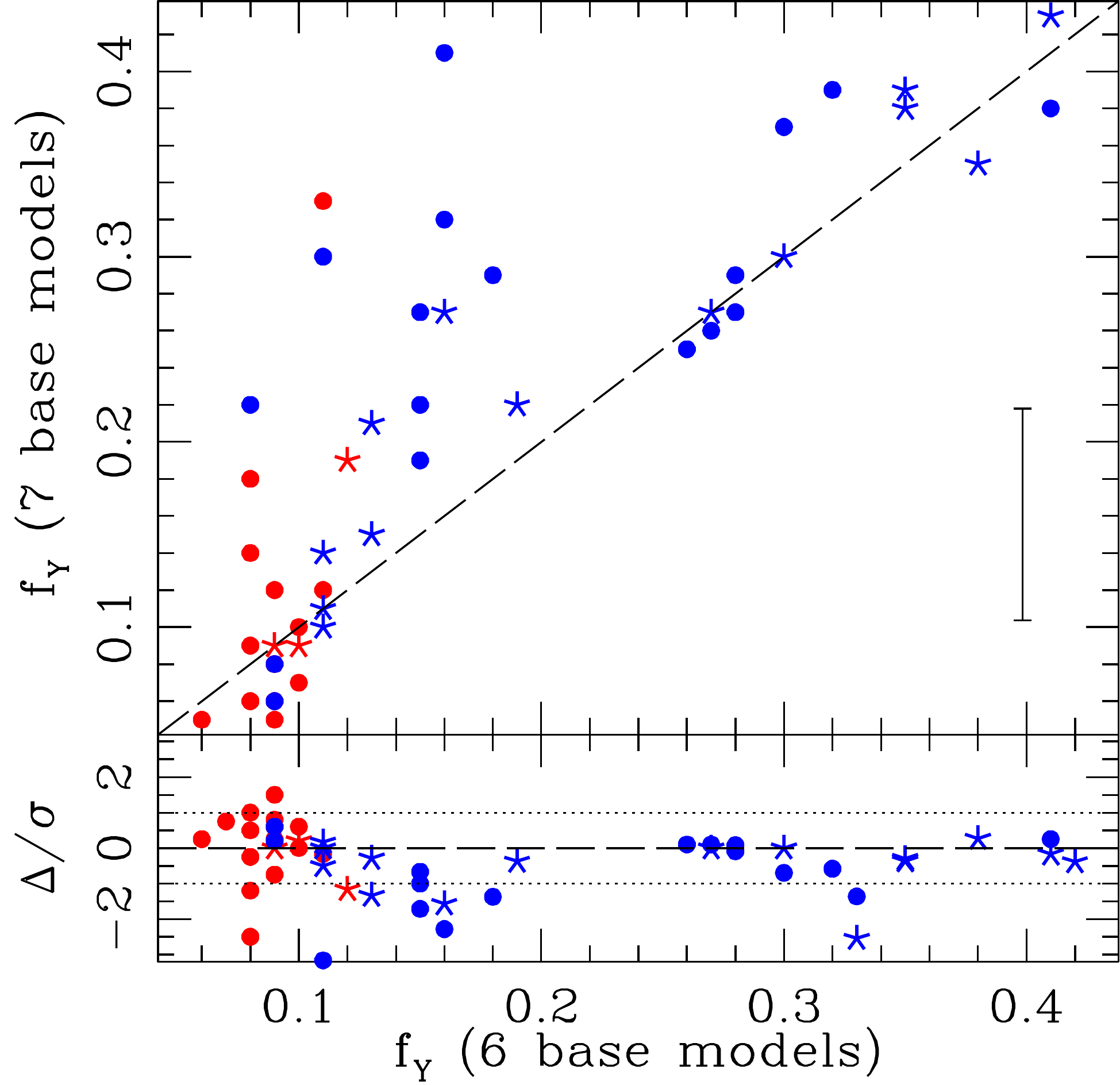}
\includegraphics[angle=90,width=55mm]{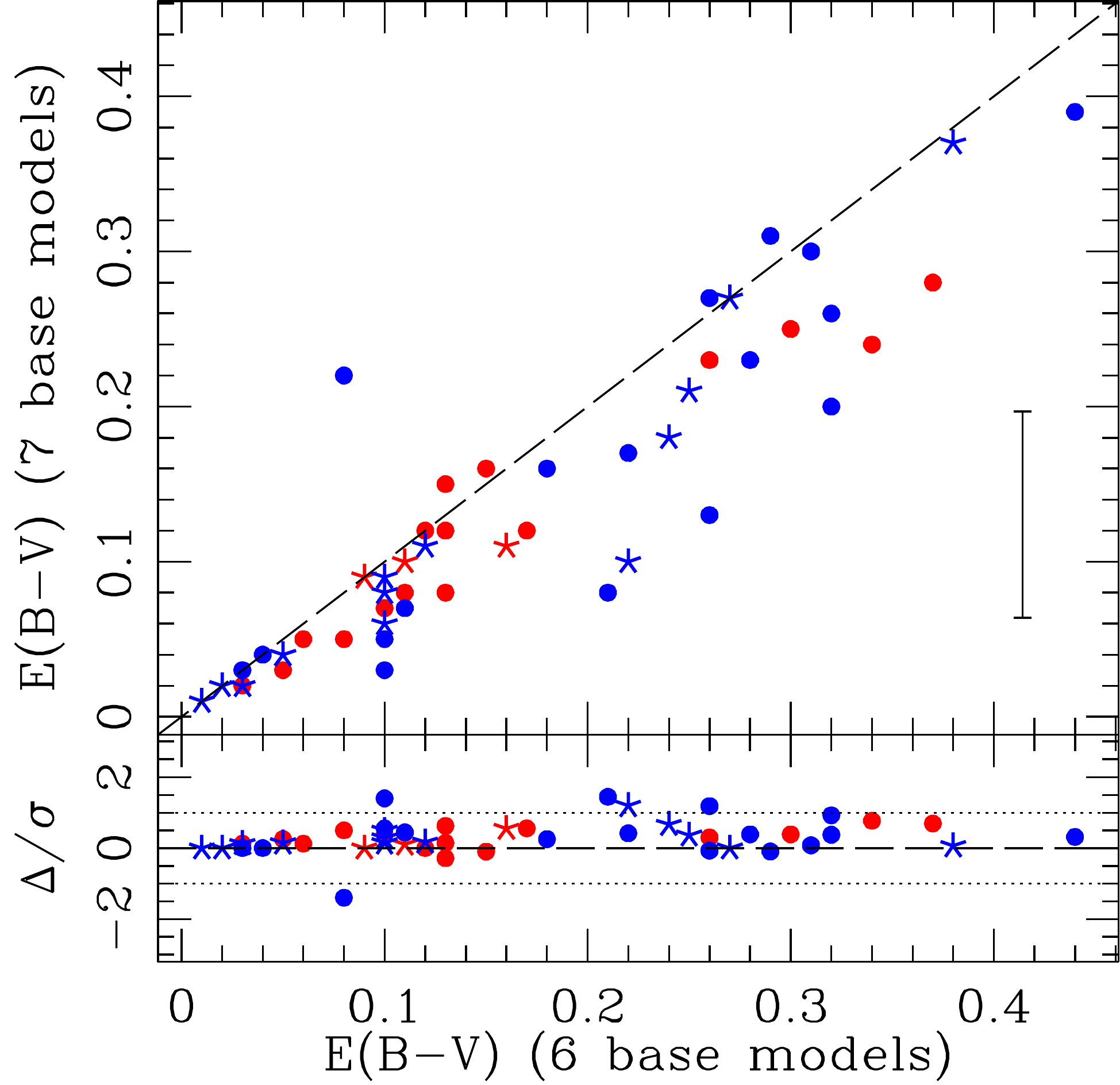}\\
\end{center}
\caption{Comparison of parameter fits by use of two different sets
of models comprising 6 (our fiducial choice) and 7 base models (BM).
For each case, as labelled, we show the parameter from the 7\,BM
(vertical) vs the 6\,BM analysis (horizontal). The bottom subpanels
show the difference ($\Delta$) of a given parameter between the 6\,BM
and the 7\,BM fits, measured as a fraction of the quoted uncertainty
($\sigma$). A typical error bar is shown in each case (it is roughly
the same in both, 6 and 7 base model runs).  The sample is split with
respect to visual morphology, with early-types shown as filled dots
and late-types as star symbols. The sample is colour coded, with red
(blue) galaxies representing quiescent (star-forming) galaxies, as
shown in Fig.~\ref{fig:age}. The panel with the rest-frame colours
 -- labeled (Clr)$_0$ -- include both $(U-V)_0$ and $(V-J)_0$, where the latter -- for the 7\,BM
runs -- is shifted up by 1\,mag to avoid crowding.
}
\label{fig:f6vsf7}
\end{figure*}

In order to assess the robustness of the parameter extraction,
we compared the analysis presented in the paper -- that
combines six base models as presented in Sec.~\ref{Sec:Method}
-- with a new run where one additional base model is included.
The starting procedure is identical to the original method,
performing a trial search that gives a best fit metallicity,
used as reference for base models 1 through 6.
These base models  assume a constant star formation
rate in the following age intervals:
\begin{itemize}
\item New Base Model 1: $\log t/{\rm Gyr}\in [-2,-1]$
\item New Base Model 2: $\log t/{\rm Gyr}\in [-1,-1+\delta]$
\item New Base Model 3: $\log t/{\rm Gyr}\in [-1+\delta,-1+2\delta]$
\item New Base Model 4: $\log t/{\rm Gyr}\in [-1+2\delta,-1+3\delta]$
\item New Base Model 5: $\log t/{\rm Gyr}\in [-1+3\delta,-1+4\delta]$
\item New Base Model 6: $\log t/{\rm Gyr}\in [-1+4\delta,lt_{\rm MAX}]$
\item New Base Model 7: $\log t/{\rm Gyr}\in [-1+4\delta,lt_{\rm MAX}]$,
\end{itemize}
where $lt_{\rm MAX}$ is the log$_{10}$ of the age of the Universe (in Gyr) at
the redshift of the galaxy, and $\delta\equiv (lt_{\rm MAX}+1)/5$.
Analogously to the fiducial case, BM7 is defined over the
same interval as BM6, but corresponding to a metallicity 0.3\,dex
lower than the reference. Fig.~\ref{fig:f6vsf7} shows a comparison
of the best fit models between the 6 base model choice (horizontal axes)
and the 7 base model runs (vertical axes). The bottom subpanel for each
case shows the difference between these two models ($\Delta$), measured in units
of the individual uncertainties ($\sigma$). The figure shows that, within the
quoted error bars, the results are quite robust, especially regarding
average age, oldest age and $\Delta t$. The colour excess parameters
are also quite robust, as are the rest-frame colours adopted when
segregating the sample with respect to star formation activity
(Fig.~\ref{fig:UVJ}). An accurate estimate of the colour excess 
mainly requires good flux calibration, as dust mainly affects the
illumination source as a smooth wavelength-dependent function
\citep[see, e.g.,][]{CCM:89}. In this regard, spectral resolution is
not so important when constraining the colour excess.  Due to its
excellent flat fielding, the slitless grism data provided by the {\sl HST} 
cameras allow us to produce spectra with very accurate flux
calibration \citep[see, e.g.,][]{FIGS}.

%%%%%%     f6 and f7 comparison     %%%%%%%%%%%%
%%%%%%%%%%%%%%%%%%%%%%%%%%%%%%%%%%%%%%%%%%%%%%%%
%%%%%%%%%%%%%%%%%%%%%%%%%%%%%%%%%%%%%%%%%%%%%%%%

%\twocolumn

\label{lastpage}

\end{document}